\documentclass[showpacs,preprintnumbers,nofootinbib,amsmath,amssymb]{revtex4}
\usepackage{hyperref,slashed,color}
\usepackage{graphicx}
\usepackage{ulem}
\usepackage{longtable}
\usepackage{array}

\allowdisplaybreaks
 \def\nx{\overline{\nabla}_x}
 
 \def\pk{\partial_k}
 
 \def\sk{\Sigma_k}
 \def\skp{\Sigma_k'}

 \def\Ft{\tilde{F}}
 \def\at{\tilde{a}}
 \def\gamf{\gamma_5}

 \def\parti{\partial}
 \def\asl{a\!\!\!/\;}
 
 \def\la{\langle}
 \def\ra{\rangle}

 \def\chip{\chi_+}
 
 \def\chim{\chi_-}
 
 \def\fp{f_+}
 \def\fm{f_-}

 \def\asl{\slashed{a}}

 \def\tb{\bar{t}}
 \def\tbt{\tilde{\bar{t}}}

 \def\td{{t^{\dag}}}
 \def\st{\tilde{s}}
 \def\pt{\tilde{p}}
 \def\tp{t_+}
 \def\tm{t_-}

 \def\nt{\overline{\nabla}_t}
 
 \def\atx{a_{t}}

 \def\pUt{\frac{\parti U_t}{\parti t}}
 \def\Utd{U_t^\dag}

 \def\tptx{t_{+,t}}

 \def\tmtx{t_{-,t}}

\begin{document}
\preprint{TUHEP-TH-12176}
\title{Computation of the $p^6$ order low-energy constants with tensor sources}

\bigskip
\author{Shao-Zhou
Jiang$^1$,\footnote{Email:\href{mailto:jsz@gxu.edu.cn}{jsz@gxu.edu.cn}.}
 Ying Zhang$^2$,\footnote{Email:\href{mailto:hepzhy@mail.xjtu.edu.cn}{hepzhy@mail.xjtu.edu.cn}.} Qing Wang$^3$\footnote{Email:
\href{mailto:wangq@mail.tsinghua.edu.cn}{wangq@mail.tsinghua.edu.cn}.}\footnote{corresponding
author}\\~}

\bigskip
\affiliation{$^1$College of Physical Science and Technology, Guangxi University, Nanning, Guangxi 530004, P.R.China\\
$^2$School of Science, Xi'an Jiaotong University, Xi'an 710049, P.R.China\\
$^3$Department of Physics, Tsinghua University, Beijing 100084,
P.R.China\footnote{mailing address}}

 \begin{abstract}

 We present the results of calculations of the $p^4$ and $p^6$ order low-energy constants for
 the chiral Lagrangian with tensor sources for both two and three flavors of pseudoscalar mesons.
 This is a generalization of our previous work on similar calculations without tensor sources
 in terms of the quark self-energy $\Sigma(p^2)$,
  based on the first principle derivation of the
 low-energy effective Lagrangian and computation of the low-energy constants with some rough approximations.
 With the help of partial integration and some epsilon relations,
 we find that some $p^6$ order operators with tensor sources  appearing
 in the literature are related to each other.
 That leaves 98 independent terms for $n$-flavor, 92 terms for three-flavor, and 65 terms for two-flavor cases.
 We also find that the odd-intrinsic-parity chiral Lagrangian with tensor sources cannot independently exist
 in any order of low-energy expansion.
 \end{abstract}
 \maketitle
\section{Introduction}

 In the low-energy region of the strong interaction, conventional perturbation theory is ineffective.
 If we focus on the pseudoscalar mesons ($\pi,K,\eta$), chiral perturbation theory provides us with an effective way to deal with the system.
 It can be applied not only to the  strong interaction, but also to the weak and electromagnetic interactions.
 It was first introduced by Weinberg \cite{weinberg}. The idea is to expand the meson part Lagrangian in terms of powers of external momenta and the quark masses.
  Gasser and Leutwyler \cite{GS1,GS2} then extended it to the $p^4$ order
 and built up the path integral formalism which enables us to compute
 the various Green's functions of the light-quark
 scalar, pseudoscalar, vector, and axial-vector currents in terms of the chiral Lagrangian.
 The formulation was later generalized to the $p^6$ order.
 The form of the normal  part of the $p^6$ order chiral Lagrangian had been obtained \cite{p61,p62,p6p}
 soon after the anomalous  parts were given \cite{p6a1,p6a2}.
 The latest and general review on the topic can be found in \cite{p6r}.
  Missing in this series of work are the antisymmetric tensor currents,
 although this could be partly because of the fact that tensor currents do not appear in the Standard Model Lagrangian,
 as discussed in Ref.\cite{tensor1}. Research on hadron matrix elements
 and the study of interactions beyond the Standard Model may need  these tensor currents.
 Further, antisymmetric tensor currents not only generate the conventional $1^{--}$ vector mesons,
 but also  more exotic $1^{+-}$ mesons.
 Therefore, studies involving both of these and their interactions would bring in
  the antisymmetric currents.
 More importantly, for the structure of the general currents
 $\overline{\psi}\Gamma\psi$, the $4\times 4$
 $\Gamma$ matrices generally have 16 degrees of freedoms,
 and one usually chooses the 16 $\gamma$ matrices
 $1,\gamma_5,\gamma_\mu,\gamma_\mu\gamma_5, \sigma_{\mu\nu}$ to represent these freedoms.
 This implies that $\Gamma$ can be expanded in terms of the 16 $\gamma$ matrices,
 and one is used to calling the currents according to their $\gamma$ matrices' structures.
 Taking just  scalar, pseudoscalar,
 vector and axial-vector currents can not give the most general bilinear light-quark currents because of incompleteness.
 Adding in the tensor currents, we can get the set of the currents completely.
 The results of the Green's functions among the currents would then be  general.
  Six years ago, the form of the chiral Lagrangian involving tensor currents
  had been discussed first in Ref.\cite{tensor1}.
 The results  were the normal parts with tensor sources starting from $p^4$ order,
 and both the $p^4$ and $p^6$ order chiral Lagrangian with tensor sources were obtained. While
 the odd-intrinsic-parity parts with tensor sources were claimed to start from $p^8$ order.
 Based on these results, more progress has been made \cite{r1,r2,r3}.

 Within the chiral perturbation theory,  if the order of the momentum and the current quark masses expansion is increased,
 the number of independent terms rises rapidly. For example, in the three-flavor case,
 the $p^4$ order Lagrangian has 10 terms plus 2 contact terms,  but the $p^6$ order
 has 90 terms plus 3 contact terms. These independent terms generate a large number of unknown low-energy constants (LECs).
 A summary of numerical results for the LECs can be found in \cite{lecs},
 which makes discussions about the high order effects of the chiral Lagrangians even more difficult and complex.
 Compared to dealing with higher order chiral Lagrangians,
 adding tensor sources is relatively less heavy and realistic work is possible.
 Originally,  LECs were fixed via the experimented data.
 Now, because more LECs have appeared for high orders, and sufficient experimental data are lacking,
 we can no longer solely rely on experiment to determine these LECs.
 Calculations of LECs from various models or underlying QCD have subsequently developed and become popular.
 Indeed, not only experimented data, but theoretical calculations are also needed.
  With these calculations, we can check the correctness of the models or the theory.
 As members in the community of calculating LECs,
 we have calculated the $p^4$ order LECs in \cite{WQ1,oura1},
 and then the $p^6$ order calculations in \cite{our5,oura2}, including the normal and the anomalous parts,
 for two- and three-flavors cases.
 In this paper, we will extend our work to the tensor sources, based on the first principle derivation of the
 low-energy effective Lagrangian \cite{WQ0} and computation of the LECs with some rough
 approximations\footnote{The detailed approximations in these computations of LECs are:
 taking the ladder approximation,  modeling the low-energy behavior of the gluon propagator, neglecting angle dependence in the running coupling constant in the kernel of the Schwinger-Dyson equation for the quark self-energy,
 assuming the ansatz solution for the external source dependent Schwinger-Dyson equation
 in terms of the quark self-energy, and taking the large $N_c$ limit.
 The final effective action before taking the momentum expansion was shown to be equivalent to the result of
 a phenomenological, gauge invariant, nonlocal, and dynamical(GND) quark model\cite{GND}.}\cite{WQ1}
 and  present all  LECs up to the $p^6$ order\cite{zhanghh}.

 In Euclidean space, the chiral Lagrangian includes the real and the imaginary parts.
 The real part is related to the even-intrinsic-parity sector,
 and the imaginary part is related to the odd-intrinsic-parity sector.
  As the tensor source terms always appear with $\sigma_{\mu\nu}$,
 we can use the following Eq.(\ref{sigf}) to interchange even- and odd-intrinsic-parity sectors.
 \begin{eqnarray}
 \sigma^{\mu\nu}\gamma_5=\frac{i}{2}\epsilon^{\mu\nu\lambda\rho}\sigma_{\lambda\rho}\;,\label{sigf}
 \end{eqnarray}
 It implies that when calculating the tensor source parts, one needs to include both real and imaginary parts\cite{ball}
 \footnote{In a private communication, Y.-L. Ma in Ref.\cite{oura1} had already obtained similar but unpublished result in 2003.}.

 This paper is organized as follows: In Sec. \ref{cal},
  we review our previous calculations on the real part of the chiral Lagrangian from $p^2$ to $p^6$ order and add in the tensor sources.
  In Sec. \ref{cala}, we review our previous calculations on the imaginary part of the chiral Lagrangian up to $p^6$ order and add the tensor sources.
  In Sec. \ref{diff},
  we collect the differences in convention between our paper and Ref.\cite{tensor1} and discuss the possible dependent operators.
  In Sec. \ref{p4res}, we give  our $p^4$ order results with tensor sources, and Sec. \ref{p6res} presents our $p^6$ order results.
 Sec. \ref{summa} is a summary.

\section{Calculations involving the real part of the chiral Lagrangian for order $p^2$ up to $p^6$ with tensor sources}\label{cal}

 We have calculated the real part of chiral Lagrangian from $p^2$ to $p^6$ order without tensor sources in Ref.\cite{our5}.
 Using the same method, we can also deal with the tensor source part.
 For convenience, we give a short introduction here, while adding in these external tensor sources.

 The difference from Ref.\cite{our5} is the external tensor sources $\tb^{\mu\nu}$.
 Adding them in the original external sources $v^{\mu},a^{\mu},s,p$ denoted as scalar, pseudoscalar, vector and axial-vector sources,
 respectively, we get the complete sources set as follows
 \begin{eqnarray}
 J=\slashed{v}+\slashed{a}\gamma_5-s+ip\gamma_5+\sigma_{\mu\nu}\tb^{\mu\nu}.
 \end{eqnarray}
 From the original QCD, the Lagrangian can be written as the QCD Lagrangian, $\mathcal{L}_{\mathrm{QCD}}^0$,
 plus the external sources part,
 and the generating functional reads
 \begin{eqnarray}
 Z[J]&=&\int\mathcal{D}\psi\mathcal{D}\bar\psi\mathcal{D}\Psi\mathcal{D}\bar\Psi\mathcal{D}A_\mu
 \exp\left\{i\int d^4x[\mathcal{L}_{\mathrm{QCD}}^0+\bar\psi J\psi]\right\}\notag\\
 &=&\int\mathcal{D}U\exp\left\{i\int d^4x\mathcal{L}_{\mathrm{ChPT}}(U,J)\right\}
 =\int\mathcal{D}Ue^{iS_{\mathrm{eff}}},\label{effa}
 \end{eqnarray}
 where $\psi,\Psi$ and $A_\mu$ are light-quark, heavy-quark, and gluon fields, respectively;
 $U$ is the pseudoscalar meson field;
 $\mathcal{L}_{\mathrm{ChPT}}$ is the chiral Lagrangian; and $S_\mathrm{eff}$ is the effective action.
 Because this form of the chiral Lagrangian is explicitly $U$ dependent at the high momentum orders,
 and is hard to investigate \cite{p61,p62} due to its complex $U$ and $J$ structures,
 we are used to making the chiral rotation to simplify the Lagrangian as \cite{WQ0,WQ1,our5}
 \begin{eqnarray}
 &&J_{\Omega}=[\Omega P_R+\Omega^{\dag}P_L][J+i\slashed{\partial}][\Omega P_R+\Omega^\dag P_L]
 =\slashed{v}_\Omega+\slashed{a}_\Omega\gamma_5-s_\Omega+ip_\Omega\gamma_5+\sigma_{\mu\nu}\tb_\Omega^{\mu\nu},\label{JOmega}\\
 &&U=\Omega^2,\hspace{1cm}P_L=\frac{1-\gamf}{2},\hspace{1cm}P_R=\frac{1+\gamf}{2}\;.
 \end{eqnarray}
 To separate the tensor sources into  even and odd parities, $t_{\pm}^{\mu\nu}$,
 we need the following tensor chiral projectors as Ref.\cite{tensor1}
 \begin{eqnarray}
 P_R^{\mu\nu\lambda\rho}&=&\frac{1}{4}(g^{\mu\lambda}g^{\nu\rho}-g^{\nu\lambda}g^{\mu\rho}+i\epsilon^{\mu\nu\lambda\rho}),\\
 P_L^{\mu\nu\lambda\rho}&=&(P_R^{\mu\nu\lambda\rho})^\dag=\frac{1}{4}(g^{\mu\lambda}g^{\nu\rho}-g^{\nu\lambda}g^{\mu\rho}-i\epsilon^{\mu\nu\lambda\rho}),\\
 t^{\mu\nu}&=&\frac{1}{2}(\tp^{\mu\nu}+\tm^{\mu\nu}),\hspace{1cm}t^{\dag\mu\nu}=\frac{1}{2}(\tp^{\mu\nu}-\tm^{\mu\nu}),\\
 \tb^{\mu\nu}&=&P_L^{\mu\nu\lambda\rho}t_{\lambda\rho}+P_R^{\mu\nu\lambda\rho}\td_{\lambda\rho},\\
 \sigma_{\mu\nu}\tb^{\mu\nu}&=&\frac{1}{2}\sigma_{\mu\nu}\tp^{\mu\nu}-\frac{i}{4}\sigma_{\mu\nu}\epsilon^{\mu\nu\lambda\rho}t_{-,\lambda\rho}
 =\frac{1}{2}\sigma_{\mu\nu}(\tp^{\mu\nu}-\tm^{\mu\nu}\gamf)=\sigma_{\mu\nu}\tb'^{\mu\nu},\label{sigtb}\\
 \tb'^{\mu\nu}&\equiv&\frac{1}{2}(\tp^{\mu\nu}-\tm^{\mu\nu}\gamf).\label{sigr}
 \end{eqnarray}
 To obtain Eq.(\ref{sigtb}), we have used the $\gamma$-matrix identity Eq.(\ref{sigf}),
 and introduced $\tb'^{\mu\nu}$ for calculational convenience.
 After this operation, our definitions have the simple relations as found in \cite{p62,tensor1} (see Appendix \ref{symrel}).
 Using the same method as presented in \cite{WQ0,our5},
 we can obtain the effective action $S_{\mathrm{eff}}$ introduced in Eq.(\ref{effa}) from the first principle of QCD,
 \begin{eqnarray}
 S_\mathrm{eff}&=&-iN_c\mathrm{Tr}\ln[i\slashed{\partial}+J_{\Omega}-\Pi_{\Omega c}]
 +iN_c\mathrm{Tr}\ln[i\slashed{\partial}+J_{\Omega}]-iN_c\mathrm{Tr}\ln[i\slashed{\partial}+J]
 +N_c\mathrm{Tr}[\Phi_{\Omega c}\Pi^T_{\Omega c}]\label{Seff0}\\
 &&+N_c\sum^{\infty}_{n=2}{\int}d^{4}x_1\cdots
 d^4x_{n}'\frac{(-i)^{n}(N_c g_s^2)^{n-1}}{n!}\bar{G}^{\sigma_1\cdots\sigma_n}_{\rho_1 \cdots\rho_n}(x_1,x'_1,\cdots,x_n,x'_n)
 \Phi^{\sigma_1\rho_1}_{\Omega  c}(x_1 ,x'_1)\cdots \Phi^{\sigma_n\rho_n}_{\Omega c}(x_n,x'_n)+O(\frac{1}{N_c}).\nonumber
 \end{eqnarray}
 Eq.(\ref{Seff0}) is the same as Eq.(1) in \cite{our5},
 but $J_{\Omega}$ (the external source $J$, including currents and
 densities after Goldstone fields dependent chiral rotation $\Omega$) includes the tensor sources.
 $\Phi_{\Omega c}$ and $\Pi_{\Omega c}$ are, respectively, the two-point rotated quark
 Green's function and the interaction part of the two-point rotated quark
 vertices in the presence of the external sources; $\Pi_{\Omega c}$ is defined by
 \begin{eqnarray}
 \Phi_{\Omega c}^{\sigma\rho}(x,y)\equiv
 \frac{1}{N_c}\langle\overline{\psi}_{\Omega}^{\sigma}(x)\psi_{\Omega}^{\rho}(y)\rangle
 =-i[(i\slashed{\partial}+J_{\Omega}-\Pi_{\Omega c})^{-1}]^{\rho\sigma}(y,x),\hspace{1cm}
 \psi_{\Omega}(x)\equiv[\Omega(x)P_L+\Omega^\dag(x)P_R]\psi(x),\label{PhiPi}
 \end{eqnarray}
 with subscript $_c$ denoting the corresponding classical field.
 $\bar{G}^{\sigma_1\cdots\sigma_n}_{\rho_1 \cdots\rho_n}(x_1,x'_1,\cdots,x_n,x'_n)$ is the effective gluon $n$-point
 Green's function including gluon and heavy-quark contributions and $g_s$ is the strong coupling constant of QCD.
 Note that the last two terms in the
 rhs. of Eq.(\ref{Seff0}) can be shown to be independent of the
 pseudoscalar meson field $U$ or $\Omega$ and therefore are just
 irrelevant constants in the effective action, while the second and
 third terms represent the variations of
 the path integral measure of the light-quark field $\psi$. The remaining
 first term  relies on
 $\Pi_{\Omega c}$. $\Phi_{\Omega c}$ and $\Pi_{\Omega c}$ are
 related by the first equation of (\ref{PhiPi}) and determined by
 \begin{eqnarray}
 &&[\Phi_{\Omega c}+\tilde{\Xi}]^{\sigma\rho}+\sum^{\infty}_{n=1}{\int}d^4x_1d^4x'_1\cdots{d^4}x_n
 d^4x'_n\frac{(-i)^{n+1}(N_c g_s^2)^n}{n!}\overline{G}^{\sigma\sigma_1\cdots\sigma_n}_{\rho\rho_1
 \cdots\rho_n}(x,y,x_1,x'_1,\cdots,x_n,x'_n)\nonumber\\
 &&\times \Phi_{\Omega c}^{\sigma_1\rho_1}(x_1 ,x'_1)\cdots
 \Phi_{\Omega c}^{\sigma_n\rho_n}(x_n,x'_n)=O(\frac{1}{N_c}),\label{SDE} \label{fineqNc}
 \end{eqnarray}
 where $\tilde{\Xi}$ is a Lagrangian multiplier which ensures the
 constraint $\mathrm{tr}_l[\gamma_5\Phi_{\Omega c}^T(x,x)]=0$.
 Eq.(\ref{fineqNc}) is the Schwinger-Dyson equation (SDE) in the
 presence of the rotated external source. In Ref.\cite{WQ1}, we have
 assumed the ansatz solution of (\ref{fineqNc}) to be in the approximate form
 \begin{eqnarray}
 \Pi^{\sigma\rho}_{\Omega c}(x,y)=
 [\Sigma(\overline{\nabla}^2_x)]^{\sigma\rho}\delta^4(x-y)\hspace{3cm}
 \overline{\nabla}^{\mu}_x=\partial^{\mu}_x-iv_{\Omega}^{\mu}(x)\;,
 \end{eqnarray}
 where $\Sigma$ is the quark self-energy which satisfies the SDE
 (\ref{SDE}) with vanishing rotated external source. Under the ladder
 approximation, this SDE in Euclidean space-time is reduced to
 the standard form of
 \begin{eqnarray}
 \Sigma(p^2)-3C_2(R)\int\frac{d^4q}{4\pi^3}
 \frac{\alpha_s[(p-q)^2]}{(p-q)^2}
 \frac{\Sigma(q^2)}{q^2+\Sigma^2(q^2)}=0\;, \label{eq0}
 \end{eqnarray}
 where $C_2(R)$ is the second order Casimir operator of the quark
 representation $R$. In our case, quarks belong the $SU(N_c)$ fundamental
 representation; therefore $C_2(R)=(N_c^2-1)/2N_c$, and in the large $N_c$ limit,
 we will neglect the second term of it. $\alpha_s(p^2)$ is the
 running coupling constant of QCD which depends on $N_c$ and the number of quark flavors. With
 these approximations, the action (\ref{Seff0}) of the chiral Lagrangian becomes
 \begin{eqnarray}
 S_\mathrm{eff}\approx-iN_c\mathrm{Tr}\ln[i\slashed{\partial}+J_{\Omega}-\Sigma(\nx^2)]
 +iN_c\mathrm{Tr}\ln[i\slashed{\partial}+J_{\Omega}]-iN_c\mathrm{Tr}\ln[i\slashed{\partial}+J]+O(\frac{1}{N_c})\;.~~~~~~
 \label{Seff1}
 \end{eqnarray}
We have proved that
 the second and the third terms in (\ref{Seff1}) only provide the contribution correlating to Wess-Zumino-Witten term.
 In the large $N_c$ limit, if we do not focus on Wess-Zumino-Witten terms, we can neglect them.
 \begin{eqnarray}
 S_\mathrm{eff}\approx-iN_c\mathrm{Tr}\ln[i\slashed{\partial}+J_{\Omega}-\Sigma(\nx^2)]\label{Seff2}
 \end{eqnarray}

 Because in Minkowski space, it is not convenient to perform the computation,
 we perform the Wick rotation to change Eq.(\ref{Seff1}) to Euclidean space,
 with the metric tensor $g^{\mu\nu}=\mbox{diag}(1,1,1,1)$;
  \begin{eqnarray}
 &&x^0|_M\to-ix^4|_E,\hspace{1cm} x^i|_M\to x^i|_M,\notag\\
 &&\gamma^0|_M\to\gamma^4|_E,\hspace{1cm}\gamma^i|_M\to i\gamma^i|_E,\hspace{1cm}\gamma_5|_M\to\gamma_5|_E,\notag\\
 &&s_\Omega|_M\to-s_\Omega|_E,\hspace{1cm} p_\Omega|_M\to-p_\Omega|_E,\notag\\
 &&\tb_{\Omega,00}|_M\to-\tb_{\Omega,44}|_E,\hspace{1cm}\tb_{\Omega,ij}|_M\to\tb_{\Omega,ij}|_E,\hspace{1cm}
 \tb_{\Omega,0i}|_M\to i\tb_{\Omega,4i}|_E,\hspace{1cm}\tb_{\Omega,i0}|_M\to i\tb_{\Omega,i4}|_E.
 \end{eqnarray}
 Here $v^\mu_\Omega, a^\mu_\Omega$ transform as $x^\mu$,
  whereas $\tb_{\Omega,\mu\nu}$ are considered as (axial-)vector-(axial-)vector combined.
 Eq.(\ref{Seff1}) in Euclidean space is
 \begin{eqnarray}
 S_\mathrm{eff}&\approx& N_c\mathrm{Tr}\ln[\slashed{\partial}+J_{\Omega,E}+\Sigma(-\nx^2)]\notag\\
 &=&N_c\mathrm{Tr}\ln[\slashed{\partial}-i\slashed{v}_\Omega-i\slashed{a}_\Omega\gamma_5-s_\Omega+ip_\Omega\gamma_5
 +\sigma_{\mu\nu}\tb'^{\dag\mu\nu}_\Omega+\Sigma(-\nx^2)]\label{Seff2E}
 \end{eqnarray}
 With the help of the Schwinger proper time method \cite{det0},
 the real part (or, equivalently, the even-intrinsic-parity part or the normal part) of  Trln$[\cdots]$ in Euclidean space-time can be written as
 \begin{eqnarray}
 &&\hspace{-0.5cm}\mathrm{ReTr}\ln[\slashed{\partial}-i\slashed{v}_\Omega-i\slashed{a}_\Omega\gamma_5-s_\Omega+ip_\Omega\gamma_5
 +\sigma_{\mu\nu}\tb'^{\dag\mu\nu}_\Omega+\Sigma(-\nx^2)] \nonumber\\
 &\equiv&\mathrm{ReTr}\ln[D-\sigma_{\mu\nu}\tb'^{\mu\nu}_\Omega+\Sigma(-\nx^2)] \nonumber\\
 &=&\frac{1}{2}\mathrm{Tr}\ln\Big[[D^{\dagger}+\sigma_{\mu\nu}\tb'^{\dag\mu\nu}_\Omega+\Sigma(-\nx^2)]
 [D+\sigma_{\mu\nu}\tb'^{\mu\nu}_\Omega+\Sigma(-\nx^2)]\big]\notag\\
 &=&\frac{1}{2}\mathrm{Tr}\ln[O+N]\notag\\
 &=&-\frac{1}{2}\lim_{\Lambda \to \infty}
 \int^\infty_{\frac{1}{\Lambda^2}}\frac{d\tau}{\tau}\mathrm{Tr}e^{-\tau(O+N)} (\mbox{remove const term})\nonumber\\
 &=&-\frac{1}{2}\lim_{\Lambda \to \infty}
 \int^\infty_{\frac{1}{\Lambda^2}}\frac{d\tau}{\tau}\int{d^4x}\
 \mathrm{tr}_f\langle x|e^{-\tau(O+N)}|x\rangle\label{Trln}\\
 &&D\equiv\slashed{\partial}-i\slashed{v}_\Omega-i\slashed{a}_\Omega\gamma_5-s_\Omega+ip_\Omega\gamma_5.
 \end{eqnarray}
 Where a cutoff $\Lambda$ is introduced into the theory to regularize the possible ultraviolet divergences.
 $O$ is the old operator without tensor sources in \cite{our5}, and $N$ is the new operator with tensor sources
 \begin{eqnarray}
 O&=&[D^{\dagger}+\Sigma(-\nx^2)][D+\Sigma(-\nx^2)],\\
 N&=&-\nx^{\lambda}\tb^{\mu\nu}_\Omega\gamma^{\lambda}\sigma^{\mu\nu}+\tb^{\dag\mu\nu}_\Omega\nx^{\lambda}\sigma^{\mu\nu}\gamma^{\lambda}
 -i\asl_\Omega\tb^{\mu\nu}_\Omega\sigma^{\mu\nu}\gamf-i\tb^{\dag\mu\nu}_\Omega\sigma^{\mu\nu}\asl_\Omega\gamf
 -s_\Omega\tb^{\mu\nu}_\Omega\sigma^{\mu\nu}-\tb^{\dag\mu\nu}_\Omega s_\Omega\sigma^{\mu\nu}\notag\\
 &&-ip_\Omega\tb^{\mu\nu}_\Omega\sigma^{\mu\nu}\gamf+i\tb^{\dag\mu\nu}_\Omega p_\Omega\sigma^{\mu\nu}\gamf
 +\Sigma(-\nx^2)\tb^{\mu\nu}_\Omega\sigma^{\mu\nu}+\tb^{\dag\mu\nu}_\Omega\Sigma(-\nx^2)\sigma^{\mu\nu}
 +\tb^{\dag\mu\nu}_\Omega\tb^{\lambda\rho}_\Omega\sigma_{\mu\nu}\sigma_{\lambda\rho}.
 \end{eqnarray}

 If we calculate Eq.(\ref{Trln}) directly, it is not explicitly chiral covariant for each term in the calculation.
 In order to recover the chiral covariant form to get the LECs,
 we would need to collect the relevant terms together by hand, which consumes
 too much time.  Fortunately,
 we found a method of keeping the chiral covariance at each step in the low-energy expansion computation\cite{covariant2},
 and we used it successfully to obtain the  LECs of the real part without tensor sources\cite{our5}.
 We introduce it here briefly. Use the relations
 \begin{eqnarray}
 k^\mu+i\bar{\nabla}_x^\mu&=&e^{i\bar{\nabla}_{x}\cdot\frac{\partial}{\partial k}}
 \Big(k^\mu+\tilde{F}^\mu(\bar{\nabla},\frac{\partial}{\partial k})\Big)
 e^{-i\bar{\nabla}_{x}\cdot\frac{\partial}{\partial k}}\;,\label{differential}\\
 \Ft^{\mu}&=&\frac{1}{2} [\nx^{\nu},\nx^{\mu}]\pk^{\nu}
 -\frac{i}{3} [\nx^{\lambda},[\nx^{\nu},\nx^{\mu}]]\pk^{\lambda}\pk^{\nu}
 -\frac{1}{8} [\nx^{\rho},[\nx^{\lambda},[\nx^{\nu},\nx^{\mu}]]]
 \pk^{\rho}\pk^{\lambda}\pk^{\nu}\notag\\
 &&+\frac{i}{30}[\nx^\sigma,[\nx^{\rho},[\nx^{\lambda},[\nx^{\nu},\nx^{\mu}]]]]
 \pk^\sigma\pk^{\rho}\pk^{\lambda}\pk^{\nu}\notag\\
 &&+\frac{1}{144}[\nx^\delta,[\nx^\sigma,[\nx^{\rho},[\nx^{\lambda},[\nx^{\nu},\nx^{\mu}]]]]]
 \pk^\delta\pk^\sigma\pk^{\rho}\pk^{\lambda}\pk^{\nu}+O(p^7).
 \end{eqnarray}
 Substituting (\ref{differential}) into the integrand in (\ref{Trln}), we change it to
  \begin{eqnarray}
 &&\mathrm{tr}_f\langle x|e^{-\tau(O(i\nx)+N(i\nx))}|x\rangle \nonumber\\
 &&=\mathrm{tr}_f\int\frac{d^4k}{(2\pi)^4}\int\frac{d^4k'}{(2\pi)^4}e^{ik'\cdot x}\la k'|e^{-\tau(O(i\nx)+N(i\nx))}|k\ra e^{-ik\cdot x}\notag\\
 &&=\mathrm{tr}_f\int\frac{d^4k}{(2\pi)^4}e^{-\tau(O(k+i\nx)+N(k+i\nx))}\notag\\
 &&=\mathrm{tr}_f\int\frac{d^4k}{(2\pi)^4}e^{-\tau(e^{i\bar{\nabla}_{x}\cdot\frac{\partial}{\partial k}}\tilde{O}(k+i\nx)
 e^{-i\bar{\nabla}_{x}\cdot\frac{\partial}{\partial k}}
 +e^{i\bar{\nabla}_{x}\cdot\frac{\partial}{\partial k}}\tilde{N}(k+i\nx)
 e^{-i\bar{\nabla}_{x}\cdot\frac{\partial}{\partial k}})}\notag\\
 &&=\mathrm{tr}_f\int\frac{d^4k}{(2\pi)^4}e^{i\bar{\nabla}_{x}\cdot\frac{\partial}{\partial k}}e^{-\tau(\tilde{O}(k+i\nx)
 +\tilde{N}(k+i\nx))}e^{-i\bar{\nabla}_{x}\cdot\frac{\partial}{\partial k}}\notag\\
 &&=\mathrm{tr}_f\int\frac{d^4k}{(2\pi)^4}e^{-\tau(\tilde{O}(k+i\nx)
 +\tilde{N}(k+i\nx))}.\label{dB0}
 \end{eqnarray}
 where $\tilde{O}$ is the original exponent without tensor sources, which can be found in Eqs.(14) and (15) in Ref.\cite{our5},
 and $\tilde{N}$ is the new operator with tensor sources. We have
 \begin{eqnarray}
 \tilde{N}&=&i(k^{\lambda}+\Ft^{\lambda})\tbt^{\mu\nu}_\Omega\gamma^{\lambda}\sigma^{\mu\nu}
 -i\tbt^{\dag\mu\nu}_\Omega(k^{\lambda}+\Ft^{\lambda})\sigma^{\mu\nu}\gamma^{\lambda}
 -i\at^{\lambda}_\Omega\tbt^{\mu\nu}\gamma^{\lambda}\sigma^{\mu\nu}_\Omega\gamf
 -i\tbt^{\dag\mu\nu}_\Omega\at^{\lambda}_\Omega\sigma^{\mu\nu}\gamma^{\lambda}\gamf
 -\st_\Omega\tbt^{\mu\nu}_\Omega\sigma^{\mu\nu}-\tbt^{\dag\mu\nu}_\Omega\st_\Omega\sigma^{\mu\nu}\notag\\
 &&-i\pt_\Omega\tbt^{\mu\nu}_\Omega\sigma^{\mu\nu}\gamf+i\tbt^{\dag\mu\nu}_\Omega\pt_\Omega\sigma^{\mu\nu}\gamf
 +\Sigma([k^{\mu}+\Ft^{\mu}]^2)\tbt^{\mu\nu}_\Omega\sigma^{\mu\nu}+\tbt^{\dag\mu\nu}_\Omega\Sigma([k^{\mu}+\Ft^{\mu}]^2)\sigma^{\mu\nu}
 +\tbt^{\dag\mu\nu}_\Omega\tbt^{\lambda\rho}_\Omega\sigma_{\mu\nu}\sigma_{\lambda\rho},\\
 \mathcal{\tilde O}&=&\mathcal{O}-i [\nx^{\nu},\mathcal{O}]\pk^{\nu}
 -\frac{1}{2} [\nx^{\lambda},[\nx^{\nu},\mathcal{O}]]\pk^{\nu}\pk^{\lambda}
 +\frac{i}{6} [\nx^{\rho},[\nx^{\lambda},[\nx^{\nu},\mathcal{O}]]]\pk^{\nu}\pk^{\lambda}\pk^{\rho}\notag\\
 &&+\frac{1}{24}[\nx^\sigma,[\nx^{\rho},[\nx^{\lambda},[\nx^{\nu},\mathcal{O}]]]]
 \pk^{\sigma}\pk^{\nu}\pk^{\lambda}\pk^{\rho}
 -\frac{i}{120}[\nx^\delta,[\nx^\sigma,[\nx^{\rho},[\nx^{\lambda},[\nx^{\nu},\mathcal{O}]]]]]
 \pk^\delta\pk^{\sigma}\pk^{\nu}\pk^{\lambda}\pk^{\rho}+O(p^7).
 \end{eqnarray}
 where $\mathcal{\tilde O}\equiv(\tilde{a}^\mu,\tilde{s},\tilde{p},\tbt^{\alpha\beta})^T$ and
 $\mathcal{O}\equiv(a^\mu_\Omega,s_\Omega,p_\Omega,\tb'^{\alpha\beta}_\Omega)^T$.
 With Eqs.(\ref{Trln}) and (\ref{dB0}), we get
 \begin{eqnarray}
 \mathrm{ReTr}\ln[\slashed{\partial}-i\slashed{v}_\Omega-i\slashed{a}_\Omega\gamma_5
 -s_\Omega+ip_\Omega\gamma_5+\sigma_{\mu\nu}\tb^{\mu\nu}_\Omega+\Sigma(-\bar{\nabla}^2)] =-\frac{1}{2}\lim_{\Lambda \to \infty}
 \int^\infty_{\frac{1}{\Lambda^2}}\frac{d\tau}{\tau}\int{d^4x}\int\frac{d^4k}{(2\pi)^4}\mathrm{tr}~e^{B+B_t}\cdot 1\;.
 ~~~~~\label{dB1}
 \end{eqnarray}
 $B$ can be found in Eq.(17) of Ref.\cite{our5}, and $B_t=-\tau\tilde{N}$.
  Expanding Eq.(\ref{dB1}) in powers of momentum,
 theoretically, we can get all orders of the chiral Lagrangian.
 Before giving our result, we need to discuss the difference between our paper and Ref.\cite{tensor1};
 this will be done in Sec. \ref{diff}.

\section{Calculations involving the imaginary part of the chiral Lagrangian for order $p^2$ up to $p^6$ with tensor sources}\label{cala}

 Because of Eq.(\ref{sigf}) or, equivalently (\ref{epsilon4}) below in the next section,
 all the odd-intrinsic-parity sectors of the chiral Lagrangian can be changed to the
 even-intrinsic-parity sectors. If one keeps $t_+$ and $t_-$ explicit, then the odd-intrinsic-parity sector is
 redundant and can be shown to be equivalent to even-intrinsic-parity operators.
 If instead only one of the two ($t_+$ or $t_-$) is used, then odd-intrinsic-parity operators are present.
 In Euclidean space, the odd-intrinsic-parity sectors belong to the imaginary parts.
 In other words,  in Euclidean space, we can interchange the imaginary and real tensor source dependent parts.
 But in this section, we also call the parts without using Eq.(\ref{sigf}) imaginary parts.
 Now we deal with the imaginary parts of Eq.(\ref{Seff2E}) as a compensation of real parts discussion in Sec.\ref{cal}. We have calculated the $p^6$ order imaginary part's LECs without tensor sources in Ref.\cite{oura2}.
 Using the same method, we can also  compute the contributions involving the tensor source part. We repeat the process here,
 adding the tensor source.

 First, to confirm  the Wess-Zumino-Witten terms, we need to introduce a fifth-dimension integral.
 We now write $\Omega$, in Eq.(\ref{JOmega}), as $\Omega=e^{-i\beta}$
 and further introduce a parameter $t$ dependent rotation element $\Omega(t)=e^{-i\beta t}$.
 With the help of relations $\Omega(1)=\Omega$ and $\Omega(0)=1$, Eq.(\ref{Seff2E}) becomes
 \begin{eqnarray}
 S_\mathrm{eff}[U(t),J(t)]\approx N_c\mathrm{Tr}\ln[\slashed{\partial}+J_{\Omega(t)}+\Sigma(-\overline{\nabla}_t^2)]_{t=1},\label{Seff1Ea}
 \end{eqnarray}
 with $\overline{\nabla}_t^\mu=\partial^\mu-iv_{\Omega(t)}^\mu$.
 $J_{\Omega(t)}$ is $J_{\Omega}$ with $\Omega$ replaced by $\Omega(t)$.

 Second, because we only focus on the meson terms,
 adding in an extra pure source makes no sense for the results.
 We insert an extra pure source term, setting $t=0$ in (\ref{Seff1Ea}), with the help of
 \begin{eqnarray}
 \frac{\partial J_{\Omega(t)}}{\partial t}=\frac{1}{2}\bigg[\frac{\partial U}{\partial t}U^\dag_t\gamf
 ,\slashed{\partial}+J_{\Omega(t)}\bigg]_+,\hspace{0.5cm}U_t=\Omega^2(t),
 \end{eqnarray}
 we can further proceed to express the chiral Lagrangian in terms of an integration over the parameter $t$.
 \begin{eqnarray}
 S_\mathrm{eff}[U(t),J(t)]&=& N_c\mathrm{Tr}\ln[\slashed{\partial}+J_{\Omega(t)}+\Sigma(-\overline{\nabla}_t^2)]_{t=0,{\Sigma~\mbox{\tiny dependent}}}^{t=1}\notag\\
 &=&N_c\int_0^1dt~\frac{d}{dt}\mathrm{Tr}\ln[\slashed{\partial}+J_{\Omega(t)}+\Sigma(-\nabla_t^2)]\bigg|_{\Sigma~\mbox{\tiny dependent}}\nonumber\\
 &=& N_c\int_0^1dt~\mathrm{Tr}\bigg[\bigg[\frac{\partial J_{\Omega(t)}}{\partial t}
 +\frac{\partial\Sigma(-\nabla_t^2)}{\partial t}\bigg][\slashed{\partial}+J_{\Omega(t)}
 +\Sigma(-\nabla_t^2)]^{-1}\bigg]_{\Sigma~\mbox{\tiny dependent}}\nonumber\\
 &=&N_c \int_0^1dt~\mathrm{Tr}\bigg[\bigg(\frac{1}{2}\bigg[\frac{\partial U_t}{\partial t}U^\dag_t\gamma_5,\slashed{\partial}
 +J_{\Omega(t)}\bigg]_++\frac{\partial\Sigma(-\nabla_t^2)}{\partial t}\bigg)
 [\slashed{\partial}+J_{\Omega(t)}+\Sigma(-\nabla_t^2)]^{-1}\bigg]_{\Sigma~\mbox{\tiny dependent}}. \label{Seff2Ea}
 \end{eqnarray}
 We only need to calculate the $\Sigma$ dependent terms in Eq.(\ref{Seff2Ea}),
 because the $\Sigma$ independent terms are related to the contact terms\cite{our5}.

 Finally, as in \cite{oura2},  expanding (\ref{Seff2Ea}) to $p^4$ order, we can get the Wess-Zumino-Witten term
 and terms related to the tensor source.
 Furthermore, to the $p^6$ order, including the tensor source, we can get the imaginary part we need.
 The particular calculation after expanding (\ref{Seff2Ea}) will be introduced in Sec.\ref{imp4}.

 \section{Convention differences and independent operators}\label{diff}

 To accord with our original results and for  computational convenience,
 we make the following changes in this paper:
 \begin{itemize}
 \item To match Ref.\cite{p62} and our original results in Ref.\cite{our5}, we define
 \begin{eqnarray}
 \chi_{\pm,\mu}=\nabla_{\mu}\chi_{\pm}-\frac{i}{2}\{\chi_{\mp},u_{\mu}\},
 \end{eqnarray}
 comparing this with
  \begin{eqnarray}
 \chi_{\pm,\mu}=\nabla_{\mu}\chi_{\pm},
 \end{eqnarray}
  of Ref.\cite{tensor1}. Here $\chip,\chim$ and $u_\mu$ are analogous to $s_\Omega, p_\Omega$ and $a_\Omega$
 in our  notation (see details in Appendix \ref{symrel}).

 \item To match the coefficients' dimensions for a given order,
 i.e., all the coefficients in a given order have the same dimensions,
 we change $t_{\pm}^{\mu\nu}$ in Table 2 in \cite{tensor1} to $b_0t_{\pm}^{\mu\nu}$,
 the analog of $\tau^{\mu\nu}$ defined in \cite{tensor1}.  Now, all the coefficients in the $p^4$ order are dimensionless,  whereas in the $p^6$ order, their dimensions are GeV$^{-2}$.

 \item Via partial integration and the application of the equations of motion,
 $Y_{115}$ and $Y_{116}$ are not independent. We list the new relations in Appendix \ref{newrelation}.

 \item Ref.\cite{tensor1} does not consider the epsilon relations
 \begin{eqnarray}
 \epsilon^{\lambda\rho\delta\mu}{\epsilon_{\lambda\rho\delta}}^{\nu}&=&-6g^{\mu\nu},\label{epsilon1}\\
 \epsilon^{\sigma\delta\mu\nu}{\epsilon_{\sigma\delta}}^{\lambda\rho}
 &=&-2g^{\mu\lambda}g^{\nu\rho}+2g^{\mu\rho}g^{\nu\lambda},\label{epsilon2}\\
 \epsilon^{\alpha\mu\nu\lambda}{\epsilon_{\alpha}}^{\rho\sigma\delta}
 &=&-g^{\mu\rho}g^{\nu\sigma}g^{\lambda\delta}+g^{\mu\rho}g^{\nu\delta}g^{\lambda\sigma}
 -g^{\mu\sigma}g^{\nu\delta}g^{\lambda\rho}\notag\label{epsilon3}\\
 &&+g^{\mu\sigma}g^{\nu\rho}g^{\lambda\delta}
 -g^{\mu\delta}g^{\nu\rho}g^{\lambda\sigma}+g^{\mu\delta}g^{\nu\sigma}g^{\lambda\rho}.
 \end{eqnarray}
 Combining with Eq.(5.3) in \cite{tensor1},
 \begin{eqnarray}
 \epsilon_{\mu\nu\alpha\beta}t_{\pm}^{\alpha\beta}=2it_{\mp,\mu\nu}\label{epsilon4},
 \end{eqnarray}
 one can reduce $\tm\tm\to\tp\tp$ and $\tm\tp\to\tp\tm$, i.e.,
 changing even $\tm$ to the corresponding $\tp$, and exchanging the order of $\tp$ and $\tm$. For example
 \begin{eqnarray}
 \tm^{\mu\nu} t_{-,\mu\lambda}
 =\frac{1}{2}{g^{\nu}}_{\lambda}t_{+}^{\mu\rho}t_{+,\mu\rho}
 -{{t_{+}^\mu}}_{\lambda}{t_{+,\mu}}^{\nu}.
 \end{eqnarray}
 Hence even $\tm$ terms and some $\la\cdots\tm\cdots\tp\cdots\ra$ terms are not independent
 \footnote{To avoid confusion of our notation with that used in Ref.\cite{tensor1},
 and for more convenience  both in  calculation and  displaying results,
 we use  both tr$_f[\cdots]$ in the calculation
 and $\la\cdots\ra$ in the result to represent the tracing over flavor indices.}.
  We substitute the epsilon relations in
 $Y_{i},i=23-30,53,56,81,83,89,91,93,104,105,109-111,118,119$,
  finding that most terms lead to new relations.
 We list all the new relations in Appendix \ref{newrelation}.
 All the terms in the lhs of (\ref{dependoperators}) are considered to be dependent
 and reducible.
 We find that, in total, there are 22 additional dependent operators
 in the $n$-flavors case, 21 in the three-flavor and 13 in the two-flavor cases,
  leaving 98 independent operators for $n$ flavors, 92 for three flavors,
 and 65 for two flavors.
 In  Sec. IV of Ref.\cite{our5}, we found  the result that without quark self-energy,
 all the coefficients, except contact terms, must vanish. Now in the present work,
 if we similarly ignore the quark self-energy, without relations (\ref{epsilon1})-(\ref{epsilon4}),
 we cannot obtain these zero results. Instead, with  relations (\ref{epsilon1})-(\ref{epsilon4}),
 we do reproduce the vanishing result. This shows the importance of
 relations (\ref{epsilon1})-(\ref{epsilon4}) in the computation.

 \item Also, with (\ref{epsilon1})-(\ref{epsilon4}), adding (\ref{epsilon5})
 \begin{eqnarray}
 \epsilon^{\mu\nu\lambda\rho}\epsilon^{\mu'\nu'\lambda'\rho'}=-\det(g^{\alpha\alpha'}),
 \hspace{1cm}\alpha=\mu,\nu,\lambda,\rho\hspace{1cm}\alpha'=\mu',\nu',\lambda',\rho',\label{epsilon5}
 \end{eqnarray}
 one can remove all  epsilons for the tensor source terms in the chiral Lagrangian as follows.
 First, even epsilons can be changed to $g_{\mu\nu}$, and odd epsilons can be reduced to 1.
 Second, in one-epsilon terms, one can change $\tp$ to $\tm$ or $\tm$ to $\tp$ with the help of (\ref{epsilon4}),
  leaving only two epsilons.
 Finally, using (\ref{epsilon1})-(\ref{epsilon3}) and (\ref{epsilon5}), all the epsilons can be removed.
 In other words, there do not exist the odd-intrinsic-parity parts with tensor sources
 at any order of the low-energy expansion, if one keeps $\tp$ and $\tm$ as building blocks and
 changes them with Eq.(\ref{sigf}).
 \end{itemize}

 \section{the $p^4$ order chiral Lagrangian with tensor sources}\label{p4res}
 \subsection{Real part}

  With the same method used in Ref.\cite{our5}, we can expand the exponent in Eq.(\ref{dB1}) to the order $p^4$.
  Ref.\cite{tensor1} had given us the $p^4$ order Lagrangian of the form
 \begin{eqnarray}
 \mathcal{L}_{4,t}=\Lambda_1\la\tp^{\mu\nu}f_{+\mu\nu}\ra-i\Lambda_2\la\tp^{\mu\nu}u_{\mu}u_{\nu}\ra
 +\Lambda_3\la\tp^{\mu\nu}t_{\mu\nu}^+\ra+\Lambda_4\la\tp^{\mu\nu}\ra^2
 \equiv\sum_{n=1}^4\Lambda_n X_n.\label{Sp4r}
 \end{eqnarray}
 Considering that our computation is done under the large $N_c$ limit,
 if we only expand Eq.(\ref{dB1}), but do not consider the equations of motion,
 \begin{eqnarray}
 \nabla_\mu u^{\mu}=\frac{i}{2}\bigg(\chim-\frac{\la\chim\ra}{N_f}\bigg),
 \end{eqnarray}
 terms in the chiral Lagrangian with two
 or more traces vanish. To avoid unnecessary complications, in this
 paper we  retain in the calculation only those terms with one-trace:
  \begin{eqnarray}
 \mathcal{L}_{4,n,t}=\lambda_1\mathrm{tr}_f[t_{+,\Omega,\mu\nu}t_{+,\Omega}^{\mu\nu}]
 +i\lambda_2 \mathrm{tr}_f[a_{\Omega,\mu}a_{\Omega,\nu}t_{+,\Omega}^{\mu\nu}]
 +\lambda_3\mathrm{tr}_f[V_{\Omega,\mu\nu}t_{+,\Omega}^{\mu\nu}]+O\bigg(\frac{1}{N}\bigg)
 \equiv\sum_{n=1}^3\lambda_n x_n+O\bigg(\frac{1}{N}\bigg).\label{Sp4ours}
 \end{eqnarray}
  Expanding Eq.(\ref{dB1}),  we get the analytic results as
 \begin{eqnarray}
 \lambda_1&=&N_C\int^\infty_{\frac{1}{\Lambda^2}}\frac{d\tau}{\tau}\int\frac{d^4k}{(2\pi)^4}e^{-\tau(k^2+\sk^2)}
 (-2 \tau^2\sk^2),\\
 \lambda_2&=&N_C\int^\infty_{\frac{1}{\Lambda^2}}\frac{d\tau}{\tau}\int\frac{d^4k}{(2\pi)^4}e^{-\tau(k^2+\sk^2)}
 (12 \tau^2\sk -4 \tau^3 k^2\sk -8 \tau^3\sk^3),\\
 \lambda_3&=&N_C\int^\infty_{\frac{1}{\Lambda^2}}\frac{d\tau}{\tau}\int\frac{d^4k}{(2\pi)^4}e^{-\tau(k^2+\sk^2)}
 (-2 \tau^2\sk +2 \tau^2 k^2\skp),
 \end{eqnarray}

 and the relations between $\lambda_n$ and $\Lambda_n$ are
 \begin{eqnarray}
 \Lambda_1&=&\frac{1}{2b_0}\lambda_3,\\
 \Lambda_2&=&-\frac{1}{4b_0}\lambda_2-\frac{1}{2b_0}\lambda_3,\label{Sp4l}\\
 \Lambda_3&=&\frac{1}{b_0^2}\lambda_1,\\
 \Lambda_4&=&0.
 \end{eqnarray}
 The numerical results are listed in the second and sixth columns in Table \ref{p4r}.

 \subsection{Imaginary part}\label{imp4}
  As in Ref.\cite{oura2}, we can expand the exponent in Eq.(\ref{Seff2Ea}) to the $p^4$ order.
 Using (\ref{epsilon4}), we change $t_{\pm}$ to $t_{\mp}$, absorb the $\epsilon$ factors,
 and finally, get the results
 \begin{eqnarray}
 \mathcal{L}_{4,i,t}=\sum_{n=1}^8 z_n \bar{o}_n.\label{Sp4wours}
 \end{eqnarray}
 The terms $\bar{o}_n$ are listed in Table \ref{p4o}, and $z_n$ are listed in Eq. (\ref{Sp4rw1}).
 \begin{table*}[h]
 \caption{\label{p4o}The obtained operators of the $p^4$ order.}
 \begin{eqnarray}
 \extrarowheight 10pt
 \begin{array}{cl}
 \hline\hline n & \hspace{2.5cm}\bar{o}_n\\
 \hline  1 &  \displaystyle \la\pUt\Utd\tmtx^{\mu\nu}\atx^{\mu}\atx^{\nu} + \pUt\Utd\atx^{\mu}\atx^{\nu}\tmtx^{\mu\nu}\ra \\
 2 &  \displaystyle \la\pUt\Utd\atx^{\mu}\tmtx^{\mu\nu}\atx^{\nu}\ra \\
 3 &  \displaystyle \la\pUt\Utd\atx^{\mu}\tptx^{\mu\nu}\nt^{\nu} - \pUt\Utd\nt^{\mu}\tptx^{\mu\nu}\atx^{\nu}\ra \\
 4 &  \displaystyle \la\pUt\Utd\nt^{\mu}\tmtx^{\mu\nu}\nt^{\nu}\ra \\
 5 &  \displaystyle \la\pUt\Utd\tmtx^{\mu\nu}\nt^{\mu}\nt^{\nu} + \pUt\Utd\nt^{\mu}\nt^{\nu}\tmtx^{\mu\nu}\ra \\
 6 &  \displaystyle \la\pUt\Utd\tptx^{\mu\nu}\atx^{\mu}\nt^{\nu} - \pUt\Utd\nt^{\mu}\atx^{\nu}\tptx^{\mu\nu}\ra \\
 7 &  \displaystyle \la\pUt\Utd\tptx^{\mu\nu}\nt^{\mu}\atx^{\nu} - \pUt\Utd\atx^{\mu}\nt^{\nu}\tptx^{\mu\nu}\ra \\
 8 &  \displaystyle \la\pUt\Utd\tmtx^{\mu\nu}\tptx^{\mu\nu} + \pUt\Utd\tptx^{\mu\nu}\tmtx^{\mu\nu}\ra \\
 \hline\hline
 \end{array}\notag
 \end{eqnarray}
 \end{table*}
 \begin{eqnarray}
 &&z_{1}=N_C\int\frac{d^4k}{(2\pi)^4}(2 \sk X^2
 -4 \sk^3 X^3
 ),\notag\\
 &&z_{2}=0,\notag\\
 &&z_{3}=N_C\int\frac{d^4k}{(2\pi)^4}(2i \skp X
 -2i \sk^2\skp X^2
 -4i \sk^3 X^3
 ),\notag\\
 &&z_{4}=0,\notag\\
 &&z_{5}=N_C\int\frac{d^4k}{(2\pi)^4}(-2 \sk X^2
 +2 \skp X
 -2 \sk^2\skp X^2
 ),\notag\\
 &&z_{6}=N_C\int\frac{d^4k}{(2\pi)^4}(-2i \sk X^2
 +4i \sk^3 X^3
 ),\notag\\
 &&z_{7}=N_C\int\frac{d^4k}{(2\pi)^4}(-2i \sk X^2
 +2i \skp X
 -2i \sk^2\skp X^2
 ),\notag\\
 &&z_{8}=N_C\int\frac{d^4k}{(2\pi)^4}(2i \sk^2 X^2
 ),\label{Sp4rw1}
 \end{eqnarray}
 where $X\equiv 1/(k^2+\sk^2)$. Theoretically, we can integrate (\ref{Sp4wours}) to confirm (\ref{Sp4r}),
 but it is too hard to do the integral, and even worse to extend to the $p^6$ order.
 Oppositely, we differentiate (\ref{Sp4r}) to compare with Eq.(\ref{Sp4wours}).
 With the relation in Table \ref{symb}, $X_n$ in Eq.(\ref{Sp4r}) can be represented by $\Omega$ fields denoted in Eq.(\ref{JOmega}).
 Introducing a parameter $t$ as in Eq.(\ref{Seff1Ea}), we change $X_n\to X_n(t)$, and $X_n=X_n(1)$.
 Under the large $N_C$ limit, our results only relate to the one-trace terms $X_{1,2,3}$.
 With the help of Eq(\ref{difft}), we can differentiate one-trace term $X_{1,2,3}(t)$ to get Eq.(\ref{p4rw0}).
 \begin{eqnarray}
 &&\hspace{-0.3cm}\frac{\partial a_t^\mu}{\partial t}
 =\frac{i}{2}[\nabla^\mu_t\frac{\partial U_t}{\partial t}U^\dag_t
 -\frac{\partial U_t}{\partial t}U^\dag_t\nabla^\mu_t]\hspace{4cm}\frac{\partial v_t^\mu}{\partial t}
 =\frac{1}{2}[a^\mu_t\frac{\partial U_t}{\partial t}U^\dag_t-\frac{\partial U_t}{\partial t}U^\dag_ta^\mu_t]\nonumber\\
 &&\hspace{-0.3cm}\frac{\partial s_t}{\partial t}
 =-\frac{i}{2}[p_t\frac{\partial U_t}{\partial t}U^\dag_t+\frac{\partial U_t}{\partial t}U^\dag_tp_t]
 \hspace{4.2cm}\frac{\partial p_t}{\partial t}
 =\frac{i}{2}[s_t\frac{\partial U_t}{\partial t}U^\dag_t+\frac{\partial U_t}{\partial t}U^\dag_ts_t]\notag\\
 &&\hspace{-0.3cm}\frac{\partial d^{\mu}a_t^\nu}{\partial t}\!
 =\frac{i}{2}[(\nabla^\mu_t\nabla^\nu_t\!\!+\!a^\nu_ta^\mu_t)\frac{\partial U_t}{\partial t}U^\dag_t\!
 -\!\nabla^\nu_t\frac{\partial U_t}{\partial t}U^\dag_t\nabla^\mu_t\!
 -\!\nabla^\mu_t\frac{\partial U_t}{\partial t}U^\dag_t\nabla^\nu_t\!
 -\!a^\nu_t\frac{\partial U_t}{\partial t}U^\dag_ta^\mu_t\!
 -\!a^\mu_t\frac{\partial U_t}{\partial t}U^\dag_ta^\nu_t\!+\!\frac{\partial U_t}{\partial t}U^\dag_t(\nabla^\nu_t\nabla^\mu_t\!\!
 +\!a^\mu_ta^\nu_t)]~~~~\nonumber\\
 &&\hspace{-0.3cm}\frac{\partial V^{\mu\nu}_t}{\partial t}
 =\frac{1}{2}[-\nabla^\mu_t\frac{\partial U_t}{\partial t}U^\dag_ta^\nu_t
 +\frac{\partial U_t}{\partial t}U^\dag_t\nabla^\mu_ta^\nu_t
 -\frac{\partial U_t}{\partial t}U^\dag_t d_t^{\mu}a^\nu_t+d_t^{\mu}a^\nu_t\frac{\partial U_t}{\partial t}U^\dag_t
 +a^\nu_t\nabla^\mu_t\frac{\partial U_t}{\partial t}U^\dag_t -a^\nu_t\frac{\partial U_t}{\partial t}U^\dag_t\nabla^\mu_t\nonumber\\
 &&\hspace{1cm}+\nabla^\nu_t\frac{\partial U_t}{\partial t}U^\dag_ta^\mu_t
 -\frac{\partial U_t}{\partial t}U^\dag_t\nabla^\nu_ta^\mu_t +\frac{\partial U_t}{\partial t}U^\dag_t d_t^{\nu}a^\mu_t
 -d_t^{\nu}a^\mu_t\frac{\partial U_t}{\partial t}U^\dag_t -a^\mu_t\nabla^\nu_t\frac{\partial U_t}{\partial t}U^\dag_t
 +a^\mu_t\frac{\partial U_t}{\partial t}U^\dag_t\nabla^\nu_t]\nonumber\\
 &&\hspace{-0.3cm}\frac{\partial t_+^{\mu\nu}}{\partial t}=-\frac{1}{2}\pUt\Utd\tmtx^{\mu\nu}-\frac{1}{2}\tmtx^{\mu\nu}\pUt\Utd
 \hspace{1cm}\frac{\partial t_-^{\mu\nu}}{\partial t}=-\frac{1}{2}\pUt\Utd\tptx^{\mu\nu}-\frac{1}{2}\tptx^{\mu\nu}\pUt\Utd,
 \label{difft}
 \end{eqnarray}
 \begin{eqnarray}
 &&(X_{1,t},X_{2,t},X_{3,t})^T=A_4(\bar{o}_1,\bar{o}_2,\bar{o}_3,\bar{o}_4,\bar{o}_5,\bar{o}_6,\bar{o}_7)^T,
 \hspace{0.5cm}X_{i,t}=d X_i(t)/dt,\label{p4rw0}\\
 &&A_4=\left(\begin{array}{cccccccc}
 2i & 0 & 0 & 0 & -2i & 2 & 2 & 0\\
 2i & 0 & -2 & 0 & 0 & 2 & 0 & 0\\
 0 & 0 & 0 & 0 & 0 & 0 & 0 & -1\\
 \end{array}\right)\label{p4rw1}
 \end{eqnarray}
 Combined with Eqs.(\ref{Sp4r}),(\ref{Sp4wours}),(\ref{p4rw0}),(\ref{p4rw1}), we get
 \begin{eqnarray}
 &&\hspace*{-0.5cm}(\Lambda_1,\Lambda_2,\Lambda_3)(X_{1,t},X_{2,t},X_{3,t})^T
 =(\Lambda_1,\Lambda_2,\Lambda_3)A_4(\bar{o}_1,\bar{o}_2,\bar{o}_3,\bar{o}_4,\bar{o}_5,\bar{o}_6,\bar{o}_7,\bar{o}_8)^T\notag\\
 &=&(z_1,z_2,z_3,z_4,z_5,z_6,z_7,z_8)(\bar{o}_1,\bar{o}_2,\bar{o}_3,\bar{o}_4,\bar{o}_5,\bar{o}_6,\bar{o}_7,\bar{o}_8)^T\\
 &\Rightarrow&(\Lambda_1,\Lambda_2,\Lambda_3)A_4=(z_1,z_2,z_3,z_4,z_5,z_6,z_7,z_8).\label{azr}
 \end{eqnarray}
 In Eq.(\ref{azr}), $z_i$ and $A_4$ are obtained; they are eight linear equations with three unknown variables.
 Calculating the reduced row echelon form of $A_4$, we get
 \begin{eqnarray}
 A'_4=\left(\begin{array}{cccccccc}
 1 & 0 & 0 & 0 & -1 & -i & -i & 0\\
 0 & 0 & 1 & 0 & -i & 0 & 1 & 0\\
 0 & 0 & 0 & 0 & 0 & 0 & 0 & 1\\
 \end{array}\right).\label{p4rw1p}
 \end{eqnarray}
 We choose $\bar{o}_{1,3,8}$ as independent operators to calculate $\Lambda_n$,
  leaving the other five equations as strict constraints.
 The analytical results are

 \begin{eqnarray}
 \Lambda_{1}&=&\frac{N_C}{b_0}\int\frac{d^4k}{(2\pi)^4}\bigg[ \skp X
 - \sk X^2
 - \sk^2\skp X^2
 \bigg],\notag\\
 \Lambda_{2}&=&\frac{N_C}{b_0}\int\frac{d^4k}{(2\pi)^4}\bigg[- \skp X
 + \sk^2\skp X^2
 +2 \sk^3 X^3
 \bigg],\notag\\
 \Lambda_{3}&=&\frac{N_C}{b_0^2}\int\frac{d^4k}{(2\pi)^4}\bigg[-2 \sk^2 X^2
 \bigg],\notag\\
 \Lambda_{4}&=&0.\label{p4rw2}
 \end{eqnarray}
 In Eq.(\ref{p4rw2}), we find that  unlike those terms without tensors, which
 result in a fifth-dimension integral as Wess-Zumino-Witten terms,
 the tensor source terms can be fully worked out in the fourth dimension for the $p^4$ order.
 The numerical results are listed in the third and seventh columns in Table \ref{p4r}.

 As for the definition of $b_0$, we take the same spirit as that for $B_0$, where we have taken
  $\chi=2B_0(s+ip)$ with $\chi_\pm=u^\dag\chi u^\dag\pm u\chi^\dag u$,
  and let the final $p^2$ order of the chiral Lagrangian appear as $\mathcal{L}_2=\frac{F_0^2}{4}(u_{\mu}u^\mu+\chi_+)$ \cite{p62}.
  In other words, with dimensional coefficient $B_0$ in front of operator $u^\dag(s+ip)u^\dag\pm u(s-ip)u$,
  the LEC of this operator becomes $\frac{F_0^2}{4}$
  which is the same as that of another $p^2$ order operator $u_{\mu}u^\mu$.
  Now we choose $b_0$ in such a way that it makes $\Lambda_2$ equal to the value of $\hat{L_3}$.
  Here, $\hat{L_3}$ is the LEC of another $p^4$ order operator $\la(u^{\mu}u_{\mu})^2\ra$
  and it is the $p^4$ order analogue of dimension coefficient $F_0^2$ in the $p^2$ order chiral Lagrangian.
  Although there may exist ambiguities for this kind of definition(because instead of $\Lambda_2$,
  one can also choose $\Lambda_1$ by fixing its values to $\hat{L_3}$ or other $\hat{L_i}$ to define $b_0$),
  we emphasize that no matter what value of $b_0$ we choose, the final interaction strength is the same.
  For example, for operator $\langle t_+^{\mu\nu}u_{\mu}u_\nu\rangle$, the coefficient is $\Lambda_1b_0$,
  but from Eq.(\ref{p4rw2}), we know $\Lambda_1$ is proportional to $1/b_0$;
  therefore, the interaction strength $\Lambda_1b_0$ is independent of the value of $b_0$.

 \subsection{Numerical results}\label{p4ra}

 In Table \ref{p4r}, we list our $p^4$ order LECs with tensor sources for cutoff
 $\Lambda=$1000$^{+100}_{-100}$MeV in Eq.(\ref{Trln}). The $10\%$ variation of the
 cutoff is considered in our calculation to examine the effects of
 cutoff dependence and the result change can be treated as the error of
 our calculations. The results are taken as the values at
 $\Lambda=1$GeV. The superscript is the difference caused at
 $\Lambda=1.1$GeV and the subscript is the difference caused at
 $\Lambda=0.9$GeV, i.e.\footnote{Notice that $\Lambda$ with subscript $n$,
 $\Lambda_n$, means the $p^4$ order coefficients in Eq.(\ref{Sp4r}),
 but $\Lambda$ without subscript is the cutoff in our calculation introduced in Eq.(\ref{Trln})},
 \begin{eqnarray}
 \Lambda_{n,\Lambda=1\mathrm{GeV}}\bigg|^{\Lambda_{n,\Lambda=1.1\mathrm{GeV}}
 -\Lambda_{n,\Lambda=1\mathrm{GeV}}}_{\Lambda_{n,\Lambda=0.9\mathrm{GeV}}-\Lambda_{n,\Lambda=1\mathrm{GeV}}}.
 \end{eqnarray}
 The LECs include three and two flavors both from the real part, $\Lambda_{r,n}$,
 and the imaginary part, $\Lambda_{i,n}$. We also list the sum $\Lambda_{n}=\Lambda_{r,n}+\Lambda_{i,n}$.
 We get $b_0=1.32^{-0.04}_{+0.06}$GeV in the three-flavor case, and $b_0=2.01^{-0.06}_{+0.09}$GeV in the two-flavor case.
 $b_0$ in the two-flavor cases seems larger, because of the Cayley-Hamilton relation giving more relations.

 \begin{table*}[h]
 \caption{\label{p4r}The obtained values of the $p^4$ order coefficients.
 The definition has some difference from \cite{tensor1}.  The details can be found in Sec. \ref{diff}.
 $\Lambda_{r,n}$ come from the real part, and $\Lambda_{i,n}$ come from the imaginary part,
 $\Lambda_{n}$ are their sum.}
 {\extrarowheight 3pt
 \begin{tabular}{rrrr|rrrr}
 \hline\hline \multicolumn{4}{c|}{$N_f=3$}&\multicolumn{4}{c}{$N_f=2$}\\
 \hline $n$ & $10^3\Lambda_{r,n}$ & $10^3\Lambda_{i,n}$ & $10^3\Lambda_{n}$ & $m$ & $10^3\Lambda_{r,m}$ & $10^3\Lambda_{i,m}$ & $10^3\Lambda_{m}$\\
 \hline 1 &$ -7.62^{-0.44}_{+0.62}$ & $-9.01^{-0.14}_{+0.17}$ & $-16.62^{-0.58}_{+0.79}$ & 1 & $-4.98^{-0.30}_{+0.42}$ & $-5.93^{-0.10}_{+0.12}$ & $-10.92^{-0.40}_{+0.53}$ \\
 2 &$ 6.85^{+0.14}_{-0.21}$ & $6.94^{+0.11}_{-0.13}$ & $13.79^{+0.25}_{-0.34}$ & 2 & $4.46^{+0.10}_{-0.14}$ & $4.52^{+0.07}_{-0.09}$ & $8.99^{+0.17}_{-0.23}$ \\
 3 &$ -1.42^{-0.09}_{+0.12}$ & $-1.55^{-0.05}_{+0.06}$ & $-2.97^{-0.14}_{+0.18}$ & 3 & $-0.60^{-0.04}_{+0.05}$ & $-0.66^{-0.02}_{+0.03}$ & $-1.26^{-0.06}_{+0.08}$ \\
 4 & $\equiv 0$& $\equiv 0$&$\equiv 0$ &4 & $\equiv 0$&$\equiv 0$ &$\equiv 0$ \\
 \hline\hline
 \end{tabular}
 }
 \end{table*}

 To compare with our original results,
 the parameters we use to get Table \ref{p4r} are the same as Refs.\cite{our5,oura2}
 \footnote{In order to match the $N_f=2$ result,
 we choose a heavy quark  number $4$ here,
 instead of $2$ in Ref.\cite{our5,oura2}. (Heavy-quark fields are integrated out and absorbed into effective gluon propagator)}.
 We choose the running coupling constant from Ref.\cite{alphas} to
 solve Eq.(\ref{eq0}), and we get the same quark self-energy  as Fig. 2 in Ref.\cite{WQ1},
 but adding the two-flavor case.
 Except for the quark self-energy, we need another input parameter $F_0$,
 the $p^2$ order coefficient in the chiral Lagrangian.
 We set $F_0=87$MeV to get $F_\pi$ of about 93 MeV \cite{our5}.

 To examine our numerical result, we compute the magnetic susceptibility of the quark condensate,
 which we will show is proportional to $\Lambda_1$.
 We first introduce an external electromagnetic field $A_{\mathrm{em}}^\mu$ into the generating functional (\ref{effa})
 by adding in an external field term $-\overline{\psi}q\slashed{A}_{\mathrm{em}}\psi$
 into the Lagrangian on the exponential integrand.
 The generating functional is changed from $Z[J]$ to $Z[J-q\slashed{A}_{\mathrm{em}}]$.
 The magnetic susceptibility of the quark condensate $\chi$ is
  \begin{eqnarray}
 \frac{e}{3}\chi\langle\bar{\psi}\psi\rangle F^{\mu\nu}(x)=\langle 0|\bar{\psi}(x)\sigma^{\mu\nu}\psi(x)|0\rangle=
 -i\frac{1}{Z[J-iq\slashed{A}_{\mathrm{em}}]}\frac{\delta Z[J-iq\slashed{A}_{\mathrm{em}}]}{\delta\bar{t}_{\mu\nu}(x)}\bigg|_{J=0}\label{MatrixDef}
 \end{eqnarray}
 which leads to $\chi\langle\bar{\psi}\psi\rangle=-4\Lambda_1b_0$.
 In the case of $\Lambda=1$GeV,  we obtain
 \begin{eqnarray}\chi=-\frac{4\Lambda_1b_0}{\la\bar\psi\psi\ra}=\frac{4\Lambda_1b_0}{N_f F_0^2B_0}
 =-7.3\mbox{GeV}^{-2}.
 \end{eqnarray}
 Comparing this result with those in Refs.\cite{e3,e1,e2},
 gives the results, $-(8.16\pm0.41)\mbox{GeV}^{-2}$, $-2.7\mbox{GeV}^{-2}$
 and $-3.3\mbox{GeV}^{-2}$, respectively. There is a factor of $1\sim 3$
 difference between our result and the references. Although
 we choose $F_0=87$MeV instead of $F_\pi=92.4$MeV in the quark condensate, it
 will decrease the absolute value of our result a little bit,
 but the correction  is not enough.  We leave the investigation of this discrepancy to future studies.

 \section{the $p^6$ order chiral Lagrangian with tensor sources}\label{p6res}
 \subsection{Real part}

 Continuing our process in Sec. \ref{p4res}, we can obtain the $p^6$ order results directly.
 Before listing our results, we first introduce the existing results.
 Ref.\cite{tensor1}  gave the $p^6$ order Lagrangian as follows,
  \begin{eqnarray}
 S_{\mathrm{eff}}|_{p^6,\mathrm{tensor~sources}}=\int d^4x\left\{
 \begin{array}{ll}
 \displaystyle\sum_{n=1}^{117}K^T_n Y_n+3~\mbox{contact terms} & n \mbox{ flavors}\\
 \displaystyle\sum_{n=1}^{110}C^T_n O_n+3~\mbox{contact terms} & \mbox{three flavors}\\
 \displaystyle\sum_{n=1}^{75}c^T_n P_n+3~\mbox{contact terms} & \mbox{two flavors}
 \end{array}\right. .
 \end{eqnarray}
  Here, we use the  notation $Y_n,O_n,P_n$ to denote $n$, three and two flavors' independent monomials,
 which can be found in Table 2 in Ref.\cite{tensor1},
 and $K^T_n,C^T_n,c^T_n$ for their coefficients.  For reasons in Sec. \ref{diff} and Appendix \ref{newrelation},
 some of them are not independent, but we use the same numbers. If one monomial is not independent, we just neglect it.

 In our calculation, expanding Eq.(\ref{Trln})
 as the $p^4$ order, we only get one trace terms without the equation of motion
  \begin{eqnarray}
 S_{\mathrm{eff}}|_{p^6,\mathrm{tensor~sources}}=\int d^4x\bigg[\sum_{n=1}^{77}\mathcal{Z}^T_n\mathrm{tr}_f[\bar{O}_n]
 +O\bigg(\frac{1}{N_c}\bigg)\bigg].\label{Sp6ours}
 \end{eqnarray}
 The terms $\bar{O}_n$ are the $p^6$ order operators we can obtain from our calculation,
 and $\mathcal{Z}^T_n$ are the corresponding coefficients. For those operators with more than one derivative, for example
 $\bar{O}_{66}=d_{\mu}{V_{\Omega}^{\mu}}_{\nu}d_{\lambda}t_{+,\Omega}^{\nu\lambda}$, the derivatives are arranged in such a way that
 each ${V_{\Omega,\mu}}^{\nu}$ and $t_{+,\Omega,\nu\lambda}$ has a derivative and we do not put two derivatives in one operator.
 We list all operators in Table \ref{indter}.
 \begin{table*}[h]
 \caption{\label{indter}The $p^6$ order operators $\bar{O}_n$}
 \begin{eqnarray}
 \begin{array}{rc|rc|rc}
 \hline\hline n & \bar{O}_n & n & \bar{O}_n & n & \bar{O}_n \\
 \hline 1 & i \{a_{\Omega,\mu}a_\Omega^{\mu},a_{\Omega,\nu}a_{\Omega,\lambda}\}t_{+,\Omega}^{\nu\lambda}
  & 27 & i a_{\Omega,\mu}[d_{\nu}a_{\Omega,\lambda},d^{\mu}t_{+,\Omega}^{\nu\lambda}]
  & 53 & i V_{\Omega,\mu\nu}\{d_\lambda a_\Omega^{\mu},t_{\Omega,-}^{\nu\lambda}\}
 \\
 2 & i a_{\Omega,\mu}a_{\Omega,\nu}(a_\Omega^{\mu}a_{\Omega,\lambda}t_{+,\Omega}^{\nu\lambda}
 +a_{\Omega,\lambda}a_\Omega^{\nu}t_{+,\Omega}^{\mu\lambda})
  & 28 & i a_{\Omega,\mu[}d_\nu a_{\Omega,\lambda},d^\nu t_{\Omega,+}^{\mu\lambda}]
  & 54 & i V_{\Omega,\mu\nu}\{d_\lambda a_\Omega^{\lambda},t_{-,\Omega}^{\mu\nu}\}
 \\
 3 & i a_{\Omega,\mu}a_{\Omega,\nu}a_\Omega^{\nu}a_{\Omega,\lambda}t_{+,\Omega}^{\mu\lambda}
  & 29 & i a_{\Omega,\mu}[d_\nu a_{\Omega,\lambda},d^\lambda t_{+,\Omega}^{\mu\nu}]
  & 55 &  s_\Omega [d_\mu a_{\Omega,\nu},t_{-,\Omega}^{\mu\nu}]
 \\
 4 & i a_{\Omega,\mu}a_{\Omega,\nu}a_{\Omega,\lambda}a_\Omega^{\mu}t_{+,\Omega}^{\nu\lambda}
  & 30 & i a_{\Omega,\mu}\{d^\nu t_{+,\Omega,\nu\lambda},t_{-,\Omega}^{\mu\lambda}\}
  & 56 & i p_\Omega [t_{+,\Omega}^{\mu\nu},d_\mu a_{\Omega,\nu}]
 \\
 5 &  a_{\Omega,\mu}a_\Omega^{\mu}[d_\nu a_{\Omega,\lambda},t_{-,\Omega}^{\nu\lambda}]
  & 31 & i a_{\Omega,\mu}\{d_\nu {t_{+,\Omega}^{\mu}}_{\lambda},t_{-,\Omega}^{\nu\lambda}\}
  & 57 & i V_{\Omega,\mu\nu}{V_\Omega^{\mu}}_{\lambda}t_{+,\Omega}^{\nu\lambda}
 \\
 6 &  a_{\Omega,\mu}a_{\Omega,\nu}(d^\mu a_{\Omega,\lambda}t_{-,\Omega}^{\nu\lambda}
 - {t_{-,\Omega}^{\mu}}_{\lambda}d^\nu a_\Omega^{\lambda})
  & 32 & i a_{\Omega,\mu}\{d^{\mu}t_{+,\Omega,\nu\lambda},t_{-,\Omega}^{\nu\lambda}\}
  & 58 & i V_{\Omega,\mu\nu}{t_{+,\Omega}^{\mu}}_{\lambda}t_{+,\Omega}^{\nu\lambda}
 \\
 7 &  a_{\Omega,\mu}a_{\Omega,\nu}(d^{\nu}a_{\Omega,\lambda}t_{-,\Omega}^{\mu\lambda}
 -{t_{-,\Omega}^{\nu}}_{\lambda}d^\mu a_{\Omega}^{\lambda})
  & 33 & i a_{\Omega,\mu}a_{\Omega,\nu}\{t_{+,\Omega}^{\mu\nu},s_\Omega\}
  & 59 & i V_{\Omega,\mu\nu} {t_{-,\Omega}^{\mu}}_{\lambda}t_{-,\Omega}^{\nu\lambda}
 \\
 8 &  a_{\Omega,\mu}a_{\Omega,\nu}(d_\lambda a_\Omega^{\mu}t_{-,\Omega}^{\nu\lambda}
 - {t_{-,\Omega}^{\mu}}_{\lambda}d^\lambda a_\Omega^{\nu})
  & 34 & i a_{\Omega,\mu}s_\Omega a_{\Omega,\nu}t_{+,\Omega}^{\mu\nu}
  & 60 & i V_{\Omega,\mu\nu}\{p_\Omega,t_{-,\Omega}^{\mu\nu}\}
 \\
 9 &  a_{\Omega,\mu}a_{\Omega,\nu}(d_\lambda a_\Omega^{\nu}t_{-,\Omega}^{\mu\lambda}
 -{t_{-\Omega}^{\nu}}_{\lambda}d^\lambda a_\Omega^{\mu})
  & 35 & i d_\mu a_\Omega^{\mu}[d_\mu a_{\Omega,\lambda},t_{+,\Omega}^{\nu\lambda}]
  & 61 &  V_{\Omega,\mu\nu}\{t_{+,\Omega}^{\mu\nu},s_\Omega\}
 \\
 10 &  a_{\Omega,\mu}a_{\Omega,\nu}\{d_\lambda a_\Omega^{\lambda},t_{-,\Omega}^{\mu\nu}\}
  & 36 & i d_\mu a_{\Omega,\nu}d^\mu a_{\Omega,\lambda}t_{+,\Omega}^{\nu\lambda}
  & 62 & i t_{\Omega,\mu\nu}{t_{+,\Omega}^{\mu}}_{\lambda}t_{+,\Omega}^{\nu\lambda}
 \\
 11 &  a_{\Omega,\mu}(d^\mu a_{\Omega,\nu}a_{\Omega,\lambda}t_{-,\Omega}^{\nu\lambda}
 + d_\nu a_{\Omega,\lambda}a_\Omega^{\nu}t_{-,\Omega}^{\mu\lambda})
  & 37 & i d_\mu a_{\Omega,\nu}[d^\nu a_{\Omega,\lambda},t_{+,\Omega}^{\mu\lambda}]
  & 63 & i t_{\Omega,\mu\nu}{t_{-,\Omega}^{\mu}}_{\lambda}t_{-,\Omega}^{\nu\lambda}
 \\
 12 &  a_{\Omega,\mu}(d_\nu a_\Omega^{\mu}a_{\Omega,\lambda}t_{-,\Omega}^{\nu\lambda}
 + d_\nu a_{\Omega,\lambda}a_\Omega^{\lambda}t_{-,\Omega}^{\mu\nu})
  & 38 & i d_\mu a_{\Omega,\nu}d_\lambda a_{\Omega}^{\nu}t_{+,\Omega}^{\mu\lambda}
  & 64 &  s_\Omega t_{+,\Omega,\mu\nu}t_{+,\Omega}^{\mu\nu}
 \\
 13 &  a_{\Omega,\mu}d_\nu a_\Omega^{\nu}a_{\Omega,\lambda}t_{-,\Omega}^{\mu\lambda}
  & 39 &  V_{\Omega,\mu\nu}(a_{\Omega}^{\mu}a_{\Omega,\lambda}t_{+,\Omega}^{\nu\lambda}
 - {t_{+,\Omega}^{\mu}}_{\lambda}a_\Omega^{\lambda}a_\Omega^{\nu})
  & 65 &  s_\Omega t_{-,\Omega,\mu\nu}t_{-,\Omega}^{\mu\nu}
 \\
 14 &  a_{\Omega,\mu}a_\Omega^\mu t_{+,\Omega,\nu\lambda}t_{+,\Omega}^{\nu\lambda}
  & 40 &  V_{\Omega,\mu\nu}(a_{\Omega,\lambda}a_\Omega^{\mu}t_{+,\Omega}^{\nu\lambda}
 - {t_{+,\Omega}^{\mu}}_{\lambda}a_\Omega^{\nu}a_\Omega^{\lambda})
  & 66 &  d_\mu {V_{\Omega}^{\mu}}_{\nu}d_\lambda t_{+,\Omega}^{\nu\lambda}
 \\
 15 &  a_{\Omega,\mu}a_{\Omega,\nu} {t_{+,\Omega}^{\mu}}_{\lambda}t_{+,\Omega}^{\nu\lambda}
  & 41 &  V_{\Omega,\mu\nu}\{a_{\Omega,\lambda}a_\Omega^\lambda,t_{+,\Omega}^{\mu\nu}\}
  & 67 &  d_\mu V_{\Omega,\nu\lambda}d^\mu t_{+,\Omega}^{\nu\lambda}
 \\
 16 &  a_{\Omega,\mu}a_{\Omega,\nu}{t_{+,\Omega}^{\nu}}_{\lambda}t_{+,\Omega}^{\mu\lambda}
  & 42 &  V_{\Omega,\mu\nu}(a_\Omega^{\mu}{t_{+,\Omega}^{\nu}}_{\lambda}a_\Omega^{\lambda}
 - a_{\Omega,\lambda}t_{+,\Omega}^{\mu\lambda}a_\Omega^{\nu})
  & 68 &  d_\mu V_{\Omega,\nu\lambda}d^\nu t_{+,\Omega}^{\nu\mu\lambda}
 \\
 17 &  a_{\Omega,\mu}a_\Omega^\mu t_{-,\Omega,\nu\lambda}t_{-,\Omega}^{\nu\lambda}
  & 43 &  V_{\Omega,\mu\nu}a_{\Omega,\lambda}t_{+,\Omega}^{\mu\nu}a_\Omega^{\lambda}
  & 69 &  d_\mu {t_{+,\Omega}^{\mu}}_{\nu}d_\lambda t_{+,\Omega}^{\nu\lambda}
 \\
 18 &  a_{\Omega,\mu}a_{\Omega,\nu}{t_{-,\Omega}^{\mu}}_{\lambda}t_{-,\Omega}^{\nu\lambda}
  & 44 & i a_{\Omega,\mu}[{t_{+,\Omega}^{\mu}}_{\nu},d^\nu p_\Omega]
  & 70 &  d_\mu t_{+,\Omega,\nu\lambda}d^\mu t_{+,\Omega}^{\nu\lambda}
 \\
 19 &  a_{\Omega,\mu}a_{\Omega,\nu}{t_{-,\Omega}^{\nu}}_{\lambda}t_{-,\Omega}^{\mu\lambda}
  & 45 &  a_{\Omega,\mu}[{t_{-,\Omega}^{\mu}}_{\nu},d^\nu s_\Omega]
  & 71 &  d_\mu t_{+,\Omega,\nu\lambda}d^\nu t_{+,\Omega}^{\mu\lambda}
 \\
 20 &  a_{\Omega,\mu}({t_{+,\Omega}^{\mu}}_{\nu}a_{\Omega,\lambda}t_{+,\Omega}^{\nu\lambda}
 - t_{\Omega,\nu\lambda}a_\Omega^{\nu}t_{+,\Omega}^{\mu\lambda})
  & 46 & i d_\mu a_\Omega^{\mu}\{t_{-,\Omega,\nu\lambda},t_{+,\Omega}^{\nu\lambda}\}
  & 72 &  d_\mu {t_{-,\Omega}^{\mu}}_{\nu}d_\lambda t_{-,\Omega}^{\nu\lambda}
 \\
 21 &  a_{\Omega,\mu}t_{+,\Omega,\nu\lambda}a_\Omega^{\mu}t_{+,\Omega}^{\nu\lambda}
  & 47 & i d_\mu a_{\Omega,\nu}\{{t_{-,\Omega}^{\mu}}_{\lambda},t_{+,\Omega}^{\nu\lambda}\}
  & 73 &  d_\mu t_{-,\Omega,\nu\lambda}d^\mu t_{-,\Omega}^{\nu\lambda}
 \\
 22 &  a_{\Omega,\mu}({t_{-,\Omega}^{\mu}}_{\nu}a_{\Omega,\lambda}t_{-,\Omega}^{\nu\lambda}
 - t_{\Omega,\nu\lambda}a_\Omega^{\nu}t_{-,\Omega}^{\mu\lambda})
  & 48 & i d_\mu a_{\Omega,\nu}\{{t_{-,\Omega}^{\nu}}_{\lambda},t_{+,\Omega}^{\mu\lambda}\}
  & 74 &  d_\mu t_{-,\Omega,\nu\lambda}d^\nu t_{-,\Omega}^{\mu\lambda}
 \\
 23 &  a_{\Omega,\mu}t_{-,\Omega,\nu\lambda}a_\Omega^{\mu}t_{-,\Omega}^{\nu\lambda}
  & 49 & i V_{\Omega,\mu\nu}\{a_\Omega^{\mu},d_\lambda t_{-,\Omega}^{\nu\lambda}\}
  & 75 &  a_{\Omega,\mu}a_{\Omega,\nu}\{t_{-,\Omega}^{\mu\nu},p_\Omega\}
 \\
 24 & i a_{\Omega,\mu}[d^\mu a_{\Omega,\nu}d_\lambda, t_{+,\Omega}^{\nu\lambda}]
  & 50 & i V_{\Omega,\mu\nu}\{a_{\Omega,\lambda},d^\mu t_{-,\Omega}^{\nu\lambda}\}
  & 76 &  a_{\Omega,\mu}p_\Omega a_{\Omega,\nu}t_{-,\Omega}^{\mu\nu}
 \\
 25 & i a_{\Omega,\mu}[d_\nu a_\Omega^{\mu},d_\lambda t_{+,\Omega}^{\nu\lambda}]
  & 51 & i V_{\Omega,\mu\nu}\{a_{\Omega,\lambda},d^\lambda t_{-,\Omega}^{\mu\nu}\}
  & 77 &  i p_\Omega \{t_{+,\Omega\mu\nu},t_{-,\Omega}^{\mu\nu}\}
 \\
 26 & i a_{\Omega,\mu}[d_\nu a_\Omega^{\nu},d_\lambda t_{+,\Omega}^{\mu\lambda}]
  & 52 & i V_{\Omega,\mu\nu}\{d^\mu a_{\Omega,\lambda},t_{-,\Omega}^{\nu\lambda}\}
  &  &
 \\
 \hline\hline
 \end{array}\notag
 \end{eqnarray}
 \end{table*}
 With the help of a computer, we can get the coefficients $\mathcal{Z}^T_n$, listed in Appendix \ref{Zcs}.
 Making use of Table \ref{symb},
  relations for our coefficients $\mathcal{Z}^T_n$ and $K^T_n$, listed in Appendix \ref{ZKrs}
 can be obtained directly.
 Combining Appendixes \ref{newrelation}, \ref{Zcs}, and \ref{ZKrs} and using the parameters in Sec. \ref{p4ra},
 we obtain both two and three flavors numerical results,
 and list them in the second and sixth columns in Table \ref{Ccs6}.

 \subsection{Imaginary part}
 As in Sec. \ref{imp4}, continuing our process to the $p^6$ order, we can  derive the $p^6$ order LECs too.
 In the large $N_C$ limits, without using the equations of motion, and removing the contact terms, we get 56 independent terms, $\bar{O}^{T,W}_n$,
 and list them in Table \ref{indter6w}. We also list their relation to $Y_i$ of Ref.\cite{tensor1}.
 \begin{table*}[h]
 \caption{\label{indter6w}The $p^6$ order operators in the large $N_C$ limits, $\bar{O}^{T,W}_n$}
 \begin{eqnarray}
 \begin{array}{rclrcl}
 \hline\hline
 n&\bar{O}_n^{T,W} & Y_i &n&\bar{O}_n^{T,W} & Y_i\\
 \hline
 1&i\la t_{+\mu\nu}\{u_{\lambda}u^{\lambda},u^{\mu}u^{\nu}\}\ra/b_0 &Y_{1}/b_0 & 29&\la  t_{-\mu\nu}[\fm^{\mu\nu},u_{\lambda}u^{\lambda}]\ra/b_0 &Y_{69}/b_0\\
 2&i\la t_{+\mu\nu} u^{\lambda}u^{\mu}u^{\nu}u_{\lambda}\ra/b_0 &Y_{2}/b_0 & 30&\la  t_{-\mu\nu}(\fm^{\mu\lambda}u^{\nu}u_{\lambda}-u_{\lambda}u^{\nu}\fm^{\mu\lambda})\ra/b_0 &Y_{70}/b_0\\
 3&i\la t_{+\mu\nu} u^{\mu}u_{\lambda}u^{\lambda}u^{\nu}\ra/b_0 &Y_{3}/b_0 & 31&\la t_{-\mu\nu}(\fm^{\mu\lambda}u_{\lambda}u^{\nu}-u^{\nu}u_{\lambda}\fm^{\mu\lambda})\ra/b_0 &Y_{71}/b_0\\
 4&i\la t_{+\mu\nu} \{u_{\lambda},u^{\mu}u^{\lambda}u^{\nu}\}\ra/b_0 &Y_{4}/b_0 & 32&\la t_{-\mu\nu}(u^{\nu}\fm^{\mu\lambda}u_{\lambda}-u_{\lambda}\fm^{\mu\lambda}u^{\nu})\ra/b_0 &Y_{72}/b_0\\
 5&\la t_{+\mu\nu}\tp^{\mu\nu} u_{\lambda}u^{\lambda}\ra/b_0^2 &Y_{9}/b_0^2 & 33&\la t_{+\mu\nu}\{\fp^{\mu\nu},\chip\}\ra/B_0b_0 &Y_{74}/B_0b_0\\
 6&\la t_{+\mu\nu}u_{\lambda}\tp^{\mu\nu} u^{\lambda}\ra/b_0^2 &Y_{10}/b_0^2 & 34&\la t_{-\mu\nu}\{\fp^{\mu\nu},\chim\}\ra/B_0b_0 &Y_{75}/B_0b_0\\
 7&\la t_{+\mu\nu}\tp^{\mu\lambda} u_{\lambda}u^{\nu}\ra/b_0^2 &Y_{11}/b_0^2 & 35&\la t_{+\mu\nu}[\fm^{\mu\nu},\chim]\ra/B_0b_0 &Y_{76}/B_0b_0\\
 8&\la t_{+\mu\nu}\tp^{\mu\lambda} u^{\nu}u_{\lambda}\ra/b_0^2 &Y_{12}/b_0^2 & 36&i\la t_{-\mu\nu}\{\fp^{\mu\nu},{h^{\lambda}}_{\lambda}\}\ra/b_0 &-Y_{75}/b_0+\frac{2}{N_f b_0}Y_{80}\\
 9&\la t_{+\mu\nu}(u^{\nu}\tp^{\mu\lambda} u_{\lambda}+u_{\lambda}\tp^{\mu\lambda} u^{\nu})\ra/b_0^2 &Y_{13}/b_0^2 & 37&i\la t_{+\mu\nu}\{\tm^{\nu\lambda},{h^{\mu}}_{\lambda}\ra/b_0^2 &Y_{81}/b_0^2\\
 10&\la t_{+\mu\nu}\tp^{\mu\nu}\chip\ra/B_0 b_0^2 &Y_{31}/B_0 b_0^2 & 38&i\la t_{+\mu\nu}\fm^{\mu\lambda}f^{\nu}_{-\lambda}\ra/b_0 &Y_{84}/b_0\\
 11&\la t_{+\mu\nu}\tm^{\mu\nu}\chim+\tm^{\mu\nu}\tp^{\mu\nu}\chim\ra/B_0 b_0^2 &2Y_{32}/B_0 b_0^2 & 39&\la i t_{+\mu\nu}\fp^{\mu\lambda}f^{\nu}_{+\lambda}\ra/b_0 &Y_{85}/b_0\\
 12&i\la t_{+\mu\nu}\{\tm^{\mu\nu},{h^{\lambda}}_{\lambda}\}\ra/B_0 b_0 &-2Y_{32}/B_0 b_0+\frac{2}{N_f B_0 b_0}Y_{35} & 40&i\la t_{-\mu\nu}\{\fm^{\nu\lambda},f^{\mu}_{+\lambda}\}\ra/b_0 &Y_{86}/b_0\\
 13&i\la t_{+\mu\nu}\{\chip,u^{\mu}u^{\nu}\}\ra/B_0b_0 &Y_{39}/B_0b_0 & 41&i\la t_{+\mu\nu}\tp^{\mu\lambda}t^{\nu}_{+\lambda}\ra/b_0^3 &Y_{88}/b_0^3\\
 14&i\la t_{+\mu\nu}u^{\mu}\chip u^{\nu}\ra/B_0b_0 &Y_{40}/B_0b_0 & 42& i\la f_{+\mu\nu}\tp^{\nu\lambda}t^{\mu}_{+\lambda}\ra/b_0^2 &Y_{90}/b_0^2\\
 15&i\la t_{-\mu\nu}\{\chim,u^{\mu}u^{\nu}\}\ra/B_0b_0 &Y_{43}/B_0b_0 & 43&\la\nabla_{\mu}\tp^{\mu\nu}\nabla^{\lambda}f_{+\lambda\nu}\ra/b_0 &Y_{94}/b_0\\
 16&\la i t_{-\mu\nu}u^{\mu}\chim u^{\nu}\ra/B_0b_0 &Y_{44}/B_0b_0 & 44&i\la\nabla_{\lambda}t_{+\mu\nu}[h^{\mu\lambda},u^{\nu}]\ra/b_0 &Y_{95}/b_0\\
 17&\la t_{-\mu\nu}u^{\mu}{h^{\lambda}}_{\lambda} u^{\nu}\ra/b_0 &Y_{44}/b_0-\frac{1}{N_f B_0}Y_{45} & 45&i\la\nabla^{\mu}t_{+\mu\nu}[h^{\nu\lambda},u_{\lambda}]\ra/b_0 &Y_{96}/b_0\\
 18&\la t_{-\mu\nu}\{u^{\mu}u^{\nu},{h^{\lambda}}_{\lambda}\}\ra/b_0 &Y_{43}/b_0-\frac{2}{N_f B_0}Y_{45} & 46&i\la\nabla^{\mu}t_{+\mu\nu}[\fm^{\nu\lambda},u_{\lambda}]\ra/b_0 &Y_{97}/b_0\\
 19&\la t_{-\mu\nu}(h^{\nu\lambda}u_{\lambda}u^{\mu}-u^{\mu}u_{\lambda}h^{\nu\lambda})\ra/b_0 &Y_{47}/b_0 & 47&i\la\nabla^{\mu}t_{+\nu\lambda}[\fm^{\nu\lambda},u_{\mu}]\ra/b_0 &Z_{1}/b_0
 \footnote{$Z_1=-\frac{1}{2}Y_{1}+Y_{2}-Y_{59}+Y_{61}+\frac{1}{2}Y_{76}-Y_{84}-Y_{96}-Y_{97}+Y_{114},$}\\
 20&\la t_{-\mu\nu}(h^{\nu\lambda}u^{\mu}u_{\lambda}-u_{\lambda}u^{\mu}h^{\nu\lambda})\ra/b_0 &Y_{48}/b_0 & 48&i\la\nabla_{\mu}t_{+\nu\lambda}[\fm^{\mu\nu},u^{\lambda}]\ra/b_0 &Z_{2}/b_0
 \footnote{$Z_2=\frac{1}{2}Y_{1}+Y_{3}-Y_{4}+\frac{1}{2}Y_{39}+Y_{40}+Y_{60}-Y_{61}+Y_{84}-Y_{95}+Y_{113}-Y_{114},$}\\
 21&\la t_{-\mu\nu}(u_{\lambda}h^{\nu\lambda}u^{\mu}-u^{\mu}h^{\nu\lambda}u_{\lambda})\ra/b_0 &Y_{49}/b_0 & 49&i\la\nabla^{\mu}t_{-\mu\nu}\{\fp^{\nu\lambda},u_{\lambda}\}\ra/b_0 &Y_{98}/b_0\\
 22&\la \nabla_{\lambda}t_{\mu\nu}\nabla^{\lambda}\tp^{\mu\nu}\ra/b_0^2 &Y_{51}/b_0^2 & 50&i\la\nabla_{\lambda}t_{-\mu\nu}\{\fp^{\mu\nu},u^{\lambda}\}\ra/b_0 &Y_{99}/b_0\\
 23&\la \nabla_{\mu}\tp^{\mu\nu}\nabla^{\lambda}t_{+\lambda\nu}\ra/b_0^2 &Y_{52}/b_0^2 & 51&i\la\nabla_{\lambda}t_{-\mu\nu}\{\fp^{\mu\lambda},u^{\nu}\}\ra/b_0 &Y_{100}/b_0\\
 24&\la t_{+\mu\nu}\{\fp^{\mu\nu},u_{\lambda}u^{\lambda}\}\ra/b_0 &Y_{57}/b_0 & 52&i\la\{\nabla_{\mu}\tp^{\mu\nu},t_{-\nu\lambda}\}u^{\lambda}\ra/b_0^2 &Y_{105}/b_0^2\\
 25&\la t_{+\mu\nu}u_{\lambda}\fp^{\mu\nu}u^{\lambda}\ra/b_0 &Y_{58}/b_0 & 53&i\la\nabla^{\mu}\tp^{\nu\lambda}\{t_{-\mu\lambda},u_{\nu}\}\ra/b_0^2 &Z_{3}/b_0^2
 \footnote{$Z_3=-\frac{1}{4}Y_{9}-\frac{1}{4}Y_{10}+Y_{11}+\frac{1}{2}Y_{13}+\frac{1}{4}Y_{32}-\frac{1}{4N_f}Y_{35}-\frac{1}{2}Y_{51}+2Y_{52}-Y_{90}+Y_{105}-4Y_{118}+2Y_{119}.$}\\
 26&\la t_{+\mu\nu}(\fp^{\mu\lambda}u^{\nu}u_{\lambda}+u_{\lambda}u^{\nu}\fp^{\mu\lambda})\ra/b_0 &Y_{59}/b_0 & 54&\la \tm^{\mu\nu}[\chi_{+\mu},u_\nu]\ra/B_0b_0 &Y_{112}/B_0b_0\\
 27&\la t_{+\mu\nu}(\fp^{\mu\lambda}u_{\lambda}u^{\nu}+u^{\nu}u_{\lambda}\fp^{\mu\lambda})\ra/b_0 &Y_{60}/b_0 & 55&\la \tp^{\mu\nu}[\chi_{-\mu},u_\nu]\ra/B_0b_0 &Y_{113}/B_0b_0\\
 28&\la t_{+\mu\nu}(u^{\nu}\fp^{\mu\lambda}u_{\lambda}+u_{\lambda}\fp^{\mu\lambda}u^{\nu})\ra/b_0 &Y_{61}/b_0 & 56&i\la t_{+\mu\nu}h^{\mu\lambda}{h^{\nu}}_{\lambda}\ra/b_0 &Y_{114}/b_0\\
 \hline\hline
 \end{array}\notag
 \end{eqnarray}
 \end{table*}
 As in Eq.(\ref{Sp4ours}), in $p^6$ order, we get
 \begin{eqnarray}
 \mathcal{L}_{6,i,t}=\sum_{n=1}^{56} \widetilde{K}^{T,W}_n \widetilde{O}^{T,W}_n.\label{Sp6ours}
 \end{eqnarray}
 where $\widetilde{K}^{T,W}_n$ are the coefficients related to $\widetilde{O}^{T,W}_n$.
 Introducing a parameter $t$ to $\widetilde{O}^{T,W}_n$, we change $\widetilde{O}^{T,W}_n\to \widetilde{O}^{T,W}_n(t)$.
 Then differentiating $\widetilde{O}^{T,W}_n(t)$,
 we  get in like manner to (\ref{p4rw0}),
 \begin{eqnarray}
 &&(\widetilde{O}^{T,W}_{1,t},\widetilde{O}^{T,W}_{2,t},\cdots,\widetilde{O}^{T,W}_{56,t})^T
 =A_6(\bar{O}^{T,W}_1,\bar{O}^{T,W}_2,\cdots,\bar{O}^{T,W}_{339})^T,
 \hspace{0.5cm}\widetilde{O}^{T,W}_{i,t}=d\widetilde{O}^{T,W}_{i}(t)/dt\label{p6rw1}
 \end{eqnarray}
 and
 \begin{eqnarray}
 &&(\widetilde{K}^{T,W}_1,\widetilde{K}^{T,W}_2,\cdots,\widetilde{K}^{T,W}_{56})(\widetilde{O}^{T,W}_{1,t},\widetilde{O}^{T,W}_{2,t},\cdots,\widetilde{O}^{T,W}_{56,t})^T
 =(\widetilde{K}^{T,W}_1,\widetilde{K}^{T,W}_2,\cdots,\widetilde{K}^{T,W}_{56})A_6(\bar{O}^{T,W}_1,\bar{O}^{T,W}_2,\cdots,\bar{O}^{T,W}_{339})^T\notag\\
 &=&(\bar{K}^{T,W}_1,\bar{K}^{T,W}_2,\cdots,\bar{K}^{T,W}_{339})(\bar{K}^{T,W}_1,\bar{K}^{T,W}_2,\cdots,\bar{K}^{T,W}_{339})^T\\
 &\Rightarrow&(\widetilde{K}^{T,W}_1,\widetilde{K}^{T,W}_2,\cdots,\widetilde{K}^{T,W}_{56})A_6
 =(\bar{K}^{T,W}_1,\bar{K}^{T,W}_2,\cdots,\bar{K}^{T,W}_{339}).\label{p6rw2}
 \end{eqnarray}
 $\bar{O}^{T,W}_n$  in $p^6$ order are the same as $\bar{o}_n$ in $p^4$ order,
 and $\bar{K}^{T,W}_n$ are the same as $z_n$ in $p^4$ order.
 There  are 339 $\bar{O}^{T,W}_n$ and $\bar{K}^{T,W}_n$,
  which are too many to be listed here.
 Moreover $A_6$ is the same as $A_4$ in $p^4$ order too.
 Using (\ref{p6rw2}), we can also get $\widetilde{O}^{T,W}_n$,
  which are listed in Appendix \ref{tkwcs}.
 Combined with Table \ref{indter6w},
 we can get the analytical results from the imaginary part.
 The numerical results are listed in the third and seventh columns in Table \ref{Ccs6}.

 \subsection{Numerical results}
 In Table \ref{Ccs6}, we list our $p^6$ order LECs with tensor sources, including the three- and two-flavor case,
 and the results for the real and the imaginary parts.
 Similar to the $p^4$ order,
 we calculated the values with $\Lambda=1$GeV,
 and use the superscript and subscript to denote the differences caused
  with $\Lambda=1.1$GeV and $\Lambda=0.9$GeV, respectively:
 \begin{eqnarray}
 &&C^T_{n,\Lambda=1\mathrm{GeV}}\bigg|^{C^T_{n,\Lambda=1.1\mathrm{GeV}}-C^T_{n,\Lambda=1\mathrm{GeV}}}_{C^T_{n,\Lambda=0.9\mathrm{GeV}}-C^T_{n,\Lambda=1\mathrm{GeV}}}\hspace{1cm}
 c^T_{n,\Lambda=1\mathrm{GeV}}\bigg|^{c^T_{n,\Lambda=1.1\mathrm{GeV}}-c^T_{n,\Lambda=1\mathrm{GeV}}}_{c^T_{n,\Lambda=0.9\mathrm{GeV}}-c^T_{n,\Lambda=1\mathrm{GeV}}}\;.
 \end{eqnarray}
 $C^T_{n,\Lambda}$ and $c^T_{n,\Lambda}$ denote the three- and two-flavors cases, respectively.
 Because of the relations given in Appendix \ref{newrelation},
 some terms are not independent, we denote their coefficients by the symbol ``$-$".
 In Ref.\cite{tensor1},
 the coefficients were multiplied by a suitable power of $b_0$
 to  express these with the same dimensional units.
 We choose $b_0$ at the end in Sec.\ref{imp4}.

 {\extrarowheight 5pt
 \begin{longtable}{rrrrrrrrr}
 \caption{\label{Ccs6}The obtained values of the $p^6$ order LECs. $C^T_n$ denote the three-flavor coefficients,
 and $c^T_m$ denote two flavors. $n,m$ are the number of independent monomials in \cite{tensor1} with some difference.
 The subscripts $r,i$ denote the LECs from the real part and imaginary part of the chiral Lagrangian.
  As some monomials are not independent, we denote their coefficients by a preceding symbol "$-$".
 The details can be found in Sec. \ref{diff}.
 The value $\equiv0$ means that the constants vanish  in the large $N_C$ limit.}\\

 \hline\hline $n$ & $10^3$GeV$^2C^T_{r,n}$ & $10^3$GeV$^2C^T_{i,n}$ & $10^3$GeV$^2C^T_{n}$
        & $m$ & $10^3$GeV$^2c^T_{r,m}$ & $10^3$GeV$^2c^T_{i,m}$ & $10^3$GeV$^2c^T_{m}$\\
 \hline\endfirsthead

 \hline\hline $n$ & $10^3$GeV$^2C^T_{r,n}$ & $10^3$GeV$^2C^T_{i,n}$ & $10^3$GeV$^2C^T_{n}$
        & $m$ & $10^3$GeV$^2c^T_{r,m}$ & $10^3$GeV$^2c^T_{i,m}$ & $10^3$GeV$^2c^T_{m}$\\
 \hline\endhead

 \hline\hline 
 \endfoot

 \hline\endlastfoot

 1 & $2.20^{+0.05}_{-0.06}$ & $2.06^{+0.09}_{-0.14}$ & $4.26^{+0.14}_{-0.20}$ & & &\\
 2 &$ 3.93^{+0.18}_{-0.26}$ & $3.94^{+0.18}_{-0.26}$ & $7.87^{+0.35}_{-0.52}$ & 1 & $3.53^{+0.11}_{-0.15}$ & $3.38^{+0.16}_{-0.23}$ & $6.91^{+0.26}_{-0.38}$ \\
 3 &$ -3.30^{-0.24}_{+0.37}$ & $-3.55^{-0.16}_{+0.24}$ & $-6.85^{-0.40}_{+0.61}$ & 2 & $1.51^{-0.12}_{+0.20}$ & $0.99^{+0.05}_{-0.07}$ & $2.50^{-0.07}_{+0.14}$ \\
 4 & $-1.71^{+0.01}_{-0.03}$ & $-1.46^{-0.07}_{+0.10}$ & $-3.17^{-0.05}_{+0.06}$ & & &\\
 5 & $\equiv 0$~~~~~ & $\equiv 0$~~~~~ & $\equiv 0$~~~~~ & & &\\
 6 & $\equiv 0$~~~~~ & $\equiv 0$~~~~~ & $\equiv 0$~~~~~ & & &\\
 7 &$ -0.29^{-0.03}_{+0.05}$ & $-0.94^{-0.06}_{+0.08}$ & $-1.23^{-0.09}_{+0.13}$ & 3 & $-0.13^{-0.02}_{+0.02}$ & $-0.41^{-0.03}_{+0.04}$ & $-0.53^{-0.04}_{+0.06}$ \\
 8 &$ -0.08^{-0.02}_{+0.04}$ & $0.18^{+0.01}_{-0.02}$ & $0.10^{-0.01}_{+0.02}$ & 4 & $-0.04^{-0.01}_{+0.02}$ & $0.07^{+0.00}_{-0.01}$ & $0.04^{-0.01}_{+0.01}$ \\
 9 &$ 1.24^{+0.03}_{-0.02}$ & $-0.70^{-0.04}_{+0.06}$ & $0.55^{-0.02}_{+0.04}$ & 5 & $0.49^{+0.01}_{-0.01}$ & $-0.33^{-0.02}_{+0.03}$ & $0.16^{-0.01}_{+0.02}$ \\
 10 &$ -4.91^{-0.29}_{+0.39}$ & $-0.61^{-0.04}_{+0.05}$ & $-5.52^{-0.32}_{+0.44}$ & 6 & $-2.05^{-0.13}_{+0.17}$ & $-0.22^{-0.01}_{+0.02}$ & $-2.28^{-0.14}_{+0.19}$ \\
 11 &$ 1.14^{+0.10}_{-0.16}$ & $0.61^{+0.04}_{-0.05}$ & $1.76^{+0.14}_{-0.21}$ & 7 & $0.49^{+0.05}_{-0.07}$ & $0.27^{+0.02}_{-0.02}$ & $0.76^{+0.06}_{-0.09}$ \\
 12 & $\equiv 0$~~~~~ & $\equiv 0$~~~~~ & $\equiv 0$~~~~~ & & &\\
 13 & $\equiv 0$~~~~~ & $\equiv 0$~~~~~ & $\equiv 0$~~~~~ & & &\\
 14 & $\equiv 0$~~~~~ & $\equiv 0$~~~~~ & $\equiv 0$~~~~~ & & &\\
 15 & $\equiv 0$~~~~~ & $\equiv 0$~~~~~ & $\equiv 0$~~~~~ & & &\\
 16 & $\equiv 0$~~~~~ & $\equiv 0$~~~~~ & $\equiv 0$~~~~~ & & &\\
 17 & $\equiv 0$~~~~~ & $\equiv 0$~~~~~ & $\equiv 0$~~~~~ & 8 & $\equiv 0$~~~~~ & $\equiv 0$~~~~~ & $\equiv 0$~~~~~\\
 18 & $\equiv 0$~~~~~ & $\equiv 0$~~~~~ & $\equiv 0$~~~~~ & 9 & $\equiv 0$~~~~~ & $\equiv 0$~~~~~ & $\equiv 0$~~~~~\\
 19 & -~~~~~~ & -~~~~~~ & -~~~~~~ & 10 & -~~~~~~ & -~~~~~~ & -~~~~~~\\
 20 & -~~~~~~ & -~~~~~~ & -~~~~~~ & 11 & -~~~~~~ & -~~~~~~ & -~~~~~~\\
 21 & -~~~~~~ & -~~~~~~ & -~~~~~~ & 12 & -~~~~~~ & -~~~~~~ & -~~~~~~\\
 22 & -~~~~~~ & -~~~~~~ & -~~~~~~ & & & \\
 23 & -~~~~~~ & -~~~~~~ & -~~~~~~ & & & \\
 24 & -~~~~~~ & -~~~~~~ & -~~~~~~ & & & \\
 25 & -~~~~~~ & -~~~~~~ & -~~~~~~ & 13 & -~~~~~~ & -~~~~~~ & -~~~~~~\\
 26 &$ -1.44^{-0.06}_{+0.13}$ & $-2.02^{+0.08}_{-0.09}$ & $-3.47^{+0.02}_{+0.03}$ & 14 & $-0.63^{-0.03}_{+0.06}$ & $-0.89^{+0.04}_{-0.04}$ & $-1.52^{+0.01}_{+0.01}$ \\
 27 &$ 1.46^{-0.12}_{+0.09}$ & $2.96^{-0.21}_{+0.24}$ & $4.41^{-0.32}_{+0.34}$ & 15 & $0.61^{-0.05}_{+0.04}$ & $1.26^{-0.09}_{+0.11}$ & $1.88^{-0.14}_{+0.14}$ \\
 28 & $\equiv 0$~~~~~ & $\equiv 0$~~~~~ & $\equiv 0$~~~~~ & 16 & $\equiv 0$~~~~~ & $\equiv 0$~~~~~ & $\equiv 0$~~~~~\\
 29 & $\equiv 0$~~~~~ & $\equiv 0$~~~~~ & $\equiv 0$~~~~~ & 17 & $\equiv 0$~~~~~ & $\equiv 0$~~~~~ & $\equiv 0$~~~~~\\
 30 &$ 0.58^{+0.04}_{-0.05}$ & $0.28^{+0.02}_{-0.02}$ & $0.86^{+0.06}_{-0.08}$ & 18 & $0.18^{+0.01}_{-0.02}$ & $0.09^{+0.01}_{-0.01}$ & $0.27^{+0.02}_{-0.03}$ \\
 31 & $\equiv 0$~~~~~ & $\equiv 0$~~~~~ & $\equiv 0$~~~~~ & 19 & $\equiv 0$~~~~~ & $\equiv 0$~~~~~ & $\equiv 0$~~~~~\\
 32 & $\equiv 0$~~~~~ & $\equiv 0$~~~~~ & $\equiv 0$~~~~~ & & &\\
 33 & $\equiv 0$~~~~~ & $\equiv 0$~~~~~ & $\equiv 0$~~~~~ & & &\\
 34 &$ -0.43^{-0.44}_{+0.60}$ & $-4.08^{-0.19}_{+0.29}$ & $-4.51^{-0.63}_{+0.88}$ & 20 & $-0.41^{-0.30}_{+0.40}$ & $-2.76^{-0.13}_{+0.20}$ & $-3.17^{-0.43}_{+0.60}$ \\
 35 &$ -5.26^{-0.64}_{+0.91}$ & $-7.92^{-0.40}_{+0.59}$ & $-13.18^{-1.04}_{+1.50}$ & 21 & $-3.63^{-0.43}_{+0.62}$ & $-5.35^{-0.28}_{+0.40}$ & $-8.98^{-0.71}_{+1.02}$ \\
 36 & $\equiv 0$~~~~~ & $\equiv 0$~~~~~ & $\equiv 0$~~~~~ & & &\\
 37 & $\equiv 0$~~~~~ & $\equiv 0$~~~~~ & $\equiv 0$~~~~~ & & &\\
 38 &$ -5.34^{-0.30}_{+0.44}$ & $-4.93^{-0.33}_{+0.47}$ & $-10.27^{-0.63}_{+0.91}$ & 22 & $-2.99^{-0.18}_{+0.26}$ & $-2.73^{-0.20}_{+0.28}$ & $-5.72^{-0.38}_{+0.54}$ \\
 39 &$ 1.91^{-0.10}_{+0.11}$ & $4.82^{+0.29}_{-0.41}$ & $6.73^{+0.19}_{-0.30}$ & 23 & $2.21^{-0.02}_{+0.01}$ & $4.24^{+0.24}_{-0.35}$ & $6.45^{+0.22}_{-0.35}$ \\
 40 & $2.15^{+0.10}_{-0.15}$ & $2.15^{+0.10}_{-0.14}$ & $4.30^{+0.19}_{-0.29}$ & & &\\
 41 & $\equiv 0$~~~~~ & $\equiv 0$~~~~~ & $\equiv 0$~~~~~ & & &\\
 42 &$ -2.94^{-0.13}_{+0.20}$ & $-2.92^{-0.13}_{+0.19}$ & $-5.87^{-0.26}_{+0.40}$ & 24 & $-1.94^{-0.09}_{+0.14}$ & $-1.93^{-0.09}_{+0.13}$ & $-3.87^{-0.18}_{+0.27}$ \\
 43 &$ 3.41^{+0.16}_{-0.25}$ & $3.43^{+0.15}_{-0.23}$ & $6.84^{+0.32}_{-0.47}$ & 25 & $2.23^{+0.11}_{-0.17}$ & $2.25^{+0.10}_{-0.15}$ & $4.48^{+0.21}_{-0.32}$ \\
 44 &$ 6.55^{+0.29}_{-0.43}$ & $6.52^{+0.29}_{-0.43}$ & $13.06^{+0.58}_{-0.86}$ & 26 & $4.29^{+0.20}_{-0.29}$ & $4.27^{+0.20}_{-0.29}$ & $8.57^{+0.39}_{-0.59}$ \\
 45 & $\equiv 0$~~~~~ & $\equiv 0$~~~~~ & $\equiv 0$~~~~~ & 27 & $\equiv 0$~~~~~ & $\equiv 0$~~~~~ & $\equiv 0$~~~~~\\
 46 &$ -0.00^{-0.07}_{+0.09}$ & $-0.84^{-0.05}_{+0.07}$ & $-0.84^{-0.12}_{+0.16}$ & 28 & $0.04^{-0.03}_{+0.04}$ & $-0.31^{-0.02}_{+0.03}$ & $-0.27^{-0.05}_{+0.06}$ \\
 47 &$ -14.18^{-0.94}_{+1.33}$ & $-4.59^{-0.28}_{+0.39}$ & $-18.77^{-1.22}_{+1.72}$ & 29 & $-6.11^{-0.42}_{+0.59}$ & $-2.02^{-0.13}_{+0.17}$ & $-8.13^{-0.55}_{+0.77}$ \\
 48 & -~~~~~~ & -~~~~~~ & -~~~~~~ & 30 & -~~~~~~ & -~~~~~~ & -~~~~~~\\
 49 & $\equiv 0$~~~~~ & $\equiv 0$~~~~~ & $\equiv 0$~~~~~ & 31 & $\equiv 0$~~~~~ & $\equiv 0$~~~~~ & $\equiv 0$~~~~~\\
 50 & $\equiv 0$~~~~~ & $\equiv 0$~~~~~ & $\equiv 0$~~~~~ & 32 & $\equiv 0$~~~~~ & $\equiv 0$~~~~~ & $\equiv 0$~~~~~\\
 51 & -~~~~~~ & -~~~~~~ & -~~~~~~ & 33 & -~~~~~~ & -~~~~~~ & -~~~~~~\\
 52 &$ -0.04^{-0.06}_{+0.09}$ & $-0.19^{-0.01}_{+0.01}$ & $-0.23^{-0.07}_{+0.11}$ & 34 & $-0.05^{-0.04}_{+0.06}$ & $-0.15^{-0.01}_{+0.01}$ & $-0.21^{-0.05}_{+0.08}$ \\
 53 &$ 1.36^{+0.06}_{-0.08}$ & $1.35^{+0.06}_{-0.09}$ & $2.71^{+0.12}_{-0.17}$ & 35 & $0.91^{+0.04}_{-0.06}$ & $0.91^{+0.04}_{-0.06}$ & $1.81^{+0.08}_{-0.12}$ \\
 54 &$ -3.36^{-0.10}_{+0.15}$ & $-3.23^{-0.14}_{+0.21}$ & $-6.58^{-0.25}_{+0.36}$ & 36 & $-2.11^{-0.07}_{+0.10}$ & $-2.02^{-0.09}_{+0.14}$ & $-4.12^{-0.16}_{+0.23}$ \\
 55 &$ -1.43^{-0.20}_{+0.32}$ & $-1.80^{-0.08}_{+0.12}$ & $-3.23^{-0.28}_{+0.44}$ & 37 & $-1.07^{-0.14}_{+0.22}$ & $-1.33^{-0.06}_{+0.09}$ & $-2.40^{-0.20}_{+0.31}$ \\
 56 &$ -2.79^{-0.03}_{+0.03}$ & $-2.53^{-0.11}_{+0.17}$ & $-5.32^{-0.14}_{+0.20}$ & 38 & $-1.74^{-0.02}_{+0.02}$ & $-1.56^{-0.07}_{+0.11}$ & $-3.30^{-0.09}_{+0.12}$ \\
 57 & $\equiv 0$~~~~~ & $\equiv 0$~~~~~ & $\equiv 0$~~~~~ & & &\\
 58 & $\equiv 0$~~~~~ & $\equiv 0$~~~~~ & $\equiv 0$~~~~~ & & &\\
 59 & $\equiv 0$~~~~~ & $\equiv 0$~~~~~ & $\equiv 0$~~~~~ & & &\\
 60 & $\equiv 0$~~~~~ & $\equiv 0$~~~~~ & $\equiv 0$~~~~~ & & &\\
 61 & $\equiv 0$~~~~~ & $\equiv 0$~~~~~ & $\equiv 0$~~~~~ & & &\\
 62 &$ -2.83^{-0.15}_{+0.24}$ & $-2.89^{-0.13}_{+0.19}$ & $-5.72^{-0.28}_{+0.43}$ & 39 & $-1.87^{-0.11}_{+0.16}$ & $-1.91^{-0.09}_{+0.13}$ & $-3.78^{-0.19}_{+0.29}$ \\
 63 &$ -6.56^{-0.30}_{+0.45}$ & $-6.59^{-0.29}_{+0.44}$ & $-13.15^{-0.60}_{+0.88}$ & 40 & $-4.26^{-0.20}_{+0.30}$ & $-4.28^{-0.20}_{+0.29}$ & $-8.54^{-0.40}_{+0.59}$ \\
 64 &$ 3.74^{+0.16}_{-0.23}$ & $8.47^{+0.38}_{-0.56}$ & $12.22^{+0.54}_{-0.79}$ & 41 & $2.42^{+0.11}_{-0.15}$ & $5.55^{+0.26}_{-0.38}$ & $7.97^{+0.36}_{-0.53}$ \\
 65 &$ 0.10^{-0.01}_{+0.03}$ & $0.05^{+0.00}_{-0.00}$ & $0.15^{-0.01}_{+0.02}$ & 42 & $0.04^{-0.01}_{+0.02}$ & $0.01^{+0.00}_{-0.00}$ & $0.05^{-0.01}_{+0.02}$ \\
 66 & $\equiv 0$~~~~~ & $\equiv 0$~~~~~ & $\equiv 0$~~~~~ & 43 & $\equiv 0$~~~~~ & $\equiv 0$~~~~~ & $\equiv 0$~~~~~\\
 67 &$ 2.26^{-0.12}_{+0.12}$ & $0.35^{-0.02}_{+0.02}$ & $2.62^{-0.13}_{+0.15}$ & 44 & $1.45^{-0.07}_{+0.08}$ & $0.22^{-0.01}_{+0.02}$ & $1.67^{-0.09}_{+0.09}$ \\
 68 &$ -1.76^{-0.16}_{+0.23}$ & $-2.16^{-0.13}_{+0.19}$ & $-3.92^{-0.29}_{+0.42}$ & 45 & $-0.36^{-0.07}_{+0.09}$ & $-0.61^{-0.05}_{+0.07}$ & $-0.96^{-0.12}_{+0.16}$ \\
 69 &$ -4.23^{-0.31}_{+0.46}$ & $-4.76^{-0.25}_{+0.36}$ & $-8.99^{-0.56}_{+0.82}$ & 46 & $-2.81^{-0.21}_{+0.31}$ & $-3.15^{-0.17}_{+0.25}$ & $-5.97^{-0.38}_{+0.55}$ \\
 70 & $\equiv 0$~~~~~ & $\equiv 0$~~~~~ & $\equiv 0$~~~~~ & 47 & $\equiv 0$~~~~~ & $\equiv 0$~~~~~ & $\equiv 0$~~~~~\\
 71 & $\equiv 0$~~~~~ & $\equiv 0$~~~~~& $\equiv 0$~~~~~ & 48 & $-0.82^{-0.04}_{+0.06}$ & $-0.83^{-0.04}_{+0.06}$ & $-1.65^{-0.08}_{+0.12}$ \\
 72 & $\equiv 0$~~~~~ & $\equiv 0$~~~~~ & $\equiv 0$~~~~~ & & &\\
 73 & $1.66^{+0.08}_{-0.13}$ & $1.67^{+0.07}_{-0.11}$ & $3.34^{+0.16}_{-0.24}$ & & &\\
 74 & -~~~~~~ & -~~~~~~ & -~~~~~~ & 49 & -~~~~~~ & -~~~~~~ & -~~~~~~\\
 75 & $\equiv 0$~~~~~ & $\equiv 0$~~~~~ & $\equiv 0$~~~~~ & 50 & $\equiv 0$~~~~~ & $\equiv 0$~~~~~ & $\equiv 0$~~~~~\\
 76 & -~~~~~~ & -~~~~~~ & -~~~~~~ & & & \\
 77 &$ -1.20^{-0.07}_{+0.11}$ & $-1.24^{-0.06}_{+0.08}$ & $-2.44^{-0.13}_{+0.19}$ & 51 & $-0.76^{-0.05}_{+0.07}$ & $-0.79^{-0.04}_{+0.05}$ & $-1.56^{-0.08}_{+0.13}$ \\
 78 &$ 5.46^{+0.27}_{-0.41}$ & $5.51^{+0.25}_{-0.37}$ & $10.97^{+0.51}_{-0.78}$ & 52 & $3.45^{+0.18}_{-0.27}$ & $3.48^{+0.16}_{-0.24}$ & $6.93^{+0.34}_{-0.51}$ \\
 79 &$ -4.71^{-0.29}_{+0.44}$ & $-4.96^{-0.22}_{+0.33}$ & $-9.67^{-0.52}_{+0.77}$ & 53 & $-3.02^{-0.20}_{+0.29}$ & $-3.19^{-0.15}_{+0.22}$ & $-6.20^{-0.34}_{+0.51}$ \\
 80 & $\equiv 0$~~~~~ & $\equiv 0$~~~~~ & $\equiv 0$~~~~~ & & &\\
 81 &$ 7.99^{+0.54}_{-0.77}$ & $10.12^{+0.62}_{-0.85}$ & $18.11^{+1.16}_{-1.61}$ & 54 & $3.44^{+0.24}_{-0.34}$ & $4.37^{+0.28}_{-0.38}$ & $7.81^{+0.52}_{-0.72}$ \\
 82 & -~~~~~~ & -~~~~~~ & -~~~~~~ & 55 & -~~~~~~ & -~~~~~~ & -~~~~~~\\
 83 &$ -7.07^{-0.47}_{+0.67}$ & $-1.06^{-0.06}_{+0.09}$ & $-8.12^{-0.54}_{+0.76}$ & 56 & $-2.96^{-0.21}_{+0.29}$ & $-0.40^{-0.03}_{+0.03}$ & $-3.36^{-0.23}_{+0.32}$ \\
 84 & -~~~~~~ & -~~~~~~ & -~~~~~~ & 57 & -~~~~~~ & -~~~~~~ & -~~~~~~\\
 85 & $\equiv 0$~~~~~ & $\equiv 0$~~~~~ & $\equiv 0$~~~~~ & 58 & $\equiv 0$~~~~~ & $\equiv 0$~~~~~ & $\equiv 0$~~~~~\\
 86 & -~~~~~~ & -~~~~~~ & -~~~~~~ & & & \\
 87 &$ -22.85^{-1.32}_{+2.00}$ & $-23.69^{-1.06}_{+1.57}$ & $-46.54^{-2.38}_{+3.57}$ & 59 & $-15.16^{-0.90}_{+1.36}$ & $-15.72^{-0.73}_{+1.07}$ & $-30.88^{-1.63}_{+2.43}$ \\
 88 &$ -0.81^{+0.02}_{-0.03}$ & $-0.63^{-0.03}_{+0.04}$ & $-1.44^{-0.01}_{+0.01}$ & 60 & $-0.43^{+0.02}_{-0.03}$ & $-0.31^{-0.01}_{+0.02}$ & $-0.73^{+0.00}_{-0.01}$ \\
 89 &$ 5.07^{+0.27}_{-0.40}$ & $5.19^{+0.23}_{-0.34}$ & $10.26^{+0.50}_{-0.74}$ & 61 & $3.37^{+0.18}_{-0.27}$ & $3.45^{+0.16}_{-0.24}$ & $6.82^{+0.34}_{-0.51}$ \\
 90 &$ 1.55^{+0.11}_{-0.16}$ & $1.69^{+0.08}_{-0.11}$ & $3.24^{+0.19}_{-0.28}$ & 62 & $1.14^{+0.08}_{-0.12}$ & $1.23^{+0.06}_{-0.08}$ & $2.38^{+0.14}_{-0.20}$ \\
 91 &$ 4.82^{+0.22}_{-0.34}$ & $4.78^{+0.21}_{-0.32}$ & $9.60^{+0.43}_{-0.65}$ & 63 & $3.22^{+0.15}_{-0.23}$ & $3.19^{+0.15}_{-0.22}$ & $6.41^{+0.30}_{-0.45}$ \\
 92 &$ -4.93^{-0.29}_{+0.43}$ & $-5.12^{-0.23}_{+0.34}$ & $-10.04^{-0.51}_{+0.77}$ & 64 & $-3.19^{-0.19}_{+0.29}$ & $-3.32^{-0.15}_{+0.23}$ & $-6.51^{-0.35}_{+0.52}$ \\
 93 &$ 9.51^{+0.44}_{-0.68}$ & $9.49^{+0.42}_{-0.63}$ & $19.01^{+0.86}_{-1.31}$ & 65 & $6.31^{+0.30}_{-0.47}$ & $6.30^{+0.29}_{-0.43}$ & $12.61^{+0.60}_{-0.90}$ \\
 94 & $\equiv 0$~~~~~ & $\equiv 0$~~~~~ & $\equiv 0$~~~~~ & & &\\
 95 & $\equiv 0$~~~~~ & $\equiv 0$~~~~~ & $\equiv 0$~~~~~ & & &\\
 96 & $\equiv 0$~~~~~ & $\equiv 0$~~~~~ & $\equiv 0$~~~~~ & & &\\
 97 & -~~~~~~ & -~~~~~~ & -~~~~~~ & 66 & -~~~~~~ & -~~~~~~ & -~~~~~~\\
 98 &$ 6.92^{+0.46}_{-0.65}$ & $3.35^{+0.20}_{-0.28}$ & $10.28^{+0.67}_{-0.94}$ & 67 & $2.93^{+0.20}_{-0.29}$ & $1.41^{+0.09}_{-0.12}$ & $4.34^{+0.29}_{-0.41}$ \\
 99 & $\equiv 0$~~~~~ & $\equiv 0$~~~~~ & $\equiv 0$~~~~~ & 68 & $\equiv 0$~~~~~ & $\equiv 0$~~~~~ & $\equiv 0$~~~~~\\
 100 & $\equiv 0$~~~~~ & $\equiv 0$~~~~~ & $\equiv 0$~~~~~ & 69 & $\equiv 0$~~~~~ & $\equiv 0$~~~~~ & $\equiv 0$~~~~~\\
 101 & $\equiv 0$~~~~~ & $\equiv 0$~~~~~ & $\equiv 0$~~~~~ & 70 & $\equiv 0$~~~~~ & $\equiv 0$~~~~~ & $\equiv 0$~~~~~\\
 102 & -~~~~~~ & -~~~~~~ & -~~~~~~ & & & \\
 103 & -~~~~~~ & -~~~~~~ & -~~~~~~ & & & \\
 104 & -~~~~~~ & -~~~~~~ & -~~~~~~ & & & \\
 105 &$ -3.06^{+0.16}_{-0.17}$ & $0.00^{+0.00}_{+0.00}$ & $-3.06^{+0.16}_{-0.17}$ & 71 & $-1.96^{+0.10}_{-0.11}$ & $0.00^{+0.00}_{+0.00}$ & $-1.96^{+0.10}_{-0.11}$ \\
 106 &$ -2.07^{-0.78}_{+1.04}$ & $-6.97^{-0.45}_{+0.65}$ & $-9.04^{-1.23}_{+1.69}$ & 72 & $-1.58^{-0.53}_{+0.70}$ & $-4.74^{-0.31}_{+0.45}$ & $-6.32^{-0.84}_{+1.15}$ \\
 107 &$ -0.15^{+0.06}_{-0.10}$ & $0.06^{+0.00}_{-0.00}$ & $-0.09^{+0.07}_{-0.10}$ & 73 & $0.01^{+0.05}_{-0.07}$ & $0.15^{+0.01}_{-0.01}$ & $0.15^{+0.05}_{-0.08}$ \\
 108 & -~~~~~~ & -~~~~~~ & -~~~~~~ & 74 & -~~~~~~ & -~~~~~~ & -~~~~~~\\
 109 & -~~~~~~ & -~~~~~~ & -~~~~~~ & 75 & -~~~~~~ & -~~~~~~ & -~~~~~~\\
 110 & $\equiv 0$~~~~~ & $\equiv 0$~~~~~ & $\equiv 0$~~~~~ & & &\\
 \hline\hline
 \end{longtable}
 }

 The calculation process is too complicated; to avoid  possible mistakes,
 the expansion in Eqs.(\ref{dB1}) and (\ref{Seff2Ea}) and most of the other calculations
 are done by computer.
 To check the correctness of our results, we examine them in various alternative ways.
 First, because these results contain the original results in \cite{our5,oura2}, if we switch off the tensor sources,
 as a check, we must recover the original results. Second, some terms in Table \ref{indter} and the $p^6$ order
 operators in Table \ref{Ccs6} have two parts;
 we calculate them separately. $C,P$ and Hermitian invariance constrain the two parts of the coefficients
 to be equal(or with a minus sign difference).
 Our analytical results for the separate parts must give the same coefficients. Third,
 if we switch off the quark self-energy, all the LECs, except the contact terms', must be zeros\cite{our5}.
 This places a strong restriction  on our results.
 We found that this restriction can be realized only when we use the new relations given
 in Appendix \ref{newrelation}.
 Fourth, in the $p^6$ order,  because of the strict constraint conditions in Eq.(\ref{p4rw1p}),
 we have $339-56=283$ constraint conditions. They are also a strong restriction on our results.
 With all the above  assessments, we are confident of the reliability of our numerical results for LECs.

 Ref.\cite{tensor1} told us that operators contributing to the odd-intrinsic-parity part with tensor fields start from the $p^8$ order,
 and we showed in Sec. \ref{diff} that the odd-intrinsic-parity parts with tensor fields cannot independently exist.
 So we have obtained all the LECs to the $p^6$ order, with scalar, pseudoscalar, vector, axial-vector and tensor sources,
 including the normal and anomalous parts, and two- and three- flavor cases.
 We found that in our method, all the contact terms' coefficients are divergent,
 except $H_1$ in the $p^4$ order normal part.

 \section{summary}\label{summa}

 To summarize our results, we extended our previous computation for LECs in Refs.\cite{our5,WQ1}
  include tensor sources, and obtain all  LECs of order $p^4$ and $p^6$ for the chiral Lagrangian.
 We find that the operators given in Ref.\cite{tensor1} are not all independent
 because of certain relations involving epsilon.
 Adding these relations, we can reduce 22 operators for $n$-flavor, 21 for three-flavor case,
 and 13 for two-flavor case,
  leaving 98 independent operators for $n$-flavor, 92 for three-flavor, and 65 for two-flavor cases.
 Our LECs are presented with numerical values for both two- and three- flavors cases.
 We also find that the odd-intrinsic-parity parts chiral Lagrangian with tensor sources cannot independently exist.
  Thus, up to the $p^6$ order, we have already given all the LECs' values,
 although,  in obtaining these values, we have made many approximations.
 As a first step in estimating values,
 these results not only  provide the sign and order of magnitude, but also the quantitative information of LECs.
  With improvements in the computation procedure,
 we expect more precise results  to be obtained in the future.
 Another direction of research is
 applying the present chiral Lagrangian with tensor sources,
 adding the known LECs to various low-energy ($\pi,K,\eta$) processes.
 We hope more physical results can be obtained.

 \section*{Acknowledgments}
 This work was supported by the National Science Foundation of China
 (NSFC) under Grants No.11147192, No. 11205034, and No. 11075085;
  the Specialized Research Fund for the Doctoral Program of High Education of
 China No. 20110002110010; the Scientific Research Foundation of GuangXi University Grant No. XBZ100686;
 and the Tsinghua University Initiative Scientific Research Program.

 \appendix
 \section{relations among our symbols and those used in Ref.\cite{tensor1}}\label{symrel}
 To help in understanding the mutual relation between the notations in our current formulation and those in
 Ref.\cite{tensor1}, we provide a comparison in Table \ref{symb}.
 \begin{table*}[h]
 \caption{\label{symb}Comparison between notations
 introduced in Ref.\cite{tensor1} (first and third columns) and those defined in the current paper (second and fourth columns).}
 \begin{eqnarray}
 \begin{array}{cc|cc}
 \hline\hline
 \mbox{Ref.~\cite{tensor1}}&\mbox{Present paper}& \mbox{Ref.~\cite{tensor1}}&\mbox{Present paper}\\
 \hline \nabla^\mu & d^\mu &\chi_-^{\mu} & 4iB_0 d^{\mu}p_\Omega-4iB_0 s_\Omega a_\Omega^{\mu}-4iB_0a_\Omega^{\mu} s_\Omega\\
 u&\Omega &f_+^{\mu\nu} & 2V_\Omega^{\mu\nu}-2i(a_\Omega^\mu a_\Omega^{\nu}-a_\Omega^{\nu}a_\Omega^{\mu}) \\
 u^\mu & 2a_\Omega^\mu& \nabla^{\lambda}f_+^{\mu\nu} & 2d^{\lambda}V_\Omega^{\mu\nu}-2id^{\lambda}(a_\Omega^\mu a_\Omega^{\nu}
 -a_\Omega^{\nu}a_\Omega^{\mu})\\
 \chi & \chi &f_-^{\mu\nu} & -2(d^\mu a_\Omega^{\nu}-d^{\nu}a_\Omega^{\mu}) \\
 \chi_+ & 4B_0s_\Omega &\nabla^{\lambda}f_-^{\mu\nu} & -2(d^{\lambda}d^\mu a_\Omega^{\nu}-d^{\lambda}d^{\nu}a_\Omega^{\mu}) \\
 \chi_+^{\mu} & 4B_0 d^{\mu}s_\Omega+4B_0 p_\Omega a_\Omega^{\mu}+4B_0a_\Omega^{\mu}p_\Omega & h^{\mu\nu} & 2(d^\mu a_\Omega^{\nu}+d^{\nu}a_\Omega^{\mu})\\
 \chi_- & 4iB_0p_\Omega &\Gamma^\mu &-iv_\Omega^\mu \\
 t_{+}^{\mu\nu} & t_{+,\Omega}^{\mu\nu} & \tm^{\mu\nu} & t_{-,\Omega}^{\mu\nu}\\
 \hline\hline
 \end{array}\notag
 \end{eqnarray}
 \end{table*}

 \section{new relations}\label{newrelation}
 In this appendix, we list the new relations when using the epsilon relations in Sec. \ref{diff}.
 The left-hand sides of (\ref{dependoperators}) are considered to be dependent and  reducible.
 \begin{eqnarray}
 Y_{23}&=&\frac{1}{2}Y_{9}-Y_{12},\notag\\
 Y_{24}&=&\frac{1}{2}Y_{9}-Y_{11},\notag\\
 Y_{25}&=&Y_{10}-Y_{13},\notag\\
 Y_{26}&=&\frac{1}{2}Y_{14}-Y_{15},\notag\\
 Y_{27}&=&\frac{1}{2}Y_{16}-Y_{18},\notag\\
 Y_{28}&=&\frac{1}{2}Y_{16}-Y_{17},\notag\\
 Y_{29}&=&Y_{19}-Y_{20},\notag\\
 Y_{30}&=&Y_{21}-Y_{22},\notag\\
 Y_{53}&=&-\frac{1}{2}Y_{11}+\frac{1}{2}Y_{12}+\frac{1}{2}Y_{51}-Y_{52}+Y_{90},\notag\\
 Y_{56}&=&\frac{1}{2}Y_{54}-Y_{55},\notag\\
 Y_{81}&=&\frac{1}{2}Y_{32}-\frac{1}{2n_f}Y_{35},\notag\\
 Y_{83}&=&\frac{1}{2}Y_{36}-\frac{1}{2n_f}Y_{38}-Y_{82},\notag\\
 Y_{89}&=&Y_{88},\notag\\
 Y_{91}&=&Y_{90},\notag\\
 Y_{93}&=&Y_{92},\notag\\
 Y_{104}&=&\frac{1}{2}Y_{32}-\frac{1}{2n_f}Y_{35},\notag\\
 Y_{109}&=&\frac{1}{2}Y_{36}-\frac{1}{2n_f}Y_{38}-Y_{106},\notag\\
 Y_{110}&=&-\frac{1}{2}Y_{82}+\frac{1}{2}Y_{92}+\frac{1}{2}Y_{106}+Y_{107},\notag\\
 Y_{111}&=&-\frac{1}{4}Y_{36}+\frac{1}{4n_f}Y_{38}+\frac{1}{2}Y_{82}+\frac{1}{2}Y_{92}+\frac{1}{2}Y_{106}+Y_{108},\notag\\
 Y_{115}&=&-\frac{1}{2}Y_{1}+Y_{2}-\frac{1}{2}Y_{57}+Y_{58}+Y_{84}-Y_{96}+Y_{97}+Y_{114},\notag\\
 Y_{116}&=&\frac{1}{2}Y_{69}-Y_{70}+Y_{72}+\frac{1}{2}Y_{75}-\frac{1}{n_f}Y_{80}
 -Y_{86}-2Y_{98}-Y_{99},\notag\\
 Y_{119}&=&0.\label{dependoperators}
 \end{eqnarray}

 \section{$\mathcal{Z}_n$ coefficients}\label{Zcs}
 \begin{eqnarray}
 \mathcal{Z}^T_{1}&=&\int dK\bigg[-10 \tau^3\sk
 +\frac{10}{3} \tau^4 k^2\sk
 +\frac{40}{3} \tau^4\sk^3
 -\frac{2}{9} \tau^5 k^4\sk
 -2 \tau^5 k^2\sk^3
 -\frac{8}{3} \tau^5\sk^5
 \bigg],\notag\\
 \mathcal{Z}^T_{2}&=&\int dK\bigg[+10 \tau^3\sk
 -\frac{8}{3} \tau^4 k^2\sk
 -\frac{40}{3} \tau^4\sk^3
 +2 \tau^5 k^2\sk^3
 +\frac{8}{3} \tau^5\sk^5
 \bigg],\notag\\
 \mathcal{Z}^T_{3}&=&\int dK\bigg[-10 \tau^3\sk
 +4 \tau^4 k^2\sk
 +\frac{40}{3} \tau^4\sk^3
 -\frac{2}{9} \tau^5 k^4\sk
 -2 \tau^5 k^2\sk^3
 -\frac{8}{3} \tau^5\sk^5
 \bigg],\notag\\
 \mathcal{Z}^T_{4}&=&\int dK\bigg[-10 \tau^3\sk
 +2 \tau^4 k^2\sk
 +\frac{40}{3} \tau^4\sk^3
 +\frac{2}{9} \tau^5 k^4\sk
 -2 \tau^5 k^2\sk^3
 -\frac{8}{3} \tau^5\sk^5
 \bigg],\notag\\
 \mathcal{Z}^T_{5}&=&\int dK\bigg[+4 \tau^3\sk
 -\frac{2}{3} \tau^4 k^2\sk
 -\frac{4}{3} \tau^4\sk^3
 -\frac{1}{9} \tau^4 k^4\skp
 \bigg],\notag\\
 \mathcal{Z}^T_{6}&=&\int dK\bigg[-\frac{22}{3} \tau^3\sk
 +\frac{1}{3} \tau^3 k^2\skp
 +2 \tau^4 k^2\sk
 +\frac{8}{3} \tau^4\sk^3
 \bigg],\notag\\
 \mathcal{Z}^T_{7}&=&\int dK\bigg[+6 \tau^3\sk
 + \tau^3 k^2\skp
 -\frac{4}{3} \tau^4 k^2\sk
 -\frac{8}{3} \tau^4\sk^3
 -\frac{2}{9} \tau^4 k^4\skp
 \bigg],\notag\\
 \mathcal{Z}^T_{8}&=&\int dK\bigg[+4 \tau^3\sk
 -\frac{4}{3} \tau^4 k^2\sk
 -\frac{4}{3} \tau^4\sk^3
 +\frac{1}{9} \tau^4 k^4\skp
 \bigg],\notag\\
 \mathcal{Z}^T_{9}&=&\int dK\bigg[-4 \tau^3\sk
 +\frac{4}{3} \tau^4 k^2\sk
 +\frac{4}{3} \tau^4\sk^3
 -\frac{1}{9} \tau^4 k^4\skp
 \bigg],\notag\\
 \mathcal{Z}^T_{10}&=&\int dK\bigg[-6 \tau^3\sk
 + \tau^3 k^2\skp
 +\frac{4}{3} \tau^4 k^2\sk
 +\frac{8}{3} \tau^4\sk^3
 -\frac{2}{9} \tau^4 k^4\skp
 \bigg],\notag\\
 \mathcal{Z}^T_{11}&=&\int dK\bigg[-\frac{20}{3} \tau^3\sk
 +\frac{2}{3} \tau^3 k^2\skp
 +\frac{4}{3} \tau^4 k^2\sk
 +\frac{8}{3} \tau^4\sk^3
 -\frac{2}{9} \tau^4 k^4\skp
 \bigg],\notag\\
 \mathcal{Z}^T_{12}&=&\int dK\bigg[+\frac{2}{3} \tau^4 k^2\sk
 -\frac{2}{9} \tau^4 k^4\skp
 \bigg],\notag\\
 \mathcal{Z}^T_{13}&=&\int dK\bigg[+4 \tau^3\sk
 +2 \tau^3 k^2\skp
 -\frac{2}{3} \tau^4 k^2\sk
 -\frac{8}{3} \tau^4\sk^3
 -\frac{4}{9} \tau^4 k^4\skp
 \bigg],\notag\\
 \mathcal{Z}^T_{14}&=&\int dK\bigg[+2 \tau^2
 -\frac{3}{2} \tau^3 k^2
 -8 \tau^3\sk^2
 +\frac{2}{9} \tau^4 k^4
 +2 \tau^4 k^2\sk^2
 +\frac{8}{3} \tau^4\sk^4
 \bigg],\notag\\
 \mathcal{Z}^T_{15}&=&\int dK\bigg[-4 \tau^2
 + \tau^3 k^2
 +16 \tau^3\sk^2
 +\frac{2}{9} \tau^4 k^4
 -4 \tau^4 k^2\sk^2
 -\frac{16}{3} \tau^4\sk^4
 \bigg],\notag\\
 \mathcal{Z}^T_{16}&=&\int dK\bigg[+4 \tau^2
 -3 \tau^3 k^2
 -16 \tau^3\sk^2
 +\frac{2}{9} \tau^4 k^4
 +4 \tau^4 k^2\sk^2
 +\frac{16}{3} \tau^4\sk^4
 \bigg],\notag\\
 \mathcal{Z}^T_{17}&=&\int dK\bigg[-2 \tau^2
 +\frac{3}{2} \tau^3 k^2
 +2 \tau^3\sk^2
 -\frac{2}{9} \tau^4 k^4
 -\frac{4}{3} \tau^4 k^2\sk^2
 \bigg],\notag\\
 \mathcal{Z}^T_{18}&=&\int dK\bigg[+4 \tau^2
 - \tau^3 k^2
 -4 \tau^3\sk^2
 -\frac{2}{9} \tau^4 k^4
 +\frac{4}{3} \tau^4 k^2\sk^2
 \bigg],\notag\\
 \mathcal{Z}^T_{19}&=&\int dK\bigg[-4 \tau^2
 +3 \tau^3 k^2
 +4 \tau^3\sk^2
 -\frac{2}{9} \tau^4 k^4
 -\frac{4}{3} \tau^4 k^2\sk^2
 \bigg],\notag\\
 \mathcal{Z}^T_{20}&=&\int dK\bigg[+2 \tau^2
 - \tau^3 k^2
 -8 \tau^3\sk^2
 +\frac{1}{9} \tau^4 k^4
 +2 \tau^4 k^2\sk^2
 +\frac{8}{3} \tau^4\sk^4
 \bigg],\notag\\
 \mathcal{Z}^T_{21}&=&\int dK\bigg[+ \tau^2
 -4 \tau^3\sk^2
 -\frac{1}{9} \tau^4 k^4
 + \tau^4 k^2\sk^2
 +\frac{4}{3} \tau^4\sk^4
 \bigg],\notag\\
 \mathcal{Z}^T_{22}&=&\int dK\bigg[-2 \tau^2
 + \tau^3 k^2
 -\frac{1}{9} \tau^4 k^4
 +\frac{2}{3} \tau^4 k^2\sk^2
 \bigg],\notag\\
 \mathcal{Z}^T_{23}&=&\int dK\bigg[- \tau^2
 +\frac{1}{9} \tau^4 k^4
 +\frac{2}{3} \tau^4 k^2\sk^2
 \bigg],\notag\\
 \mathcal{Z}^T_{24}&=&\int dK\bigg[+\frac{4}{3} \tau^3\sk
 - \tau^3 k^2\skp
 -\frac{2}{3} \tau^4 k^2\sk
 +\frac{1}{9} \tau^4 k^4\skp
 +\frac{2}{3} \tau^4 k^4\sk\skp^2
 +\frac{1}{9} \tau^5 k^4\sk
 -\frac{4}{9} \tau^5 k^4\sk^3\skp^2
 \bigg],\notag\\
 \mathcal{Z}^T_{25}&=&\int dK\bigg[-\frac{2}{3} \tau^3\sk
 +\frac{2}{3} \tau^3 k^2\skp
 \bigg],\notag\\
 \mathcal{Z}^T_{26}&=&\int dK\bigg[-\frac{2}{3} \tau^3\sk
 +\frac{2}{3} \tau^4 k^2\sk
 -\frac{4}{9} \tau^4 k^4\sk\skp^2
 -\frac{1}{9} \tau^5 k^4\sk
 +\frac{4}{9} \tau^5 k^4\sk^3\skp^2
 \bigg],\notag\\
 \mathcal{Z}^T_{27}&=&\int dK\bigg[+\frac{1}{3} \tau^3 k^2\skp
 +\frac{2}{3} \tau^4 k^2\sk
 -\frac{1}{9} \tau^4 k^4\skp
 -\frac{2}{3} \tau^4 k^4\sk\skp^2
 -\frac{1}{9} \tau^5 k^4\sk
 +\frac{4}{9} \tau^5 k^4\sk^3\skp^2
 \bigg],\notag\\
 \mathcal{Z}^T_{28}&=&\int dK\bigg[-6 \tau^3\sk
 +10 \tau^3 k^2\sk\skp^2
 +\frac{8}{3} \tau^4 k^2\sk
 +\frac{8}{3} \tau^4\sk^3
 -\frac{1}{9} \tau^4 k^4\skp
 -\frac{10}{9} \tau^4 k^4\sk\skp^2
 -\frac{40}{3} \tau^4 k^2\sk^3\skp^2
 \notag\\
 &&-\frac{2}{9} \tau^5 k^4\sk
 -\frac{2}{3} \tau^5 k^2\sk^3
 +\frac{8}{9} \tau^5 k^4\sk^3\skp^2
 +\frac{8}{3} \tau^5 k^2\sk^5\skp^2
 \bigg],\notag\\
 \mathcal{Z}^T_{29}&=&\int dK\bigg[+\frac{2}{3} \tau^3\sk
 +\frac{2}{3} \tau^4 k^2\sk
 -\frac{4}{9} \tau^4 k^4\sk\skp^2
 -\frac{1}{9} \tau^5 k^4\sk
 +\frac{4}{9} \tau^5 k^4\sk^3\skp^2
 \bigg],\notag\\
 \mathcal{Z}^T_{30}&=&\int dK\bigg[+4 \tau^2
 -\frac{4}{3} \tau^3 k^2
 -4 \tau^3\sk^2
 \bigg],\notag\\
 \mathcal{Z}^T_{31}&=&\int dK\bigg[+\frac{8}{3} \tau^2
 -\frac{2}{3} \tau^3 k^2
 -\frac{4}{3} \tau^3\sk^2
 \bigg],\notag\\
 \mathcal{Z}^T_{32}&=&\int dK\bigg[-3 \tau^2
 +\frac{7}{6} \tau^3 k^2
 +2 \tau^3\sk^2
 \bigg],\notag\\
 \mathcal{Z}^T_{33}&=&\int dK\bigg[-4 \tau^2
 + \tau^3 k^2
 +16 \tau^3\sk^2
 -\frac{8}{3} \tau^4 k^2\sk^2
 -\frac{16}{3} \tau^4\sk^4
 \bigg],\notag\\
 \mathcal{Z}^T_{34}&=&\int dK\bigg[-4 \tau^2
 +2 \tau^3 k^2
 +16 \tau^3\sk^2
 -\frac{8}{3} \tau^4 k^2\sk^2
 -\frac{16}{3} \tau^4\sk^4
 \bigg],\notag\\
 \mathcal{Z}^T_{35}&=&\int dK\bigg[-\frac{4}{3} \tau^3\sk
 + \tau^3 k^2\skp
 +\frac{2}{3} \tau^4 k^2\sk
 -\frac{1}{9} \tau^4 k^4\skp
 -\frac{2}{3} \tau^4 k^4\sk\skp^2
 -\frac{1}{9} \tau^5 k^4\sk
 +\frac{4}{9} \tau^5 k^4\sk^3\skp^2
 \bigg],\notag\\
 \mathcal{Z}^T_{36}&=&\int dK\bigg[-6 \tau^3\sk
 +10 \tau^3 k^2\sk\skp^2
 +\frac{8}{3} \tau^4 k^2\sk
 +\frac{8}{3} \tau^4\sk^3
 -\frac{2}{9} \tau^4 k^4\skp
 -\frac{4}{3} \tau^4 k^4\sk\skp^2
 -\frac{40}{3} \tau^4 k^2\sk^3\skp^2
 \notag\\
 &&-\frac{2}{9} \tau^5 k^4\sk
 -\frac{2}{3} \tau^5 k^2\sk^3
 +\frac{8}{9} \tau^5 k^4\sk^3\skp^2
 +\frac{8}{3} \tau^5 k^2\sk^5\skp^2
 \bigg],\notag\\
 \mathcal{Z}^T_{37}&=&\int dK\bigg[+\frac{1}{3} \tau^3 k^2\skp
 +\frac{2}{3} \tau^4 k^2\sk
 -\frac{1}{9} \tau^4 k^4\skp
 -\frac{2}{3} \tau^4 k^4\sk\skp^2
 -\frac{1}{9} \tau^5 k^4\sk
 +\frac{4}{9} \tau^5 k^4\sk^3\skp^2
 \bigg],\notag\\
 \mathcal{Z}^T_{38}&=&\int dK\bigg[+\frac{2}{3} \tau^3\sk
 -\frac{2}{3} \tau^3 k^2\skp
 \bigg],\notag\\
 \mathcal{Z}^T_{39}&=&\int dK\bigg[-\frac{20}{3} \tau^3\sk
 +\frac{4}{3} \tau^3 k^2\skp
 +2 \tau^4 k^2\sk
 +\frac{8}{3} \tau^4\sk^3
 -\frac{2}{9} \tau^4 k^4\skp
 \bigg],\notag\\
 \mathcal{Z}^T_{40}&=&\int dK\bigg[+6 \tau^3\sk
 +\frac{2}{3} \tau^3 k^2\skp
 -2 \tau^4 k^2\sk
 -\frac{8}{3} \tau^4\sk^3
 -\frac{2}{9} \tau^4 k^4\skp
 \bigg],\notag\\
 \mathcal{Z}^T_{41}&=&\int dK\bigg[+\frac{10}{3} \tau^3\sk
 -\frac{1}{3} \tau^3 k^2\skp
 - \tau^4 k^2\sk
 -\frac{4}{3} \tau^4\sk^3
 +\frac{2}{9} \tau^4 k^4\skp
 \bigg],\notag\\
 \mathcal{Z}^T_{42}&=&\int dK\bigg[+6 \tau^3\sk
 +\frac{2}{3} \tau^3 k^2\skp
 -2 \tau^4 k^2\sk
 -\frac{8}{3} \tau^4\sk^3
 +\frac{2}{9} \tau^4 k^4\skp
 \bigg],\notag\\
 \mathcal{Z}^T_{43}&=&\int dK\bigg[+\frac{10}{3} \tau^3\sk
 +\frac{2}{3} \tau^3 k^2\skp
 - \tau^4 k^2\sk
 -\frac{4}{3} \tau^4\sk^3
 -\frac{2}{9} \tau^4 k^4\skp
 \bigg],\notag\\
 \mathcal{Z}^T_{44}&=&\int dK\bigg[-4 \tau^2
 +2 \tau^3 k^2
 +4 \tau^3\sk^2
 \bigg],\notag\\
 \mathcal{Z}^T_{45}&=&\int dK\bigg[+\frac{8}{3} \tau^2
 -\frac{4}{3} \tau^3 k^2
 -\frac{4}{3} \tau^3\sk^2
 \bigg],\notag\\
 \mathcal{Z}^T_{46}&=&\int dK\bigg[-2 \tau^2
 +\frac{5}{6} \tau^3 k^2
 +2 \tau^3\sk^2
 \bigg],\notag\\
 \mathcal{Z}^T_{47}&=&\int dK\bigg[-\frac{8}{3} \tau^2
 +\frac{5}{3} \tau^3 k^2
 +\frac{4}{3} \tau^3\sk^2
 \bigg],\notag\\
 \mathcal{Z}^T_{48}&=&\int dK\bigg[+6 \tau^2
 -\frac{8}{3} \tau^3 k^2
 -4 \tau^3\sk^2
 \bigg],\notag\\
 \mathcal{Z}^T_{49}&=&\int dK\bigg[+\frac{2}{3} \tau^3\sk
 -\frac{1}{9} \tau^4 k^4\skp
 -\frac{2}{9} \tau^4 k^4\sk\skp^2
 \bigg],\notag\\
 \mathcal{Z}^T_{50}&=&\int dK\bigg[+\frac{2}{3} \tau^2\skp
 +\frac{4}{3} \tau^3\sk
 -\frac{4}{3} \tau^3\sk^2\skp
 +2 \tau^3 k^2\sk\skp^2
 -\frac{1}{9} \tau^4 k^4\skp
 -\frac{2}{9} \tau^4 k^4\sk\skp^2
 \bigg],\notag\\
 \mathcal{Z}^T_{51}&=&\int dK\bigg[+ \tau^3\sk
 -\frac{7}{6} \tau^3 k^2\skp
 +\frac{1}{9} \tau^4 k^4\skp
 +\frac{2}{9} \tau^4 k^4\sk\skp^2
 \bigg],\notag\\
 \mathcal{Z}^T_{52}&=&\int dK\bigg[+\frac{2}{3} \tau^2\skp
 +\frac{2}{3} \tau^3\sk
 +\frac{2}{3} \tau^3 k^2\skp
 -\frac{4}{3} \tau^3\sk^2\skp
 -2 \tau^3 k^2\sk\skp^2
 -\frac{1}{3} \tau^4 k^2\sk
 +\frac{1}{9} \tau^4 k^4\skp
 \notag\\
 &&+\frac{2}{9} \tau^4 k^4\sk\skp^2
 +\frac{4}{3} \tau^4 k^2\sk^3\skp^2
 \bigg],\notag\\
 \mathcal{Z}^T_{53}&=&\int dK\bigg[+\frac{4}{3} \tau^3\sk
 -\frac{5}{3} \tau^3 k^2\skp
 +\frac{1}{9} \tau^4 k^4\skp
 +\frac{2}{9} \tau^4 k^4\sk\skp^2
 \bigg],\notag\\
 \mathcal{Z}^T_{54}&=&\int dK\bigg[+\frac{1}{3} \tau^3 k^2\skp
 -\frac{1}{9} \tau^4 k^4\skp
 -\frac{2}{9} \tau^4 k^4\sk\skp^2
 \bigg],\notag\\
 \mathcal{Z}^T_{55}&=&\int dK\bigg[+\frac{8}{3} \tau^2
 -\frac{1}{3} \tau^3 k^2
 -\frac{4}{3} \tau^3\sk^2
 \bigg],\notag\\
 \mathcal{Z}^T_{56}&=&\int dK\bigg[+2 \tau^2
 \bigg],\notag\\
 \mathcal{Z}^T_{57}&=&\int dK\bigg[-2 \tau^3\sk
 +2 \tau^3 k^2\skp
 -2 \tau^3 k^2\sk\skp^2
 \bigg],\notag\\
 \mathcal{Z}^T_{58}&=&\int dK\bigg[-\frac{16}{3} \tau^2
 +3 \tau^3 k^2
 +4 \tau^3\sk^2
 \bigg],\notag\\
 \mathcal{Z}^T_{59}&=&\int dK\bigg[+\frac{4}{3} \tau^2
 - \tau^3 k^2
 \bigg],\notag\\
 \mathcal{Z}^T_{60}&=&\int dK\bigg[- \tau^2
 \bigg],\notag\\
 \mathcal{Z}^T_{61}&=&\int dK\bigg[+3 \tau^2
 - \tau^3 k^2
 -2 \tau^3\sk^2
 \bigg],\notag\\
 \mathcal{Z}^T_{62}&=&\int dK\bigg[+4 \tau^2\sk
 -2 \tau^3 k^2\sk
 -\frac{8}{3} \tau^3\sk^3
 \bigg],\notag\\
 \mathcal{Z}^T_{63}&=&\int dK\bigg[-4 \tau^2\sk
 +2 \tau^3 k^2\sk
 \bigg],\notag\\
 \mathcal{Z}^T_{64}&=&\int dK\bigg[-6 \tau^2\sk
 +2 \tau^3 k^2\sk
 +4 \tau^3\sk^3
 \bigg],\notag\\
 \mathcal{Z}^T_{65}&=&\int dK\bigg[+2 \tau^2\sk
 -2 \tau^3 k^2\sk
 \bigg],\notag\\
 \mathcal{Z}^T_{66}&=&\int dK\bigg[+\frac{2}{3} \tau^2\skp
 +\frac{1}{3} \tau^3 k^2\skp
 +\frac{2}{9} \tau^3 k^4\skp^3
 -\frac{1}{9} \tau^4 k^4\skp
 -\frac{2}{9} \tau^4 k^4\sk\skp^2
 \bigg],\notag\\
 \mathcal{Z}^T_{67}&=&\int dK\bigg[+\frac{2}{3} \tau^3\sk
 -\frac{11}{18} \tau^3 k^2\skp
 -\frac{11}{9} \tau^3 k^2\sk\skp^2
 -\frac{2}{9} \tau^3 k^4\skp^3
 -\frac{1}{6} \tau^4 k^2\sk
 +\frac{1}{9} \tau^4 k^4\skp
 +\frac{2}{9} \tau^4 k^4\sk\skp^2
 \notag\\
 &&+\frac{2}{3} \tau^4 k^2\sk^3\skp^2
 \bigg],\notag\\
 \mathcal{Z}^T_{68}&=&\int dK\bigg[-\frac{1}{9} \tau^3 k^2\skp
 +\frac{4}{9} \tau^3 k^2\sk\skp^2
 -\frac{2}{9} \tau^3 k^4\skp^3
 +\frac{1}{9} \tau^4 k^4\skp
 +\frac{2}{9} \tau^4 k^4\sk\skp^2
 \bigg],\notag\\
 \mathcal{Z}^T_{69}&=&\int dK\bigg[-\frac{1}{3} \tau^2
 +\frac{1}{3} \tau^3 k^2
 -\frac{1}{18} \tau^4 k^4
 +\frac{2}{9} \tau^4 k^4\sk^2\skp^2
 \bigg],\notag\\
 \mathcal{Z}^T_{70}&=&\int dK\bigg[+\frac{5}{6} \tau^2
 -\frac{1}{2} \tau^2 k^2\skp^2
 -\frac{1}{2} \tau^3 k^2
 -\frac{2}{3} \tau^3\sk^2
 +2 \tau^3 k^2\sk^2\skp^2
 +\frac{1}{18} \tau^4 k^4
 +\frac{1}{6} \tau^4 k^2\sk^2
 \notag\\
 &&-\frac{2}{9} \tau^4 k^4\sk^2\skp^2
 -\frac{2}{3} \tau^4 k^2\sk^4\skp^2
 \bigg],\notag\\
 \mathcal{Z}^T_{71}&=&\int dK\bigg[-\frac{1}{3} \tau^2
 -\frac{1}{3} \tau^3 k^2
 +\frac{1}{18} \tau^4 k^4
 -\frac{2}{9} \tau^4 k^4\sk^2\skp^2
 \bigg],\notag\\
 \mathcal{Z}^T_{72}&=&\int dK\bigg[+\frac{1}{3} \tau^2
 -\frac{1}{3} \tau^3 k^2
 +\frac{1}{18} \tau^4 k^4
 -\frac{2}{9} \tau^4 k^4\sk^2\skp^2
 \bigg],\notag\\
 \mathcal{Z}^T_{73}&=&\int dK\bigg[-\frac{1}{2} \tau^2
 -\frac{1}{2} \tau^2 k^2\skp^2
 +\frac{1}{3} \tau^3 k^2
 -\frac{1}{18} \tau^4 k^4
 +\frac{2}{9} \tau^4 k^4\sk^2\skp^2
 \bigg],\notag\\
 \mathcal{Z}^T_{74}&=&\int dK\bigg[+\frac{1}{3} \tau^2
 +\frac{1}{3} \tau^3 k^2
 -\frac{1}{18} \tau^4 k^4
 +\frac{2}{9} \tau^4 k^4\sk^2\skp^2
 \bigg],\notag\\
 \mathcal{Z}^T_{75}&=&\int dK\bigg[-4 \tau^2
 + \tau^3 k^2
 +4 \tau^3\sk^2
 \bigg],\notag\\
 \mathcal{Z}^T_{76}&=&\int dK\bigg[-4 \tau^2
 +2 \tau^3 k^2
 \bigg],\notag\\
 \mathcal{Z}^T_{77}&=&\int dK\bigg[+2 \tau^2\sk
 \bigg],
 \end{eqnarray}
  \begin{eqnarray}
 \int dK\equiv N_c\int\frac{d^4k}{(2\pi)^4}e^{-\tau(k^2+\sk^2)}\int^\infty_{\frac{1}{\Lambda^2}}\frac{d\tau}{\tau}.
 \end{eqnarray}

 \section{$\mathcal{Z}^T_n$ and $K^T_n$'s relations}\label{ZKrs}
 This appendix list relations between our coefficients, $\mathcal{Z}^T_n$,
 and  those in Ref.\cite{tensor1}, $K^T_n$.
 Some coefficients vanish, because of the new relations in Appendix \ref{newrelation}.
  \begin{eqnarray}
 K^T_{1}&=&+\frac{1}{16b_0}\mathcal{Z}^T_{1} -\frac{1}{16b_0}\mathcal{Z}^T_{27} -\frac{1}{16b_0}\mathcal{Z}^T_{28} +\frac{1}{16b_0}\mathcal{Z}^T_{29} -\frac{1}{32b_0}\mathcal{Z}^T_{36} +\frac{1}{32b_0}\mathcal{Z}^T_{38} +\frac{1}{8b_0}\mathcal{Z}^T_{41} +\frac{1}{16b_0}\mathcal{Z}^T_{42} \notag\\
 K^T_{2}&=&+\frac{1}{16b_0}\mathcal{Z}^T_{4} +\frac{1}{8b_0}\mathcal{Z}^T_{27} +\frac{1}{16b_0}\mathcal{Z}^T_{36} -\frac{1}{16b_0}\mathcal{Z}^T_{38} +\frac{1}{8b_0}\mathcal{Z}^T_{40} +\frac{1}{8b_0}\mathcal{Z}^T_{43} +\frac{1}{16b_0}\mathcal{Z}^T_{57} +\frac{1}{4b_0}\mathcal{Z}^T_{67} +\frac{1}{8b_0}\mathcal{Z}^T_{68} \notag\\
 K^T_{3}&=&+\frac{1}{16b_0}\mathcal{Z}^T_{3} -\frac{1}{8b_0}\mathcal{Z}^T_{28} +\frac{1}{8b_0}\mathcal{Z}^T_{29} -\frac{1}{8b_0}\mathcal{Z}^T_{39} +\frac{1}{16b_0}\mathcal{Z}^T_{57} +\frac{1}{4b_0}\mathcal{Z}^T_{67} +\frac{1}{8b_0}\mathcal{Z}^T_{68} \notag\\
 K^T_{4}&=&+\frac{1}{16b_0}\mathcal{Z}^T_{2} +\frac{1}{8b_0}\mathcal{Z}^T_{28} -\frac{1}{8b_0}\mathcal{Z}^T_{29} +\frac{1}{16b_0}\mathcal{Z}^T_{39} -\frac{1}{16b_0}\mathcal{Z}^T_{40} -\frac{1}{16b_0}\mathcal{Z}^T_{42} -\frac{1}{16b_0}\mathcal{Z}^T_{57} -\frac{1}{4b_0}\mathcal{Z}^T_{67} -\frac{1}{8b_0}\mathcal{Z}^T_{68} \notag\\
 K^T_{5}&=&0\notag\\
 K^T_{6}&=&0\notag\\
 K^T_{7}&=&0\notag\\
 K^T_{8}&=&0\notag\\
 K^T_{9}&=&+\frac{1}{4b_0^2}\mathcal{Z}^T_{14} +\frac{1}{4b_0^2}\mathcal{Z}^T_{17} +\frac{1}{8b_0^2}\mathcal{Z}^T_{18} +\frac{1}{8b_0^2}\mathcal{Z}^T_{19} -\frac{1}{8b_0^2}\mathcal{Z}^T_{31} \notag\\
 K^T_{10}&=&+\frac{1}{4b_0^2}\mathcal{Z}^T_{21} -\frac{1}{4b_0^2}\mathcal{Z}^T_{22} +\frac{1}{4b_0^2}\mathcal{Z}^T_{23} -\frac{1}{8b_0^2}\mathcal{Z}^T_{31} \notag\\
 K^T_{11}&=&+\frac{1}{4b_0^2}\mathcal{Z}^T_{16} -\frac{1}{4b_0^2}\mathcal{Z}^T_{18} +\frac{1}{2b_0^2}\mathcal{Z}^T_{31} +\frac{1}{4b_0^2}\mathcal{Z}^T_{58} +\frac{1}{4b_0^2}\mathcal{Z}^T_{59} +\frac{1}{2b_0^2}\mathcal{Z}^T_{71} +\frac{1}{2b_0^2}\mathcal{Z}^T_{72} \notag\\
 K^T_{12}&=&+\frac{1}{4b_0^2}\mathcal{Z}^T_{15} -\frac{1}{4b_0^2}\mathcal{Z}^T_{19} -\frac{1}{4b_0^2}\mathcal{Z}^T_{58} -\frac{1}{4b_0^2}\mathcal{Z}^T_{59} -\frac{1}{2b_0^2}\mathcal{Z}^T_{71} -\frac{1}{2b_0^2}\mathcal{Z}^T_{72} \notag\\
 K^T_{13}&=&-\frac{1}{4b_0^2}\mathcal{Z}^T_{20} +\frac{1}{4b_0^2}\mathcal{Z}^T_{22} +\frac{1}{4b_0^2}\mathcal{Z}^T_{31} \notag\\
 K^T_{14}&=&0\notag\\
 K^T_{15}&=&0\notag\\
 K^T_{16}&=&0\notag\\
 K^T_{17}&=&0\notag\\
 K^T_{18}&=&0\notag\\
 K^T_{19}&=&0\notag\\
 K^T_{20}&=&0\notag\\
 K^T_{21}&=&0\notag\\
 K^T_{22}&=&0\notag\\
 K^T_{31}&=&+\frac{1}{4B_0b_0^2}\mathcal{Z}^T_{64} +\frac{1}{4B_0b_0^2}\mathcal{Z}^T_{65} \notag\\
 K^T_{32}&=&+\frac{1}{8b_0^2}\mathcal{Z}^T_{31} +\frac{1}{4b_0^2}\mathcal{Z}^T_{32} -\frac{1}{2b_0^2}\mathcal{Z}^T_{46} -\frac{1}{8b_0^2}\mathcal{Z}^T_{47} -\frac{1}{8b_0^2}\mathcal{Z}^T_{48} +\frac{1}{2B_0b_0^2}\mathcal{Z}^T_{77} \notag\\
 K^T_{33}&=&0\notag\\
 K^T_{34}&=&0\notag\\
 K^T_{35}&=&-\frac{1}{8N_fb_0^2}\mathcal{Z}^T_{31} -\frac{1}{4N_fb_0^2}\mathcal{Z}^T_{32} +\frac{1}{2N_fb_0^2}\mathcal{Z}^T_{46} +\frac{1}{8N_fb_0^2}\mathcal{Z}^T_{47} +\frac{1}{8N_fb_0^2}\mathcal{Z}^T_{48} \notag\\
 K^T_{36}&=&0\notag\\
 K^T_{37}&=&0\notag\\
 K^T_{38}&=&0\notag\\
 K^T_{39}&=&+\frac{1}{16b_0}\mathcal{Z}^T_{26} -\frac{1}{16b_0}\mathcal{Z}^T_{28} +\frac{1}{16b_0}\mathcal{Z}^T_{29} +\frac{1}{16B_0b_0}\mathcal{Z}^T_{33} -\frac{1}{16B_0b_0}\mathcal{Z}^T_{44} +\frac{1}{8B_0b_0}\mathcal{Z}^T_{61} -\frac{1}{16b_0}\mathcal{Z}^T_{66} +\frac{1}{8b_0}\mathcal{Z}^T_{67} +\frac{1}{16b_0}\mathcal{Z}^T_{68} \notag\\
 K^T_{40}&=&+\frac{1}{8b_0}\mathcal{Z}^T_{26} -\frac{1}{8b_0}\mathcal{Z}^T_{28} +\frac{1}{8b_0}\mathcal{Z}^T_{29} +\frac{1}{16B_0b_0}\mathcal{Z}^T_{34} -\frac{1}{8B_0b_0}\mathcal{Z}^T_{44} -\frac{1}{8b_0}\mathcal{Z}^T_{66} +\frac{1}{4b_0}\mathcal{Z}^T_{67} +\frac{1}{8b_0}\mathcal{Z}^T_{68} \notag\\
 K^T_{41}&=&0\notag\\
 K^T_{42}&=&0\notag\\
 K^T_{43}&=&+\frac{1}{16b_0}\mathcal{Z}^T_{10} -\frac{1}{16B_0b_0}\mathcal{Z}^T_{45} +\frac{1}{16b_0}\mathcal{Z}^T_{50} +\frac{1}{8b_0}\mathcal{Z}^T_{51} -\frac{1}{8b_0}\mathcal{Z}^T_{54} +\frac{1}{8B_0b_0}\mathcal{Z}^T_{60} -\frac{1}{16B_0b_0}\mathcal{Z}^T_{75} \notag\\
 K^T_{44}&=&+\frac{1}{16b_0}\mathcal{Z}^T_{13} -\frac{1}{8b_0}\mathcal{Z}^T_{45} -\frac{1}{8b_0}\mathcal{Z}^T_{50} -\frac{1}{16B_0b_0}\mathcal{Z}^T_{76} \notag\\
 K^T_{45}&=&-\frac{1}{8N_fb_0}\mathcal{Z}^T_{10} -\frac{1}{16N_fb_0}\mathcal{Z}^T_{13} -\frac{1}{4N_fb_0}\mathcal{Z}^T_{51} +\frac{1}{4N_fb_0}\mathcal{Z}^T_{54} \notag\\
 K^T_{46}&=&0\notag\\
 K^T_{47}&=&-\frac{1}{16b_0}\mathcal{Z}^T_{7} -\frac{1}{16b_0}\mathcal{Z}^T_{9} +\frac{1}{16b_0}\mathcal{Z}^T_{49} +\frac{1}{8b_0}\mathcal{Z}^T_{50} -\frac{1}{16b_0}\mathcal{Z}^T_{52} -\frac{1}{16b_0}\mathcal{Z}^T_{53} \notag\\
 K^T_{48}&=&-\frac{1}{16b_0}\mathcal{Z}^T_{6} -\frac{1}{16b_0}\mathcal{Z}^T_{8} -\frac{1}{16b_0}\mathcal{Z}^T_{50} -\frac{1}{8b_0}\mathcal{Z}^T_{51} +\frac{1}{16b_0}\mathcal{Z}^T_{52} +\frac{1}{16b_0}\mathcal{Z}^T_{53} \notag\\
 K^T_{49}&=&-\frac{1}{16b_0}\mathcal{Z}^T_{11} -\frac{1}{16b_0}\mathcal{Z}^T_{12} +\frac{1}{16b_0}\mathcal{Z}^T_{49} -\frac{1}{16b_0}\mathcal{Z}^T_{50} -\frac{1}{8b_0}\mathcal{Z}^T_{51} \notag\\
 K^T_{50}&=&0\notag\\
 K^T_{51}&=&-\frac{1}{4b_0^2}\mathcal{Z}^T_{31} +\frac{1}{b_0^2}\mathcal{Z}^T_{70} -\frac{1}{2b_0^2}\mathcal{Z}^T_{72} +\frac{1}{b_0^2}\mathcal{Z}^T_{73} +\frac{1}{2b_0^2}\mathcal{Z}^T_{74} \notag\\
 K^T_{52}&=&+\frac{1}{b_0^2}\mathcal{Z}^T_{31} -\frac{1}{b_0^2}\mathcal{Z}^T_{69} +\frac{1}{b_0^2}\mathcal{Z}^T_{71} +\frac{1}{b_0^2}\mathcal{Z}^T_{72} -\frac{1}{b_0^2}\mathcal{Z}^T_{74} \notag\\
 K^T_{54}&=&0\notag\\
 K^T_{55}&=&0\notag\\
 K^T_{57}&=&-\frac{1}{32b_0}\mathcal{Z}^T_{36} +\frac{1}{32b_0}\mathcal{Z}^T_{38} +\frac{1}{8b_0}\mathcal{Z}^T_{41} \notag\\
 K^T_{58}&=&+\frac{1}{16b_0}\mathcal{Z}^T_{36} -\frac{1}{16b_0}\mathcal{Z}^T_{38} +\frac{1}{8b_0}\mathcal{Z}^T_{43} \notag\\
 K^T_{59}&=&-\frac{1}{8b_0}\mathcal{Z}^T_{27} -\frac{1}{8b_0}\mathcal{Z}^T_{40} -\frac{1}{8b_0}\mathcal{Z}^T_{57} -\frac{1}{2b_0}\mathcal{Z}^T_{67} -\frac{1}{4b_0}\mathcal{Z}^T_{68} \notag\\
 K^T_{60}&=&-\frac{1}{8b_0}\mathcal{Z}^T_{28} +\frac{1}{8b_0}\mathcal{Z}^T_{29} -\frac{1}{8b_0}\mathcal{Z}^T_{39} +\frac{1}{8b_0}\mathcal{Z}^T_{57} +\frac{1}{2b_0}\mathcal{Z}^T_{67} +\frac{1}{4b_0}\mathcal{Z}^T_{68} \notag\\
 K^T_{61}&=&+\frac{1}{8b_0}\mathcal{Z}^T_{27} +\frac{1}{8b_0}\mathcal{Z}^T_{28} -\frac{1}{8b_0}\mathcal{Z}^T_{29} -\frac{1}{8b_0}\mathcal{Z}^T_{42} \notag\\
 K^T_{62}&=&0\notag\\
 K^T_{63}&=&0\notag\\
 K^T_{64}&=&0\notag\\
 K^T_{65}&=&0\notag\\
 K^T_{66}&=&0\notag\\
 K^T_{67}&=&0\notag\\
 K^T_{68}&=&0\notag\\
 K^T_{69}&=&+\frac{1}{16b_0}\mathcal{Z}^T_{5} -\frac{1}{16b_0}\mathcal{Z}^T_{49} +\frac{1}{16b_0}\mathcal{Z}^T_{52} +\frac{1}{16b_0}\mathcal{Z}^T_{53} \notag\\
 K^T_{70}&=&+\frac{1}{16b_0}\mathcal{Z}^T_{6} -\frac{1}{16b_0}\mathcal{Z}^T_{8} +\frac{1}{16b_0}\mathcal{Z}^T_{50} +\frac{1}{8b_0}\mathcal{Z}^T_{51} -\frac{3}{16b_0}\mathcal{Z}^T_{52} -\frac{1}{16b_0}\mathcal{Z}^T_{53} \notag\\
 K^T_{71}&=&+\frac{1}{16b_0}\mathcal{Z}^T_{7} -\frac{1}{16b_0}\mathcal{Z}^T_{9} +\frac{1}{16b_0}\mathcal{Z}^T_{49} +\frac{1}{16b_0}\mathcal{Z}^T_{52} -\frac{1}{16b_0}\mathcal{Z}^T_{53} \notag\\
 K^T_{72}&=&-\frac{1}{16b_0}\mathcal{Z}^T_{11} +\frac{1}{16b_0}\mathcal{Z}^T_{12} -\frac{1}{16b_0}\mathcal{Z}^T_{49} -\frac{1}{16b_0}\mathcal{Z}^T_{50} -\frac{1}{8b_0}\mathcal{Z}^T_{51} +\frac{1}{8b_0}\mathcal{Z}^T_{52} +\frac{1}{8b_0}\mathcal{Z}^T_{53} \notag\\
 K^T_{73}&=&0\notag\\
 K^T_{74}&=&-\frac{1}{16B_0b_0}\mathcal{Z}^T_{55} +\frac{1}{8B_0b_0}\mathcal{Z}^T_{61} \notag\\
 K^T_{75}&=&+\frac{1}{16b_0}\mathcal{Z}^T_{52} +\frac{1}{16b_0}\mathcal{Z}^T_{53} -\frac{1}{8b_0}\mathcal{Z}^T_{54} +\frac{1}{16B_0b_0}\mathcal{Z}^T_{55} +\frac{1}{8B_0b_0}\mathcal{Z}^T_{60} \notag\\
 K^T_{76}&=&+\frac{1}{16b_0}\mathcal{Z}^T_{26} +\frac{1}{16b_0}\mathcal{Z}^T_{27} -\frac{1}{16b_0}\mathcal{Z}^T_{35} +\frac{1}{16B_0b_0}\mathcal{Z}^T_{55} -\frac{1}{16B_0b_0}\mathcal{Z}^T_{56} -\frac{1}{16b_0}\mathcal{Z}^T_{66} +\frac{1}{8b_0}\mathcal{Z}^T_{67} +\frac{1}{16b_0}\mathcal{Z}^T_{68} \notag\\
 K^T_{77}&=&0\notag\\
 K^T_{78}&=&0\notag\\
 K^T_{79}&=&0\notag\\
 K^T_{80}&=&-\frac{1}{8N_fb_0}\mathcal{Z}^T_{52} -\frac{1}{8N_fb_0}\mathcal{Z}^T_{53} +\frac{1}{4N_fb_0}\mathcal{Z}^T_{54} \notag\\
 K^T_{82}&=&0\notag\\
 K^T_{84}&=&-\frac{1}{8b_0}\mathcal{Z}^T_{27} -\frac{1}{8b_0}\mathcal{Z}^T_{28} +\frac{1}{8b_0}\mathcal{Z}^T_{29} +\frac{1}{8b_0}\mathcal{Z}^T_{36} -\frac{1}{8b_0}\mathcal{Z}^T_{37} \notag\\
 K^T_{85}&=&+\frac{1}{4b_0}\mathcal{Z}^T_{57} +\frac{1}{b_0}\mathcal{Z}^T_{67} +\frac{1}{2b_0}\mathcal{Z}^T_{68} \notag\\
 K^T_{86}&=&-\frac{1}{4b_0}\mathcal{Z}^T_{52} \notag\\
 K^T_{87}&=&0\notag\\
 K^T_{88}&=&+\frac{1}{b_0^2}\mathcal{Z}^T_{62} +\frac{1}{b_0^2}\mathcal{Z}^T_{63} \notag\\
 K^T_{90}&=&-\frac{1}{2b_0^2}\mathcal{Z}^T_{31} -\frac{1}{2b_0^2}\mathcal{Z}^T_{58} -\frac{1}{2b_0^2}\mathcal{Z}^T_{59} -\frac{1}{b_0^2}\mathcal{Z}^T_{71} -\frac{1}{b_0^2}\mathcal{Z}^T_{72} \notag\\
 K^T_{92}&=&0\notag\\
 K^T_{94}&=&-\frac{1}{2b_0}\mathcal{Z}^T_{66} +\frac{1}{b_0}\mathcal{Z}^T_{67} +\frac{1}{2b_0}\mathcal{Z}^T_{68} \notag\\
 K^T_{95}&=&+\frac{1}{4b_0}\mathcal{Z}^T_{28} \notag\\
 K^T_{96}&=&+\frac{1}{8b_0}\mathcal{Z}^T_{24} +\frac{1}{8b_0}\mathcal{Z}^T_{25} -\frac{1}{8b_0}\mathcal{Z}^T_{27} -\frac{1}{16b_0}\mathcal{Z}^T_{36} +\frac{1}{16b_0}\mathcal{Z}^T_{38} +\frac{1}{8b_0}\mathcal{Z}^T_{66} -\frac{1}{4b_0}\mathcal{Z}^T_{67} -\frac{1}{8b_0}\mathcal{Z}^T_{68} \notag\\
 K^T_{97}&=&+\frac{1}{8b_0}\mathcal{Z}^T_{24} -\frac{1}{8b_0}\mathcal{Z}^T_{25} -\frac{1}{8b_0}\mathcal{Z}^T_{27} +\frac{1}{16b_0}\mathcal{Z}^T_{36} -\frac{1}{16b_0}\mathcal{Z}^T_{38} +\frac{1}{8b_0}\mathcal{Z}^T_{66} -\frac{1}{4b_0}\mathcal{Z}^T_{67} -\frac{1}{8b_0}\mathcal{Z}^T_{68} \notag\\
 K^T_{98}&=&+\frac{1}{4b_0}\mathcal{Z}^T_{49} -\frac{1}{4b_0}\mathcal{Z}^T_{52} -\frac{1}{4b_0}\mathcal{Z}^T_{53} \notag\\
 K^T_{99}&=&+\frac{1}{4b_0}\mathcal{Z}^T_{51} -\frac{1}{8b_0}\mathcal{Z}^T_{52} -\frac{1}{8b_0}\mathcal{Z}^T_{53} \notag\\
 K^T_{100}&=&-\frac{1}{4b_0}\mathcal{Z}^T_{50} \notag\\
 K^T_{101}&=&0\notag\\
 K^T_{102}&=&0\notag\\
 K^T_{103}&=&0\notag\\
 K^T_{105}&=&-\frac{1}{2b_0^2}\mathcal{Z}^T_{30} +\frac{1}{2b_0^2}\mathcal{Z}^T_{31} \notag\\
 K^T_{106}&=&0\notag\\
 K^T_{107}&=&0\notag\\
 K^T_{108}&=&0\notag\\
 K^T_{112}&=&-\frac{1}{8B_0b_0}\mathcal{Z}^T_{45} \notag\\
 K^T_{113}&=&+\frac{1}{8b_0}\mathcal{Z}^T_{26} -\frac{1}{8b_0}\mathcal{Z}^T_{28} +\frac{1}{8b_0}\mathcal{Z}^T_{29} -\frac{1}{8B_0b_0}\mathcal{Z}^T_{44} -\frac{1}{8b_0}\mathcal{Z}^T_{66} +\frac{1}{4b_0}\mathcal{Z}^T_{67} +\frac{1}{8b_0}\mathcal{Z}^T_{68} \notag\\
 K^T_{114}&=&+\frac{1}{8b_0}\mathcal{Z}^T_{27} +\frac{1}{8b_0}\mathcal{Z}^T_{28} -\frac{1}{8b_0}\mathcal{Z}^T_{29} +\frac{1}{8b_0}\mathcal{Z}^T_{36} +\frac{1}{8b_0}\mathcal{Z}^T_{37} \notag\\
 K^T_{117}&=&0\notag\\
 K^T_{118}&=&-\frac{2}{b_0^2}\mathcal{Z}^T_{31} \notag\\
 K^T_{119}&=&+\frac{1}{b_0^2}\mathcal{Z}^T_{31} -\frac{1}{b_0^2}\mathcal{Z}^T_{47} +\frac{1}{b_0^2}\mathcal{Z}^T_{48} \notag\\
 K^T_{120}&=&+\frac{1}{4B_0b_0}\mathcal{Z}^T_{55} \notag\\
 \end{eqnarray}

 \section{$\widetilde{K}^W_{t,n}$ coefficients}\label{tkwcs}

 \begin{eqnarray}
 \widetilde{K}^{T,W}_{1}&=&N_C\int\frac{d^4k}{(2\pi)^4}\bigg[-\frac{1}{9} \skp X^2
 +\frac{1}{8} \sk\skp^2 X^2
 +\frac{1}{36} \sk X^3
 +\frac{1}{9} \sk^2\skp X^3
 -\frac{19}{24} \sk^3\skp^2 X^3
 -\frac{5}{6} \sk^3 X^4
 +\frac{4}{3} \sk^5\skp^2 X^4
 \notag\\
 &&+\frac{3}{2} \sk^5 X^5
 -\frac{2}{3} \sk^7\skp^2 X^5
 \bigg],\notag\\
 \widetilde{K}^{T,W}_{2}&=&N_C\int\frac{d^4k}{(2\pi)^4}\bigg[-\frac{1}{18} \skp^3 X
 -\frac{43}{72} \skp X^2
 -\frac{1}{6} \sk\skp^2 X^2
 +\frac{1}{9} \sk^2\skp^3 X^2
 +\frac{55}{72} \sk X^3
 +\frac{37}{72} \sk^2\skp X^3
 +3 \sk^3\skp^2 X^3
 \notag\\
 &&-\frac{1}{18} \sk^4\skp^3 X^3
 +\frac{5}{24} \sk^3 X^4
 -\frac{11}{2} \sk^5\skp^2 X^4
 +\frac{8}{3} \sk^7\skp^2 X^5
 \bigg],\notag\\
 \widetilde{K}^{T,W}_{3}&=&N_C\int\frac{d^4k}{(2\pi)^4}\bigg[-\frac{1}{18} \skp^3 X
 -\frac{11}{72} \skp X^2
 +\frac{1}{12} \sk\skp^2 X^2
 +\frac{1}{9} \sk^2\skp^3 X^2
 +\frac{11}{72} \sk X^3
 +\frac{5}{72} \sk^2\skp X^3
 +\frac{17}{12} \sk^3\skp^2 X^3
 \notag\\
 &&-\frac{1}{18} \sk^4\skp^3 X^3
 -\frac{11}{24} \sk^3 X^4
 -\frac{17}{6} \sk^5\skp^2 X^4
 + \sk^5 X^5
 +\frac{4}{3} \sk^7\skp^2 X^5
 \bigg],\notag\\
 \widetilde{K}^{T,W}_{4}&=&N_C\int\frac{d^4k}{(2\pi)^4}\bigg[+\frac{1}{18} \skp^3 X
 +\frac{5}{12} \skp X^2
 -\frac{1}{12} \sk\skp^2 X^2
 -\frac{1}{9} \sk^2\skp^3 X^2
 -\frac{5}{12} \sk X^3
 -\frac{1}{3} \sk^2\skp X^3
 -\frac{17}{12} \sk^3\skp^2 X^3
 \notag\\
 &&+\frac{1}{18} \sk^4\skp^3 X^3
 +\frac{1}{12} \sk^3 X^4
 +\frac{17}{6} \sk^5\skp^2 X^4
 -\frac{2}{3} \sk^5 X^5
 -\frac{4}{3} \sk^7\skp^2 X^5
 \bigg],\notag\\
 \widetilde{K}^{T,W}_{5}&=&N_C\int\frac{d^4k}{(2\pi)^4}\bigg[-\frac{2}{3} \sk^2\skp^2 X^2
 -\frac{11}{6} \sk^2 X^3
 +\frac{4}{3} \sk^4\skp^2 X^3
 +\frac{17}{6} \sk^4 X^4
 -\frac{2}{3} \sk^6\skp^2 X^4
 \bigg],\notag\\
 \widetilde{K}^{T,W}_{6}&=&N_C\int\frac{d^4k}{(2\pi)^4}\bigg[-\frac{2}{3} \sk^2\skp^2 X^2
 -\frac{1}{3} \sk^2 X^3
 +\frac{4}{3} \sk^4\skp^2 X^3
 +\frac{5}{6} \sk^4 X^4
 -\frac{2}{3} \sk^6\skp^2 X^4
 \bigg],\notag\\
 \widetilde{K}^{T,W}_{7}&=&N_C\int\frac{d^4k}{(2\pi)^4}\bigg[+\frac{4}{3} \sk^2\skp^2 X^2
 -\frac{4}{3} \sk^2 X^3
 -\frac{8}{3} \sk^4\skp^2 X^3
 +\frac{13}{3} \sk^4 X^4
 +\frac{4}{3} \sk^6\skp^2 X^4
 \bigg],\notag\\
 \widetilde{K}^{T,W}_{8}&=&N_C\int\frac{d^4k}{(2\pi)^4}\bigg[+\frac{4}{3} \sk^2\skp^2 X^2
 +\frac{2}{3} \sk^2 X^3
 -\frac{8}{3} \sk^4\skp^2 X^3
 -\frac{11}{3} \sk^4 X^4
 +\frac{4}{3} \sk^6\skp^2 X^4
 \bigg],\notag\\
 \widetilde{K}^{T,W}_{9}&=&N_C\int\frac{d^4k}{(2\pi)^4}\bigg[+\frac{4}{3} \sk^2\skp^2 X^2
 +\frac{5}{3} \sk^2 X^3
 -\frac{8}{3} \sk^4\skp^2 X^3
 -\frac{8}{3} \sk^4 X^4
 +\frac{4}{3} \sk^6\skp^2 X^4
 \bigg],\notag\\
 \widetilde{K}^{T,W}_{10}&=&N_C\int\frac{d^4k}{(2\pi)^4}\bigg[- \sk X^2
 +2 \sk^3 X^3
 \bigg],\notag\\
 \widetilde{K}^{T,W}_{11}&=&N_C\int\frac{d^4k}{(2\pi)^4}\bigg[+\frac{1}{2} \sk X^2
 \bigg],\notag\\
 \widetilde{K}^{T,W}_{12}&=&N_C\int\frac{d^4k}{(2\pi)^4}\bigg[+\frac{3}{8} \sk^2 X^3
 \bigg],\notag\\
 \widetilde{K}^{T,W}_{13}&=&N_C\int\frac{d^4k}{(2\pi)^4}\bigg[-\frac{3}{8} \sk^2 X^3
 + \sk^4 X^4
 \bigg],\notag\\
 \widetilde{K}^{T,W}_{14}&=&N_C\int\frac{d^4k}{(2\pi)^4}\bigg[-\frac{1}{4} \sk^2 X^3
 + \sk^4 X^4
 \bigg],\notag\\
 \widetilde{K}^{T,W}_{15}&=&N_C\int\frac{d^4k}{(2\pi)^4}\bigg[+\frac{3}{8} \sk^2 X^3
 \bigg],\notag\\
 \widetilde{K}^{T,W}_{16}&=&N_C\int\frac{d^4k}{(2\pi)^4}\bigg[-\frac{1}{4} \sk^2 X^3
 \bigg],\notag\\
 \widetilde{K}^{T,W}_{17}&=&N_C\int\frac{d^4k}{(2\pi)^4}\bigg[-\frac{7}{72} \skp X^2
 +\frac{1}{3} \sk\skp^2 X^2
 +\frac{7}{72} \sk X^3
 -\frac{11}{72} \sk^2\skp X^3
 -\frac{1}{6} \sk^3\skp^2 X^3
 +\frac{17}{24} \sk^3 X^4
 -\frac{1}{6} \sk^5\skp^2 X^4
 \bigg],\notag\\
 \widetilde{K}^{T,W}_{18}&=&N_C\int\frac{d^4k}{(2\pi)^4}\bigg[+\frac{25}{144} \skp X^2
 -\frac{1}{2} \sk\skp^2 X^2
 -\frac{25}{144} \sk X^3
 -\frac{7}{144} \sk^2\skp X^3
 +\frac{3}{4} \sk^3\skp^2 X^3
 -\frac{23}{48} \sk^3 X^4
 -\frac{1}{4} \sk^5\skp^2 X^4
 \bigg],\notag\\
 \widetilde{K}^{T,W}_{19}&=&N_C\int\frac{d^4k}{(2\pi)^4}\bigg[+\frac{1}{36} \skp X^2
 -\frac{1}{3} \sk\skp^2 X^2
 -\frac{1}{36} \sk X^3
 +\frac{7}{72} \sk^2\skp X^3
 +\frac{5}{12} \sk^3\skp^2 X^3
 -\frac{5}{12} \sk^3 X^4
 -\frac{1}{12} \sk^5\skp^2 X^4
 \bigg],\notag\\
 \widetilde{K}^{T,W}_{20}&=&N_C\int\frac{d^4k}{(2\pi)^4}\bigg[-\frac{1}{8} \skp X^2
 +\frac{5}{12} \sk\skp^2 X^2
 +\frac{1}{8} \sk X^3
 +\frac{1}{8} \sk^2\skp X^3
 -\frac{5}{6} \sk^3\skp^2 X^3
 +\frac{1}{8} \sk^3 X^4
 +\frac{5}{12} \sk^5\skp^2 X^4
 \bigg],\notag\\
 \widetilde{K}^{T,W}_{21}&=&N_C\int\frac{d^4k}{(2\pi)^4}\bigg[-\frac{17}{72} \skp X^2
 +\frac{5}{12} \sk\skp^2 X^2
 +\frac{17}{72} \sk X^3
 +\frac{1}{9} \sk^2\skp X^3
 -\frac{7}{12} \sk^3\skp^2 X^3
 +\frac{7}{24} \sk^3 X^4
 +\frac{1}{6} \sk^5\skp^2 X^4
 \bigg],\notag\\
 \widetilde{K}^{T,W}_{22}&=&N_C\int\frac{d^4k}{(2\pi)^4}\bigg[- \skp^2 X
 +5 \sk^2\skp^2 X^2
 -8 \sk^4\skp^2 X^3
 - \sk^4 X^4
 +4 \sk^6\skp^2 X^4
 \bigg],\notag\\
 \widetilde{K}^{T,W}_{23}&=&0,\notag\\
 \widetilde{K}^{T,W}_{24}&=&N_C\int\frac{d^4k}{(2\pi)^4}\bigg[-\frac{7}{144} \skp X^2
 +\frac{3}{8} \sk\skp^2 X^2
 -\frac{29}{144} \sk X^3
 +\frac{7}{144} \sk^2\skp X^3
 -\frac{47}{24} \sk^3\skp^2 X^3
 +\frac{3}{16} \sk^3 X^4
 +\frac{35}{12} \sk^5\skp^2 X^4
 \notag\\
 &&+\frac{1}{3} \sk^5 X^5
 -\frac{4}{3} \sk^7\skp^2 X^5
 \bigg],\notag\\
 \widetilde{K}^{T,W}_{25}&=&N_C\int\frac{d^4k}{(2\pi)^4}\bigg[-\frac{1}{24} \skp X^2
 -\frac{3}{4} \sk\skp^2 X^2
 +\frac{1}{8} \sk X^3
 +\frac{1}{24} \sk^2\skp X^3
 +\frac{47}{12} \sk^3\skp^2 X^3
 +\frac{11}{24} \sk^3 X^4
 -\frac{35}{6} \sk^5\skp^2 X^4
 \notag\\
 &&-\frac{2}{3} \sk^5 X^5
 +\frac{8}{3} \sk^7\skp^2 X^5
 \bigg],\notag\\
 \widetilde{K}^{T,W}_{26}&=&N_C\int\frac{d^4k}{(2\pi)^4}\bigg[-\frac{1}{9} \skp^3 X
 +\frac{49}{72} \skp X^2
 -\frac{5}{6} \sk\skp^2 X^2
 +\frac{2}{9} \sk^2\skp^3 X^2
 -\frac{43}{72} \sk X^3
 -\frac{43}{72} \sk^2\skp X^3
 +\frac{5}{3} \sk^3\skp^2 X^3
 \notag\\
 &&-\frac{1}{9} \sk^4\skp^3 X^3
 -\frac{5}{24} \sk^3 X^4
 -\frac{5}{6} \sk^5\skp^2 X^4
 \bigg],\notag\\
 \widetilde{K}^{T,W}_{27}&=&N_C\int\frac{d^4k}{(2\pi)^4}\bigg[+\frac{1}{9} \skp^3 X
 -\frac{5}{18} \skp X^2
 +\frac{1}{3} \sk\skp^2 X^2
 -\frac{2}{9} \sk^2\skp^3 X^2
 +\frac{7}{36} \sk X^3
 +\frac{7}{36} \sk^2\skp X^3
 +\frac{2}{3} \sk^3\skp^2 X^3
 \notag\\
 &&+\frac{1}{9} \sk^4\skp^3 X^3
 +\frac{7}{12} \sk^3 X^4
 -\frac{7}{3} \sk^5\skp^2 X^4
 -\frac{1}{3} \sk^5 X^5
 +\frac{4}{3} \sk^7\skp^2 X^5
 \bigg],\notag\\
 \widetilde{K}^{T,W}_{28}&=&N_C\int\frac{d^4k}{(2\pi)^4}\bigg[+\frac{1}{8} \skp X^2
 +\frac{1}{2} \sk\skp^2 X^2
 +\frac{5}{24} \sk X^3
 -\frac{1}{8} \sk^2\skp X^3
 -\frac{7}{3} \sk^3\skp^2 X^3
 -\frac{19}{24} \sk^3 X^4
 +\frac{19}{6} \sk^5\skp^2 X^4
 \notag\\
 &&+\frac{1}{3} \sk^5 X^5
 -\frac{4}{3} \sk^7\skp^2 X^5
 \bigg],\notag\\
 \widetilde{K}^{T,W}_{29}&=&N_C\int\frac{d^4k}{(2\pi)^4}\bigg[+\frac{1}{72} \skp X^2
 -\frac{11}{36} \sk X^3
 +\frac{1}{9} \sk^2\skp X^3
 -\frac{1}{4} \sk^3\skp^2 X^3
 +\frac{1}{6} \sk^3 X^4
 +\frac{1}{4} \sk^5\skp^2 X^4
 \bigg],\notag\\
 \widetilde{K}^{T,W}_{30}&=&N_C\int\frac{d^4k}{(2\pi)^4}\bigg[+\frac{37}{72} \skp X^2
 -\frac{3}{4} \sk\skp^2 X^2
 -\frac{13}{72} \sk X^3
 -\frac{55}{72} \sk^2\skp X^3
 +2 \sk^3\skp^2 X^3
 +\frac{1}{24} \sk^3 X^4
 -\frac{5}{4} \sk^5\skp^2 X^4
 \bigg],\notag\\
 \widetilde{K}^{T,W}_{31}&=&N_C\int\frac{d^4k}{(2\pi)^4}\bigg[-\frac{3}{8} \skp X^2
 +\frac{2}{3} \sk\skp^2 X^2
 +\frac{5}{24} \sk X^3
 +\frac{1}{4} \sk^2\skp X^3
 -\frac{13}{12} \sk^3\skp^2 X^3
 +\frac{3}{8} \sk^3 X^4
 +\frac{5}{12} \sk^5\skp^2 X^4
 \bigg],\notag\\
 \widetilde{K}^{T,W}_{32}&=&N_C\int\frac{d^4k}{(2\pi)^4}\bigg[-\frac{5}{36} \skp X^2
 +\frac{5}{12} \sk\skp^2 X^2
 -\frac{13}{36} \sk X^3
 +\frac{19}{72} \sk^2\skp X^3
 -\frac{13}{12} \sk^3\skp^2 X^3
 +\frac{7}{12} \sk^3 X^4
 +\frac{2}{3} \sk^5\skp^2 X^4
 \bigg],\notag\\
 \widetilde{K}^{T,W}_{33}&=&N_C\int\frac{d^4k}{(2\pi)^4}\bigg[+\frac{1}{8} \sk^2 X^3
 \bigg],\notag\\
 \widetilde{K}^{T,W}_{34}&=&N_C\int\frac{d^4k}{(2\pi)^4}\bigg[+\frac{1}{8} \sk^2 X^3
 \bigg],\notag\\
 \widetilde{K}^{T,W}_{35}&=&N_C\int\frac{d^4k}{(2\pi)^4}\bigg[+\frac{1}{8} \sk^2 X^3
 \bigg],\notag\\
 \widetilde{K}^{T,W}_{36}&=&N_C\int\frac{d^4k}{(2\pi)^4}\bigg[-\frac{1}{36} \skp X^2
 +\frac{1}{12} \sk\skp^2 X^2
 +\frac{11}{72} \sk X^3
 -\frac{7}{72} \sk^2\skp X^3
 +\frac{1}{12} \sk^3\skp^2 X^3
 +\frac{1}{24} \sk^3 X^4
 -\frac{1}{6} \sk^5\skp^2 X^4
 \bigg],\notag\\
 \widetilde{K}^{T,W}_{37}&=&N_C\int\frac{d^4k}{(2\pi)^4}\bigg[+\frac{4}{3} \sk^2\skp^2 X^2
 +\frac{5}{6} \sk^2 X^3
 -\frac{8}{3} \sk^4\skp^2 X^3
 -\frac{1}{3} \sk^4 X^4
 +\frac{4}{3} \sk^6\skp^2 X^4
 \bigg],\notag\\
 \widetilde{K}^{T,W}_{38}&=&N_C\int\frac{d^4k}{(2\pi)^4}\bigg[+\frac{1}{3} \skp X^2
 -\frac{5}{2} \sk\skp^2 X^2
 +\frac{1}{6} \sk X^3
 -\frac{1}{3} \sk^2\skp X^3
 +\frac{25}{2} \sk^3\skp^2 X^3
 + \sk^3 X^4
 -18 \sk^5\skp^2 X^4
 \notag\\
 &&-2 \sk^5 X^5
 +8 \sk^7\skp^2 X^5
 \bigg],\notag\\
 \widetilde{K}^{T,W}_{39}&=&N_C\int\frac{d^4k}{(2\pi)^4}\bigg[+\frac{2}{3} \skp^3 X
 -\frac{1}{2} \skp X^2
 + \sk\skp^2 X^2
 -\frac{4}{3} \sk^2\skp^3 X^2
 +\frac{1}{2} \sk X^3
 +\frac{1}{2} \sk^2\skp X^3
 -3 \sk^3\skp^2 X^3
 \notag\\
 &&+\frac{2}{3} \sk^4\skp^3 X^3
 -\frac{1}{2} \sk^3 X^4
 +2 \sk^5\skp^2 X^4
 \bigg],\notag\\
 \widetilde{K}^{T,W}_{40}&=&N_C\int\frac{d^4k}{(2\pi)^4}\bigg[+\frac{23}{36} \skp X^2
 -\frac{2}{3} \sk\skp^2 X^2
 -\frac{5}{36} \sk X^3
 -\frac{41}{36} \sk^2\skp X^3
 +\frac{7}{3} \sk^3\skp^2 X^3
 +\frac{5}{12} \sk^3 X^4
 -\frac{5}{3} \sk^5\skp^2 X^4
 \bigg],\notag\\
 \widetilde{K}^{T,W}_{41}&=&N_C\int\frac{d^4k}{(2\pi)^4}\bigg[+\frac{20}{3} \sk^3 X^3
 \bigg],\notag\\
 \widetilde{K}^{T,W}_{42}&=&N_C\int\frac{d^4k}{(2\pi)^4}\bigg[-2 \sk^2 X^3
 \bigg],\notag\\
 \widetilde{K}^{T,W}_{43}&=&N_C\int\frac{d^4k}{(2\pi)^4}\bigg[+\frac{8}{9} \skp^3 X
 +\frac{8}{9} \skp X^2
 -\frac{2}{3} \sk\skp^2 X^2
 -\frac{16}{9} \sk^2\skp^3 X^2
 -\frac{5}{9} \sk X^3
 -\frac{5}{9} \sk^2\skp X^3
 -\frac{2}{3} \sk^3\skp^2 X^3
 \notag\\
 &&+\frac{8}{9} \sk^4\skp^3 X^3
 -\frac{1}{3} \sk^3 X^4
 +\frac{4}{3} \sk^5\skp^2 X^4
 \bigg],\notag\\
 \widetilde{K}^{T,W}_{44}&=&N_C\int\frac{d^4k}{(2\pi)^4}\bigg[+\frac{2}{9} \skp^3 X
 +\frac{7}{24} \skp X^2
 -\frac{7}{6} \sk\skp^2 X^2
 -\frac{4}{9} \sk^2\skp^3 X^2
 -\frac{7}{24} \sk X^3
 -\frac{5}{24} \sk^2\skp X^3
 +4 \sk^3\skp^2 X^3
 \notag\\
 &&+\frac{2}{9} \sk^4\skp^3 X^3
 +\frac{5}{24} \sk^3 X^4
 -\frac{11}{2} \sk^5\skp^2 X^4
 -\frac{2}{3} \sk^5 X^5
 +\frac{8}{3} \sk^7\skp^2 X^5
 \bigg],\notag\\
 \widetilde{K}^{T,W}_{45}&=&N_C\int\frac{d^4k}{(2\pi)^4}\bigg[+\frac{5}{24} \skp X^2
 +\frac{1}{12} \sk\skp^2 X^2
 -\frac{3}{8} \sk X^3
 -\frac{5}{24} \sk^2\skp X^3
 -\frac{5}{4} \sk^3\skp^2 X^3
 -\frac{1}{24} \sk^3 X^4
 +\frac{5}{2} \sk^5\skp^2 X^4
 \notag\\
 &&+\frac{1}{3} \sk^5 X^5
 -\frac{4}{3} \sk^7\skp^2 X^5
 \bigg],\notag\\
 \widetilde{K}^{T,W}_{46}&=&N_C\int\frac{d^4k}{(2\pi)^4}\bigg[+\frac{37}{72} \skp X^2
 -\frac{17}{12} \sk\skp^2 X^2
 -\frac{25}{72} \sk X^3
 -\frac{37}{72} \sk^2\skp X^3
 +\frac{79}{12} \sk^3\skp^2 X^3
 +\frac{13}{24} \sk^3 X^4
 -\frac{55}{6} \sk^5\skp^2 X^4
 \notag\\
 &&- \sk^5 X^5
 +4 \sk^7\skp^2 X^5
 \bigg],\notag\\
 \widetilde{K}^{T,W}_{47}&=&N_C\int\frac{d^4k}{(2\pi)^4}\bigg[+\frac{2}{9} \skp^3 X
 +\frac{7}{18} \skp X^2
 +\frac{1}{6} \sk\skp^2 X^2
 -\frac{4}{9} \sk^2\skp^3 X^2
 -\frac{17}{36} \sk X^3
 -\frac{11}{36} \sk^2\skp X^3
 -\frac{13}{6} \sk^3\skp^2 X^3
 \notag\\
 &&+\frac{2}{9} \sk^4\skp^3 X^3
 -\frac{1}{4} \sk^3 X^4
 +\frac{10}{3} \sk^5\skp^2 X^4
 +\frac{1}{3} \sk^5 X^5
 -\frac{4}{3} \sk^7\skp^2 X^5
 \bigg],\notag\\
 \widetilde{K}^{T,W}_{48}&=&N_C\int\frac{d^4k}{(2\pi)^4}\bigg[+\frac{2}{9} \skp^3 X
 +\frac{11}{72} \skp X^2
 +\frac{13}{6} \sk\skp^2 X^2
 -\frac{4}{9} \sk^2\skp^3 X^2
 -\frac{47}{72} \sk X^3
 -\frac{5}{72} \sk^2\skp X^3
 -\frac{37}{3} \sk^3\skp^2 X^3
 \notag\\
 &&+\frac{2}{9} \sk^4\skp^3 X^3
 -\frac{25}{24} \sk^3 X^4
 +\frac{109}{6} \sk^5\skp^2 X^4
 +2 \sk^5 X^5
 -8 \sk^7\skp^2 X^5
 \bigg],\notag\\
 \widetilde{K}^{T,W}_{49}&=&N_C\int\frac{d^4k}{(2\pi)^4}\bigg[+\frac{1}{12} \skp X^2
 +\frac{1}{4} \sk X^3
 -\frac{7}{12} \sk^2\skp X^3
 + \sk^3\skp^2 X^3
 +\frac{1}{4} \sk^3 X^4
 - \sk^5\skp^2 X^4
 \bigg],\notag\\
 \widetilde{K}^{T,W}_{50}&=&N_C\int\frac{d^4k}{(2\pi)^4}\bigg[+\frac{5}{12} \skp X^2
 -\frac{1}{2} \sk\skp^2 X^2
 -\frac{1}{4} \sk X^3
 -\frac{2}{3} \sk^2\skp X^3
 +\frac{3}{2} \sk^3\skp^2 X^3
 +\frac{1}{4} \sk^3 X^4
 - \sk^5\skp^2 X^4
 \bigg],\notag\\
 \widetilde{K}^{T,W}_{51}&=&N_C\int\frac{d^4k}{(2\pi)^4}\bigg[+\frac{1}{36} \skp X^2
 +\frac{2}{3} \sk\skp^2 X^2
 +\frac{23}{36} \sk X^3
 -\frac{19}{36} \sk^2\skp X^3
 -\frac{1}{3} \sk^3\skp^2 X^3
 +\frac{1}{12} \sk^3 X^4
 -\frac{1}{3} \sk^5\skp^2 X^4
 \bigg],\notag\\
 \widetilde{K}^{T,W}_{52}&=&N_C\int\frac{d^4k}{(2\pi)^4}\bigg[+\frac{8}{3} \sk^2\skp^2 X^2
 +\frac{8}{3} \sk^2 X^3
 -\frac{16}{3} \sk^4\skp^2 X^3
 -\frac{2}{3} \sk^4 X^4
 +\frac{8}{3} \sk^6\skp^2 X^4
 \bigg],\notag\\
 \widetilde{K}^{T,W}_{53}&=&N_C\int\frac{d^4k}{(2\pi)^4}\bigg[-\frac{8}{3} \sk^2\skp^2 X^2
 -\frac{2}{3} \sk^2 X^3
 +\frac{16}{3} \sk^4\skp^2 X^3
 +\frac{2}{3} \sk^4 X^4
 -\frac{8}{3} \sk^6\skp^2 X^4
 \bigg],\notag\\
 \widetilde{K}^{T,W}_{54}&=&0,\notag\\
 \widetilde{K}^{T,W}_{55}&=&N_C\int\frac{d^4k}{(2\pi)^4}\bigg[+\frac{1}{2} \sk^2 X^3
 \bigg],\notag\\
 \widetilde{K}^{T,W}_{56}&=&N_C\int\frac{d^4k}{(2\pi)^4}\bigg[-\frac{1}{18} \skp X^2
 -\frac{1}{2} \sk\skp^2 X^2
 +\frac{2}{9} \sk X^3
 +\frac{1}{18} \sk^2\skp X^3
 +\frac{19}{6} \sk^3\skp^2 X^3
 +\frac{1}{6} \sk^3 X^4
 -\frac{16}{3} \sk^5\skp^2 X^4
 \notag\\
 &&-\frac{2}{3} \sk^5 X^5
 +\frac{8}{3} \sk^7\skp^2 X^5
 \bigg],\\
 X&\equiv&\frac{1}{k^2+\Sigma_k^2}.
 \end{eqnarray}

 \providecommand{\noopsort}[1]{}\providecommand{\singleletter}[1]{#1}%

\end{document}